\documentclass[5p,sort&compress]{elsarticle}
\usepackage{multicol}
\usepackage{graphicx}
\usepackage{dcolumn}% Align table columns on decimal point
\usepackage{bm}% bold math
\usepackage{amsmath}
\usepackage{amssymb}
\usepackage{gensymb}
\usepackage[version=3]{mhchem}
\usepackage[usenames, dvipsnames]{xcolor}
\usepackage{textcmds}
\usepackage{longtable}
\AtBeginDocument{\usepackage{booktabs}}
\usepackage{url}
\usepackage{textcomp}
\usepackage[colorlinks=true,
            linkcolor=blue,
            citecolor=magenta,
            urlcolor=blue]{hyperref}
\usepackage{tabularx}
\newcommand\setrow[1]{\gdef\rowmac{#1}#1\ignorespaces}
\newcommand\clearrow{\global\let\rowmac\relax}
\clearrow
\begin{document}

\title{Tailoring properties of Heusler alloys by elemental substitution and electron counting: (Co$_{2-\alpha}$Mn$_\alpha$)FeGe, Co$_2$(Fe$_{1-\beta}$Mn$_{\beta}$)Ge, and (Co$_{2-\alpha}$Fe$_\alpha$)MnGe}

\author[dpua]{Shambhu KC\corref{cor1}}
\ead{skc@crimson.ua.edu}
\cortext[cor1]{corresponding author}

\author[dpua]{R. Mahat}
\author[dpua]{S. Budhathoki}
\author[dpua]{S. Regmi}
\author[dfmc]{J. Y. Law}
\author[dfmc]{V. Franco}
\author[dpua]{W. H. Butler}
\author[dcua]{A. Gupta}
\author[dpua]{A. Hauser}
\author[dpua]{P. LeClair\corref{cor2}}
\ead{pleclair@ua.edu}
\cortext[cor2]{corresponding author}

\address[dpua]{Department of Physics and Astronomy, University of Alabama, Tuscaloosa, AL 35401, USA}
\address[dfmc]{Departmento de F\'{i}sica de la Materia Condensada ICMSE-CSIC, Universidad de Sevilla, Sevilla 41080, Spain}
\address[dcua]{Department of Chemistry, University of Alabama, Tuscaloosa, AL 35487, USA}

\date{\today}% It is always \today, today,

\begin{abstract}
Rational material design by elemental substitution is useful in tailoring materials to have desirable properties. Here we consider three non-equivalent substitutional series based on Co$_2$FeGe, viz; (Co$_{2-\alpha}$Mn$_\alpha$)FeGe, Co$_2$(Fe$_{1-\beta}$Mn$_{\beta}$)Ge, (Co$_{2-\alpha}$Fe$_\alpha$)MnGe ($0\!\le\!\alpha\!\le\!2, 0\!\le\!\beta\!\le\!1$), and study how material properties evolve with the interchange of Mn, Fe, and Co in  Co$_2$FeGe. In all three schemes, single-phase compounds can be obtained over a wide range of compositions: $0.125 < \alpha < 1.375 $ for (Co$_{2-\alpha}$Mn$_{\alpha}$)FeGe, $0 \!\le\! \beta \!\le\! 1$ for Co$_2$(Fe$_{1-\beta}$Mn$_{\beta}$)Ge, and $0 \!<\! \alpha \!<\! 1.50$ for (Co$_{2-\alpha}$Fe$_\alpha$)MnGe. All the single-phase compounds crystallise in fcc structure with chemical ordering consistent with the ``4-2'' rule of Butler et al. The compounds are soft ferromagnets with low temperature saturation magnetisation agreeing with the Slater-Pauling rule. Very high Curie temperatures are measured, with values up to 1000 K for lower Mn concentrations. First principle calculations indicate, in the most stable atomic configuration, Mn prefers sharing sublattice with Ge, also consistent with the 4-2 rule. The calculations further predict half-metallic behaviour for (Co$_{1.625}$Mn$_{0.375}$)FeGe, while finding other compositions to be nearly half-metallic. Upon comparing the results of the three series, it is found that single-phase alloys occur for a specific range of valence electrons per unit cell ($\sim\!28.5\!-\!29.75$), and that even for multi-phase samples the structural, magnetic, and electronic properties depend primarily on the number of valence electrons and not on the specific substitution scheme employed. 
\end{abstract}

%Hence, with the identification of technologically promising materials and by providing a detailed and careful investigation of a very broad set of substitution schemes, we hope that this paper may serve as an incitement for discovering additional functional materials.

\maketitle

%\tableofcontents

\section{\label{sec:level1}Introduction}
The Heusler alloys and their derivatives, due to their wide range of properties, have attracted special attention over the past few decades as being potentially ideal materials for spintronic devices\cite{chambers2003new,bai2012data,felser2013spintronics,elphick2020heusler,hirohata2020review}. The predicted potential for Heusler alloys to be half-metallic ferromagnets\cite{de1983new,galanakis2002origin,kandpal2006covalent,ishida1995search,galanakis2002slater,ozdogan2006search,lee2005magnetic,wurmehl2005geometric,dai2009new,rani2019experimental,kundu2017new,bainsla2014high} is one of the key reasons. Half-metallic ferromagnets are exotic materials in the sense that they behave like a metal in one spin-channel (spin-up) whereas other channel (spin-down) exhibits insulating behaviour (i.e., gap at the Fermi level)\cite{wolf2001spintronics,de1983new}, making them a material of interest in magnetic tunnel junctions (MTJs), spin filters, and spin valves\cite{bai2012data}. 
Numerous candidate half-metallic Heusler alloys have been identified previously based on theoretical calculations, but experimental proof of a half metallic state in these materials is generally still lacking. While part of the problem is the lack of a {\em widely available} and reliable method for directly measuring spin polarisation\cite{PhysRevB.42.1533,PhysRevLett.59.2788}, there are a number of factors leading to reduction of spin polarisation in real materials: for example, intrinsic point defects and interfacial states\cite{picozzi2007polarization}, thermal degradation of spin polarisation\cite{levzaic2006thermal,nawa2020temperature}, and surface effects\cite{hashemifar2005preserving,galanakis2002surface}. Extensive attempts have been made to understand the effects and to increase the spin-polarisation\cite{picozzi2007polarization,levzaic2006thermal,nawa2020temperature,hashemifar2005preserving,galanakis2002surface,zheng2020significant,10.1063/1.4944719,moges2016enhanced}, and finding materials with 100\% spin-polarisation at room temperature still remains an open challenge. Finding novel candidate half-metals with very high saturation magnetisation (M$_s$) and very high Curie temperature (T$_\text{C}$) should help make half-metallic behaviour robust against temperature variation, and this is one of the key objectives of this work. 

A high degree of tunability and compatibility by elemental substitution is another key reason Heusler alloys receive a great deal of attention. Elemental substitution has been proven to be successful in many ways: stabilization of a single-phase material (\textit{here by single phase we mean a single and spatially uniform atomic composition})\cite{shambhu2019tunable}, \sloppy fine tuning of magnetic properties\cite{umetsu2005half}, improvement of thermoelectric properties\cite{lue2002thermoelectric} and shifting of Fermi level (E$_F$) towards the middle of the gap thereby increasing the spin-polarisation and hence making the half-metallic character robust\cite{fecher2005design,balke2006properties,karthik2007effect,rajanikanth2008enhancement,varaprasad2012spin}. In a Heusler alloy of the X$_2$YZ type (also known as the ``full Heusler''), where X and Y are transition metal elements and Z is a main group element, one can introduce substitution of new elements in three basic schemes: (X$_{2-\alpha}$X'$_\alpha$)YZ, X$_2$(Y$_{1-\beta}$Y'$_\beta$)Z, and X$_2$Y(Z$_{1-\gamma}$Z'$_{\gamma}$). The last two schemes have been studied extensively in the literature\cite{goripati2013current,umetsu2012magnetic,ozdougan2007influence}. More complicated scenarios are of course possible, for instance, we have also recently reported on a more unusual X$_2$Y$_{1+x}$Z$_{x}$ scheme and found that Co$_2$Fe$_{1.25}$Ge$_{0.75}$ is a stable single-phase material with enhanced Curie temperature ($\sim\!1135\,$K) and magnetic moment ($\approx\!6.7\,\mu_\text{B}$/fu).\cite{shambhu2022maxmoment} However, the first mechanism, substitution on the X site, has received comparatively little attention. In our recent work\cite{shambhu2019tunable}, we reported that a new substitution scheme, which for brevity we call double sublattice substitution, can be realised if the X' element is less valence than the Y element. This substitution scheme has been further investigated in more recent studies\cite{mahat2021structural,mahat2021possible}. There are, however, few if any experimental reports where attempts have been made to understand how the properties evolve by comparing the different substitution schemes to each other within the same larger materials system. This type of study seems very useful, since the magnetic and electronic properties depend strongly on the hybridisation of nearest neighbours atomic orbitals. It is also useful to check what significance the number of valence electrons (N$_v$) has in dictating alloy properties, because it has been reported that Heusler alloys, in particular Co$_2$-based Heuslers, demonstrate linear variation of saturation magnetisation (M$_s$) and Curie temperature (T$_\text{C}$) with N$_v$\cite{fecher2006slater}. Hence, we aim at expanding the range of Heusler alloys available for study, by comparative investigation of different substitutions schemes and seek to determine the role of N$_v$ in dictating various material characteristics which will assist in more directed materials discovery.

%%%%%%%%%%%%%%%
\begin{figure}[ht]
\centering
\includegraphics[width = 0.95\columnwidth]{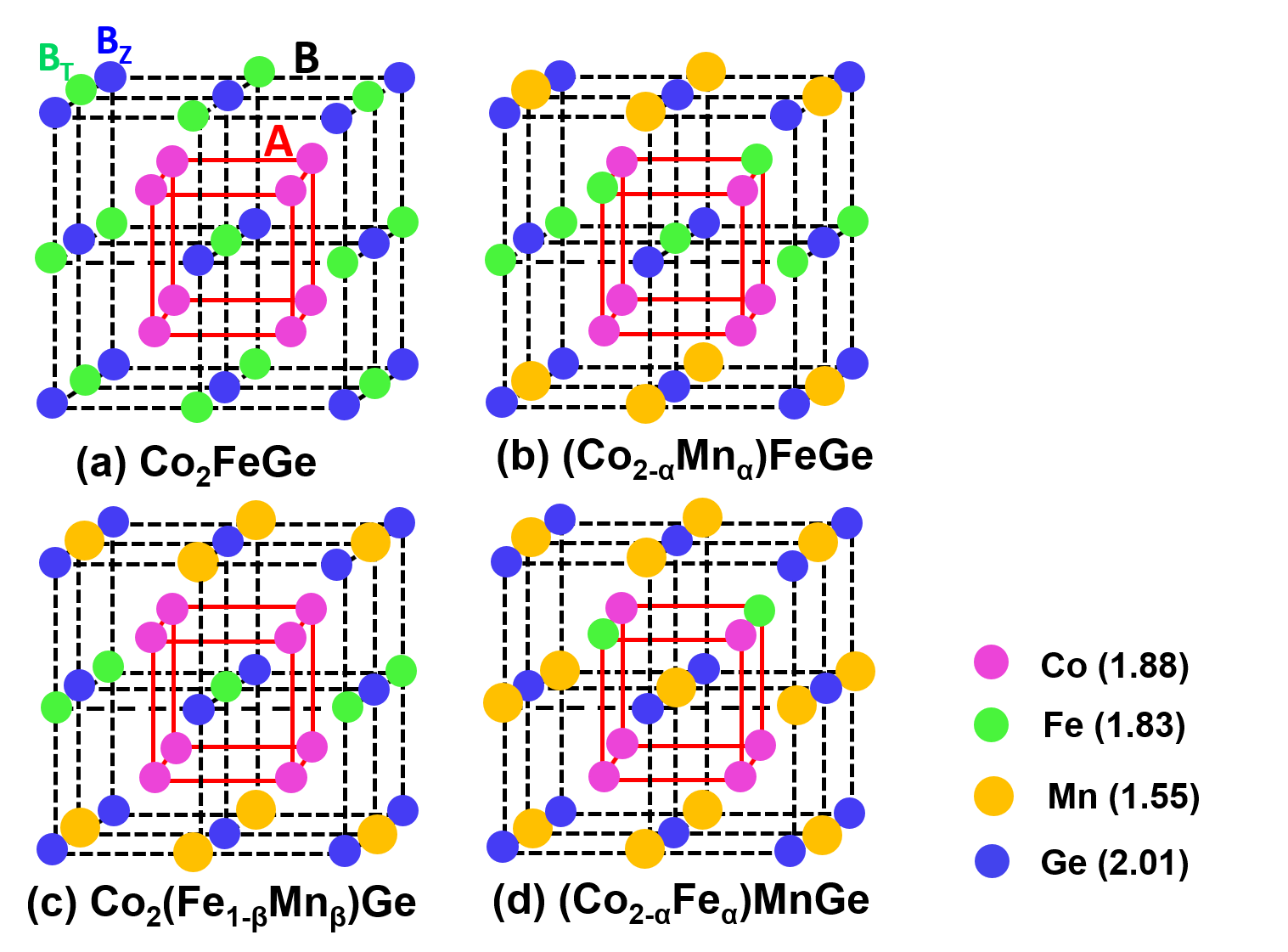}
\caption{(color online) The arrangement of atoms in the unit cell (based on site preference or Burch rule) for different substitutional series with ($\alpha, \beta$) = 0.5. Case (b) is double sublattice substitution, where Mn occupies Fe sites and displaces Fe atoms to vacant Co sites. The number in the parenthesis next to element symbol represents the electronegativity of the element\cite{li2006estimation}.}
\label{fig:unit_cell}
\end{figure}
%%%%%%%%%%%%%%%%

Co$_2$FeGe is chosen as a candidate alloy and Mn is introduced to generate different substituent schemes. There are several reasons for choosing Co$_2$FeGe and Mn. {\em First}, Co$_2$FeGe itself, for all its promise on first sight, does not generate a single-phase specimen\cite{shambhu2019tunable,varaprasad2012spin}, hence one can study the ability to stabilise a single phase alloy and over what substitutional range it may be achieved. The reason for the instability can be understood on the basis of Butler et al's ``rule of 4 and 2.''\cite{butler2011rational,adams1979} They argue on very general grounds that in B2 and related alloy systems, the minority spin channel will try to have a mean occupancy of 3 electrons to open a ``Slater-Pauling gap'' and lower its energy. The system can enhance the gap and further lower its energy even more by increasing the difference in the number of electrons for the X and Y/Z atoms on the A and B sublattices (in an L2$_1$ X$_2$YZ system; in an X$_a$ inverse Heusler system it would rather be X1/Y and X2/Z). Given the desire for an average of 3, and the fact that there are only 5 d states, a compromise is reached in having 4 minority electrons on X atoms and 2 on Y and Z atoms (or 4 on X1/Y and 2 on X2/Z in an inverse Heusler). This immediately constrains possible full Heusler half metals just by considering that the 4d and early 3d transition metals are very hard to polarise magnetically - it is no longer so surprising that most successful systems use Cr, Mn, Fe, and Co. It also explains why Si and Ge are useful Z elements - they can have 2 minority electrons without needing to carry a magnetic moment or transfer charge from another site. In the present case, applying the rule to Co$_2$FeGe would give an electron configuration of
\begin{align*}
\begin{matrix} \\ \uparrow \\ \downarrow \end{matrix}
\begin{matrix} \text{Co} \\ 5 \\4 \end{matrix}\,
\begin{matrix} \text{Fe} \\ 6 \\2 \end{matrix}\,
\begin{matrix} \text{Co} \\ 5 \\4 \end{matrix}\,
\begin{matrix} \text{Ge} \\ 2 \\2 \end{matrix}
\end{align*}

As explained by Galanakis and Dederichs\cite{galanakis2002slater}, the system can make use of lower-lying Ge $sp$ states to accommodate $d$ electrons that would normally be forced to occupy higher energy bands. That means that while Co nominally only has 8.3 $d$ electrons, it does not need to occupy its higher energy $s$ states and can use the lower energy Ge $sp$ states instead, which helps explain the stability of the Co$_2$-based compounds. (We will continue use integer number of $d$ electrons for simplicity.)  However, in this case Fe nominally takes on a moment of $4\,\mu_{\text{B}}$, nearly 2 more majority electrons than it would support in bcc Fe. Between Co and Fe, Ge needs to accommodate $\sim$2.5-3 electrons. The fact that we find Co$_2$FeGe is not a single phase material\cite{shambhu2019tunable,mahat2021possible} makes us speculate that rather than transfer significant charge between sites to reduce the Fe moment and accommodate Co, requiring higher energy Ge states, the system prefers to simply adopt a slightly different composition to reduce the number of valence electrons per unit cell. The fact that substitution with an earlier transition metal atom and thereby reducing the electron count stabilises the parent compound\cite{shambhu2019tunable,mahat2021possible} also points in this direction. This suggests that electron count is not just coincidentally important in determining the magnetic moment via the Slater-Pauling rule, but also the key to stabilising single-phase materials. The Z element plays a key but subtle role in this: it makes available low-lying $sp$ states for the X and Y elements,\cite{galanakis2002slater} and in doing so to stabilises the electron configurations that Butler et al\cite{butler2011rational} argue make the full Heusler phase stable. In that vein, the Z atom also allows the system to adopt a more open bcc-like structure where next-nearest-neighbour interactions are suppressed, so that the ``gap theorem'' of Butler et al\cite{butler2011rational} is applicable. All three substitution schemes proposed here reduce the valence electron count and make the 4-2 rule easier to follow, and therefore should all work to stabilise the parent phase, which is indeed what we observe.

{\em Second}, the Fermi level (E$_F$) of a hypothetical Co$_2$FeGe Heusler alloy in the L$2_1$ structure falls on the lower part of the conduction band in the minority channel, leading to non-half-metallic character. Hence, the thought that elemental substitution might shift E$_F$ towards the band gap and tune the system toward half-metallic behaviour.\cite{shambhu2019tunable,mahat2021possible} This is also effectively an argument for reducing the valence electron count. {\em Third}, Mn, Fe, and Co are adjacent in the periodic table, so they have nearly the same number of valence electrons (N$_v$) and are highest among the ${3d}$-transition elements. As obtaining high values of M$_s$ and T$_\text{C}$ is the main goal, and both seem to scale with N$_v$\cite{galanakis2002slater,galanakis2004appearance,fecher2006slater}, the ability to tune N$_v$ toward a maximum while maintaining phase stability should allow us to maximize both M$_s$ and T$_\text{C}$.\cite{shambhu2022maxmoment} 

One can generate three non-equivalent substitution schemes after interchanging Mn, Fe, and Co in the parent Co$_2$FeGe alloy, viz; \sloppy(Co$_{2-\alpha}$Mn$_\alpha$)FeGe, Co$_2$(Fe$_{1-\beta}$Mn$_{\beta}$)Ge, (Co$_{2-\alpha}$Fe$_\alpha$)MnGe ($0 \le \alpha \le 2, 0 \le \beta \le 1$). These three schemes can be further differentiated taking into account of expected atomic order in the unit cell. The unit cell of a full Heusler alloy, Co$_2$FeGe, contains 16 atoms, occupying four different sites. Referring to space group 216, these would be the Wyckoff sites 4a (0,0,0), 4b ($\frac{1}{2},\frac{1}{2},\frac{1}{2}$), 4c ($\frac{1}{4},\frac{1}{4},\frac{1}{4}$), and 4d ($\frac{3}{4},\frac{3}{4},\frac{3}{4}$). In the case that the 4c and 4d sites are occupied by the same atoms, they would be equivalent to the 8c (($\frac{1}{4},\frac{1}{4},\frac{1}{4}$) Wyckoff sites in space group 225 generating an L2$_1$ structure. To aid in explanation, we describe the unit cell as a two sublattice system; sublattice A formed by 4c and 4d (or 8c) sites, and sublattice B formed by 4a and 4b sites, as shown in Figure \ref{fig:unit_cell}(a). Note that in some cases, each site is considered as a separate sublattice (and hence a unit cell contains four sublattices), whereas in the present case we take the view that there are two sublattices which each contain two sites. In this scheme, nearest neighbours of atoms on one sublattice are the atoms on the next sublattice, and the atoms on the same sublattice are next nearest neighbours of each other. We label the two sites within sublattice B as B$_T$ and B$_Z$, based on the types of element occupying these sites -- T for the transition metal element on the 4b Wyckoff site and Z for the main group element on the 4a Wyckoff site. In the context of the ``4-2'' rule, this means sublattice A atoms should adopt a configuration with 4 minority electrons, and sublattice B atoms should adopt a configuration with 2 minority electrons.
 
While there is no definite rule determining chemical order in the unit cell, a common method of assigning preference is called the site-preference rule, backed by the experimental work of Burch (thus also known as the Burch rule)\cite{burch1974hyperfine}. The Burch rule states that the lower valence transition metal element shares sublattice with main group element( in this case B sublattice). A possible justification of this rule is, since the atoms on the B sublattice form rock salt-type lattice, the two sites should be occupied by the least and most electronegative elements (hence sometimes the rule is also called electronegativity rule), which, in general, are main group and less valence transition metal elements respectively\cite{graf2011simple}. This rule is further justified by the fact that many full Heusler alloys with that have the X atom valence lower than the Y atom valence are found to be more stable in the so-called inverse Heusler structure (space group 216) rather than the regular L2$_1$ structure (space group 225). \cite{ma2018computational, kreiner2014new} 

In fact, this is perhaps even more clearly understood via the ``4-2'' rule of Butler et al\cite{butler2011rational}. In a X$_2$YZ full Heusler system, the X atoms want to achieve 4 minority electrons and the Y/Z atoms 2 minority electrons. If the substituent atom cannot adopt a configuration with 4 minority electrons - certainly true for the early transition metals like Ti - but the $Y$ atom can, they will likely switch places. For the later transition metals, if an atom cannot adopt a configuration with 2 minority electrons, it will be forced to stay on the X site if the full Heusler phase to be maintained. In the present case, for Co to have 2 minority valence electrons would require 7 majority valence electrons and a moment of $5\,\mu_{\text{B}}$/atom, which is not feasible to put it mildly. Thus, Co should stay in the A sublattice. Along those lines, for an inverse Heusler structure with 2 Co atoms, both Co atoms could not be on the X sites as one of them would need to adopt the implausible 2 minority electron configuration. We {\em speculate} that if the 4-2 rule cannot be reasonably satisfied in the L2$_1$ or X$_a$ structures, the system is likely to form a different phase or decompose into multiple phases. The 4-2 rule also implies that X-Y (DO$_3$) disorder can be controlled in the L2$_1$ phase by ensuring the X atom cannot adopt a configuration with 2 minority electrons (e.g., Co) or that the Y/Z atoms cannot adopt a configuration with 4 minority electrons (e.g., Ti). The possibility of Y-Z B2-type disorder remains, but this does not preclude the formation of a half-metallic state as argued by Butler et al.\cite{butler2011rational}

In the case of (Co$_{2-\alpha}$Mn$_\alpha$)FeGe, Mn can occupy the X site with 3/4 majority/minority valence electron configuration (hence with moments antiparallel to Co) and also occupy the Y site with a 5/2 configuration. X-Y intermixing is unlikely, however, since the Co atoms cannot move from the X site as occupying the Y site would require the unphysical 7/2 configuration noted above. But, recall that Fe on the Y site is forced adopt the difficult 6/2 majority/minority configuration to maintain the 4-2 rule. If Fe moved to the X site, it could adopt the more favourable 4/4 configuration\cite{butler2011rational}, and Mn can take its place on the Y site and adopt the still {\em reasonable} 5/2 configuration with some help from Ge $sp$ states. Thus, one suspects it is favourable for to Fe move to the X site with Mn taking its place on the Y site, and in (Co$_{2-\alpha}$Mn$_\alpha$)FeGe we expect to actually observe (Co$_{2-x}$Fe$_{x}$)(Mn$_{x}$Fe$_{1-x}$)Ge. As our calculations will show, the difference in magnetism and electronic structure between the two scenarios is indeed observable. The latter arrangement also explains why Mn substitution should stabilise the unstable parent Co$_2$FeGe Heusler phase: each Mn substitution allows an Fe atom to get out of an energetically unfavourable configuration in addition to reducing the total valence electron count. In our view, the Burch rule applied to a full Heusler alloy is likely just a proxy for the 4-2 rule, resulting from the fact that reducing electron count also coincidentally reduces electronegativity, but more powerful in that it also helps explain the origin of the energy gap. 

Based on the rules described above, one expects atomic arrangement as shown in Fig.\ \ref{fig:unit_cell}(a) for Co$_2$FeGe, where 8 Co atoms occupy sublattice A whereas 4 Fe and 4 Ge atoms occupy the B$_T$ and B$_Z$ sites, on sublattice B, respectively. On Mn substitution, the Mn atoms occupy the B$_T$ site on sublattice B, as shown in Fig.\ \ref{fig:unit_cell}(b).  In case of Co$_2$(Fe$_{1-\beta}$Mn$_\beta$)Ge series on the other hand, Fig.\ \ref{fig:unit_cell}(c), we expect the substituting Mn atom directly occupies vacant Fe site on sublattice B, and it can be considered as a B sublattice substitution. In this case, while the Fe is in the relatively unfavourable 6/2 configuration, it cannot switch sites with Co without Co having an even more unfavourable configuration. Once the B$_T$ site is fully occupied by Mn atoms, for $\beta$ $\geq$  1 in Co$_2$(Fe$_{1-\beta}$Mn$_\beta$)Ge, one can then substitute Fe for Co atoms, as in (Co$_{2-\alpha}$Fe$_\alpha$)MnGe series, figure \ref{fig:unit_cell}(d). Here Fe is again favoured to directly occupy the vacant Co site so that it can adopt the 4/4 majority/minority configuration. As only sublattice A is involved in the substitution, this scheme can be represented as A sublattice substitution. Hence there are three non-equivalent substitution schemes we consider, A, B, and AB substitution. A comparative investigation of these schemes is thus useful in understanding how the different sublattice substitutions occur in principle and in practice and how they affect the alloy properties. 

Since we investigate the experimental as well as theoretical properties of the alloys over a wide composition range for each substitution series, we discuss each of the three series separately before making more general conclusions. It is worth noting here that this is the first report on the (Co$_{2-\alpha}$Mn$_\alpha$)FeGe system we are aware of, either experimentally or theoretically. There are a few reports on CoFeMnGe (which is the alloy at $\alpha = 1.0$), and this alloy has been reported to exhibit high value of spin polarisation (P) \cite{bainsla2014high}. It is then interesting to see if the P value can be further enhanced by changing Co/Mn in the vicinity of the CoFeMnGe composition. A detailed microstructure investigation of CoFeMnGe is also warranted as previous studies relied on x-ray diffraction alone to ascertain the phase-stability of the alloy. In addition, Mn$_2$FeGe, which has been predicted to exhibit half-metallic character if stabilised in cubic structure \cite{luo2008half}, has recently been reported to have a hexagonal DO$_{19}$ structure\cite{aryal2021} just as the other endpoint compound Fe$_2$MnGe\cite{keshavarz2019fe2mnge} and the parent Mn$_3$Ge\cite{qian2014} do. This implies a structural phase transition in the first and last substitutional series for some value of $\alpha$. Additionally, in contrast to (Co$_{2-\alpha}$Ti$_{\alpha}$)FeGe \cite{shambhu2019tunable} and (Co$_{2-\alpha}$Cr$_{\alpha}$)FeGe\cite{mahat2021possible} series which also generated a few promising candidate half-metals, the (Co$_{2-\alpha}$Mn$_{\alpha}$)FeGe will have advantage of extra valence electrons of Mn and hence promise of enhancing both M$_s$ and T$_\text{C}$. Thus investigation of (Co$_{2-\alpha}$Mn$_{\alpha}$)FeGe series is beneficial in many ways in addition to aiding in the comparative study of three series.

%%%%%%%%%%%%%%%%%%%%%%%%%%%%%%%%%%%
\section{\label{sec:level2}Experimental Details}
Bulk ingots of the samples were prepared by arc-melting of ultra pure (4N, Alfa Aesar) constituent elements. In case of manganese (Mn), it was necessary to remove the oxide layer at the elemental surface. This was done by pre-annealing the Mn in a vacuum-sealed quartz tube at 900$\degree$C for 9 hours. A 5 \% excess of Mn was also added in the mixture for arc-melting to compensate for its volatility. The melting was carried in an Ar atmosphere in a vacuum chamber with a base pressure of $\sim\!5.0\times10^{-5}$ Torr. To enhance the homogeneity of mixing, each sample was melted at least 4-5 times by flipping it upside down after each rounds of melting. The composition after the final round of melting was then checked by an energy dispersive x-ray spectroscopy (EDS) detector equipped in a JEOL 7000 field emission scanning electron microscope (SEM) for any elemental weight loss during the melting process. No significant loss of any of the elements was observed. To enhance homogenisation and facilitate the crystal growth, the arc-melted pieces were sealed in evacuated quartz tubes and annealed in a box furnace. A heating rate of 12 $\degree$C/min and a cooling rate of 15 $\degree$C/min were used. To make the comparison uniform across all compositions, only the samples which went through identical heat treatments (i.e., 900$\degree$C for 3 days) are chosen for this work.
 
The heat treatments were followed by metallography to produce a smooth and shiny metallic surface for microstructure analysis using optical and electron microscopes. After the heat treatment and metallography, the composition and homogeneity of the samples were again confirmed by EDS. Structural characterisation was performed  using a Bruker D8 Discover x-ray diffractometer equipped with monochromatic Co-K$_{\alpha}$ ($\lambda = 0.179\,$ nm) radiation. The experimental XRD patterns were compared with the simulated XRD patterns generated by using commercial CARINE crystallographic 3.1 software \cite{boudias2006carine} as well as custom in-house software \cite{leclair2018xrd}, and Rietveld refinement was carried out using a MATCH! software based on the FULLPROF algorithm \cite{fullprof,crystalmatch}. 

The low-temperature magnetic measurements were performed in Quantum Design Physical Properties Measurement System (PPMS), while the high temperature magnetisation was measured using the Lakeshore VSM 7400 series. For the magnetic measurements, irregularly shaped sample pieces, weighting $30-40\,$mg were utilised.
%%%%%%%%%%%%%
\section{\label{sec:level3}Computational Details}
The calculations were performed by using density functional theory (DFT) as implemented in Vienna \textit{ab initio} simulation package (VASP) code \cite{kresse1996efficiency}, which uses plane-wave basis set and projector augmented wave (PAW) pseudopotentials \cite{blochl1994projector}. In order to have consistency with Open Quantum Materials Database (OQMD) \cite{saal2013materials,kirklin2015open}, and to our in-house database, the PAW potentials for all the elements were chosen as reported in those databases. The Perdew-Burke-Ernzerhof (PBE) version of the generalised gradient approximation (GGA) was used for exchange and correlation \cite{perdew1996generalized}. The automatic mesh generation scheme within VASP with a length parameter (R$_k$) of 40 was used in order to perform integration over the irreducible Brillouin zone (IBZ). The energy cut-off for the plane wave basis set was 520 eV. The cell (initially cubic) dimensions and the atomic positions within the cell were relaxed using the conjugate-gradient method. An energy difference of $1 \times 10^{-7}$ eV between two successive ionic steps was set as a convergence criterion during ionic relaxation. Spin-orbit interaction was not included in the calculations.

%%%%%%%%%%%%%%%%%%%%
\begin{table}[htb]\scriptsize
\caption{Possible atomic arrangements for (Co$_{2-\alpha}$Mn$_\alpha$)FeGe substitutional series. Note that the site interchange; 4a $\leftrightarrow$ 4b or 4c $\leftrightarrow$ 4d, and similarly the sublattice interchange; (4a, 4b) $\leftrightarrow$ A(4c, 4d) generate equivalent structures. Also note when  $\alpha$ $>$ 1, the excess Mn atoms directly occupy the vacant 4d site.}
\renewcommand*{\arraystretch}{1.1}
\centering
\begin{tabular}{c c c c c}
\hline\hline
& \multicolumn{2}{c}{\bf Sublattice B} & \multicolumn{2}{c}{\bf Sublattice A}\\ \hline
\textbf{Type} & \multicolumn{1}{c}{\centering{\textbf{4a ($B_Z$)}}} & \multicolumn{1}{c}{\centering{\textbf{4b ($B_T$)}}} & \multicolumn{1}{c}{\centering{\textbf{4c}}} & \multicolumn{1}{c}{\centering{\textbf{4d}}}\\ \hline
 I & Ge & Fe & (1-$\alpha$)Co+$\alpha$Mn& Co \\[1ex]
 II & Ge & (1-$\alpha$)Fe+$\alpha$Mn & (1-$\alpha$)Co+$\alpha$Fe & Co\\[1.5ex]
  III& (1-$\alpha$)Ge+$\alpha$Mn & Fe & (1-$\alpha$)Co+$\alpha$Ge & Co\\[1.5ex]
\hline\hline
\end{tabular}
\label{tab:atom_config1}
\end{table}
%%%%%%%%%%%%%%%%%%%%
%%%%%%%%%%%%%
We pointed out earlier that a double sublattice substitution (i.e., the substituted Mn atom displaces Fe atom towards vacant Co site and itself occupies the Fe site) is expected here. Although the atomic arrangements based on double sublattice substitution scheme seems favourable, we also included other plausible atomic ordering in the theoretical analysis. The three atomic arrangements considered are defined as type I, II, and III. In type I, we assume the substituted Mn atoms directly occupy the vacant Co site in sublattice A, whereas type II corresponds to double sublattice substitution as explained above (atomic ordering according to Burch or the 4-2 rule). Type III is another variant of double sublattice substitution, in which we assume Mn atoms displace Ge atoms toward the vacant Co site and Mn itself occupies the B$_Z$ site. These three configurations are summarised in table \ref{tab:atom_config1}, and are also represented visually in Fig.\ \ref{fig:unit_cell} for $(\alpha,\beta) = 0.5$. Apart from these three, several other disordered atomic configurations can be generated by allowing partial or full disorder, some of them were also included in the analysis to determine the most stable configuration. Below we discuss our procedure for determining the most stable configuration out of many competing configurations.

We first obtained the relaxed structure by performing full ionic relaxations within a 16 atom unit cell (in some cases supercell with 32 atoms). For the relaxed structure, we calculated zero temperature total energy and the magnetic moment. Then, we investigated the structural stability by calculating the formation energy as follows:
\begin{equation}
\begin{split}
 {}&\Delta E_{\text{form}} (\mathrm{Co}_{2-\alpha}\mathrm{Mn}_\alpha \mathrm{FeGe}) = E (\mathrm{Co}_{2-\alpha}\mathrm{Mn}_\alpha \mathrm{FeGe})\\
   & - \frac{1}{4}[(2-\alpha)\mu_{\mathrm{Co}} + 
 \alpha \mu_{\mathrm{Mn}} + \mu_{\mathrm{Fe}} + \mu_{\mathrm{Ge}}] 
 \end{split}
 \label{eq:form_energy}
\end{equation}
where $E(\mathrm{Co}_{2-\alpha}\mathrm{Mn}_\alpha \mathrm{FeGe})$ is the total energy and $\mu_i$ is the reference chemical potential of element $i$, which is chosen from ref.\ \cite{kirklin2015open}. A negative value of $\Delta E_{\text{form}}$ indicates that the compound is more stable than its constituent elements. While this is a necessary condition for the thermodynamic stability, it does not guarantee that the phase is more stable than any other competing phases (\textit{here by ``phase'' we mean a specific atomic arrangement}) . In such cases, one must investigate the convex hull. A compound phase is said to be more thermodynamically stable at 0 K than any other phases at that chemical space if it lies on the convex hull of formation energies. If a phase does not lie on the convex hull, this indicates there are other phases or combination of phases at that region of chemical space that are lower in formation energy. This condition has, however, found not to be a strict condition -- many alloys have been found to be stable in experiment despite lying above the convex hull. This could be related to several factors, for example: degree of completeness of set of phases utilised for the construction of the convex hull, error in the calculation of the formation energy, or the phases which are not stable at 0 K can be stable at high temperature due to different entropic contributions \cite{ma2018computational}. Ma \textit{et al.} \cite{ma2018computational} investigated all inverse Heusler alloys reported as stable in the Inorganic Crystal Structure Database (ICSD) \cite{belsky2002new,bergerhoff1987crystallographic} and more recent literature. They found that DFT-calculated formation energies of the stable phases are within $0.052\,$eV/atom of the DFT-calculated convex hull.

Hence, to determine the thermal stability, we calculate the distance of the formation energy ($\Delta E_{\text{form}}$) from the convex hull energy ($E_{hull}$), also called hull distance as follows;
\begin{equation}
\Delta E_{HD} = \Delta E_{\text{form}} - E_{hull} 
\label{eq:hull_dis}
\end{equation}
The $E_{hull}$ at each composition can be obtained from the linear combination of energies of stable phases in that phase space. This is available as a look-up feature called ``grand canonical linear programming" (GCLP) in the OQMD database \cite{saal2013materials,kirklin2015open}. This database includes more than 600,000 hypothetical compounds including the stable phases that has been reported in ICSD database. Hence, we have relied upon this database for the value of $E_{hull}$ at each composition. By definition, $\Delta E_{HD}$ = 0 represents the most stable phase, whereas the larger the value of $\Delta E_{HD}$ the phase becomes more susceptible to thermal effects and decomposes or transforms into other phases. A comparison of $\Delta E_{HD}$, thus can be used as one of the ways of predicting a stable phase among many competing phases.

Further, as it will be shown that the experimental magnetic moments vary linearly over the substitution range where materials show single-phase microstructure. Hence, a comparison of experimental and theoretical moment can also be referenced to determine the stable atomic configuration.
%%%%%%%%%%%%%%%%%%%%%%%%%%%%%%%%%%
\section{\label{sec:level4}Result and Discussion: (Co$_{2-\alpha}$Mn$_\alpha$)FeGe}
%%%%%%%%%
\subsection{Findings from experiment}
\begin{figure}[ht]
\centering
\includegraphics[width =\columnwidth]{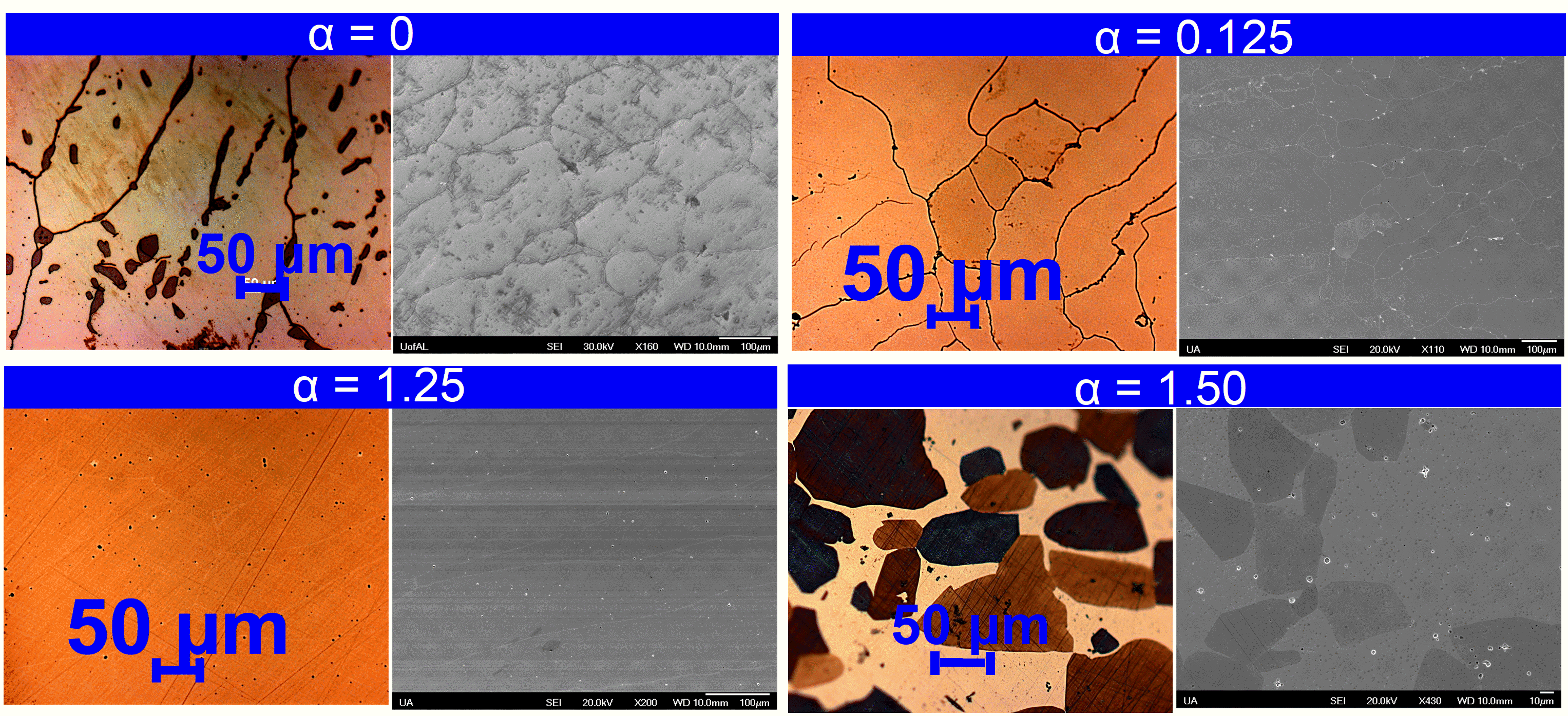}
\caption{(colour online) Optical (left) and SEM (right) micrographs at four different substitution level ($\alpha = 0, 0.125, 1.25, 1.50$). A multi-phase microstructure can be observed at $\alpha = 0$ and at $\alpha = 1.50$, whereas the intermediate compositions are single-phase. The micrographs for other intermediate compositions are provided in supplementary information.}
\label{fig:optical_Co2-aMna}
\end{figure}
%%%%%%%%%%%%%%%%%%%%%%%

%%%%%%%%%%%%%%%%%%%%

The micrographs of selected (Co$_{2-\alpha}$Mn$_\alpha$)FeGe alloys as seen in an optical microscope and an SEM are shown in figure \ref{fig:optical_Co2-aMna}. In these micrographs, one is observing the areas with different contrast, which indicates occurrence of multiple phases (\textit{in this context, by phase we mean a region of spatially uniform atomic composition}). The contrast difference between different phases in optical micrographs arises due to their differing reaction rates to the etchant thereby creating an observable height profile (all samples discussed here were etched with Adler for 10 seconds after polishing), whereas in SEM it is due to differing elemental contribution to electronic scattering phenomenon. As shown in figure \ref{fig:optical_Co2-aMna}, multi-phase behaviour is clearly visible in the parent Co$_2$FeGe ($\alpha = 0$) alloy. After Mn substitution, the micrographs show only a single contrast, aside from the grain boundaries. The grain boundaries appearing somewhat darker indicate that the composition at the grain boundaries may be slightly different than that within the grains, which we will verify with EDS analysis in the next paragraph. Nonetheless, within the grains there are no regions of differing contrast, and neighbouring grains have the same contrast, which suggests formation of a single-phase compound after replacing Co by Mn. It is interesting to note that single-phase behaviour is obtained over a rather wide substitution range, i.e., up to $\alpha$ = 1.25. At $\alpha$ = 1.375 (shown in supplementary), small dark regions are visible, but apart from those dark spots the microstructure looks single-phase. These dark spots could be pores or the regions of different composition, which will be verified from the EDS analysis in the upcoming paragraph. Above $\alpha$ = 1.375, that is, for $\alpha = 1.50$, the microstructure clearly looks multi-phase, as distinct contrast can be observed in the micrographs. A multi-phase microstructure is also observed for $\alpha = 2$ (\textit {i.e.}Mn$_2$FeGe), also shown in the supplementary information. We should note that while Aryal {\em et al.}\cite{aryal2021} found Mn$_2$FeGe to form a DO$_{19}$ structure that did not appear to have other metallic phases present, they used rather different annealing conditions than we did - 650$^\circ$C for 2 days vs our 900$^\circ$C for 3 days. Given that a cubic to hexagonal structural transition is happening as one moves toward the endpoint, it would perhaps not be surprising to have both low- and high-temperature phases of Mn$_2$FeGe accessible, as was observed in Fe$_3$Ge-based systems.\cite{mahat2020,mahat2021Cr}

\begin{figure}[ht]
\centering
\includegraphics[width =0.9\columnwidth]{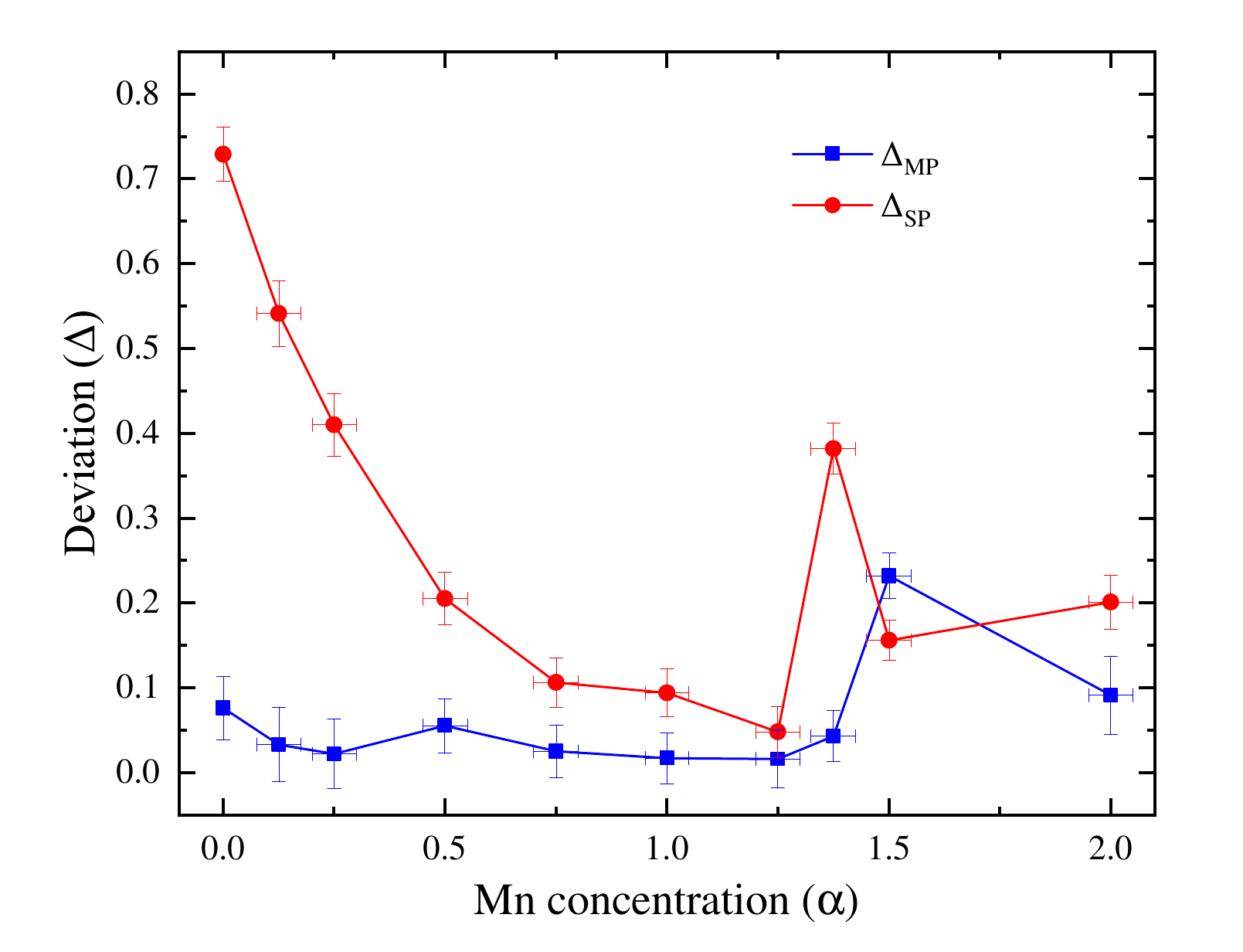}
\caption{(color online) Main phase (MP) and secondary phase (SP) composition deviation from the target composition across the whole substitution range. Note that for 0.125 $\le$ $\alpha$ $\le$ 1.25, SP represents the grain boundary composition.}
\label{fig:compo_deviation_Co2-aMna}
\end{figure}

%%%%%%%%%%%%%%%%%
In order to estimate the main and secondary phase composition with statistical confidence, at least 10 different measurements covering different regions across the sample were analysed, an example is shown in supplementary information. Here main phase represents the matrix phase or the dominant phase in terms of volume occupied, whereas the secondary phase is the impurity phase (if present) or the grain boundary composition. If a single-phase composition has formed, one would expect to measure a composition that agrees to the target composition. But as the secondary phases (with different compositions) nucleate and grow in the microstructure, the main phase composition necessarily diverges away from the target composition. In contrast, the smaller the secondary phase volume fraction, the more the main phase composition trends toward the target composition. Investigation of the composition in the two phases thus can provide useful insight regarding the formation of or {\em tendency} to form a single-phase compound. To study this trend, we defined a composition deviation parameter $\Delta$, in the spirit of error propagation, by taking the difference of each element from its targeted amount added in quadrature, i.e, $\mathrm{deviation} = \Delta \equiv \sqrt{\sum_i{(A_{i,measured} - A_{i,target})^2}}\hspace{0.5em}$ where $A_i$ is the amount of element either measured or targeted and $i = \{\mathrm{Co}, \mathrm{Fe}, \mathrm{Mn}, \mathrm{Ge}\}$. (For example, with a target of Co$_2$FeGe, $A_{Co,target}$=2 and $A_{Fe,target}=A_{Ge,target}=1$.) If the compound is single-phase $A_{i,measured} - A_{i,target}=0$ for each element and the deviation is zero. Reductions in the deviation with substitution should indicate a trend toward single-phase behaviour, while a sharp increase should suggest a tendency toward multi-phase behaviour. The deviation as a function of Mn concentration $\alpha$ is shown in figure \ref{fig:compo_deviation_Co2-aMna}. Notably, the main phase has a lower deviation than the secondary for all single phase samples as one expects, and the deviation decreases as Mn concentration increases from $\alpha=0$, suggesting a trend toward single-phase behaviour as Mn is added. But the trend breaks at $\alpha = 1.375$, above which the deviation increases. This clearly suggests that for $\alpha \geq 1.375$ the system is tending toward multi-phase behaviour, corroborated by micrographs analysis which showed distinct contrast in the images for $\alpha > 1.25$. In particular, at $\alpha=1.375$ the small dark spots indeed had slightly different composition than the surrounding matrix. Although in the ideal single-phase case one expects $\Delta = 0$, obtaining this in reality is unlikely. This is because instrument itself has an uncertainty of 5\% and even aside from that the occurrence of grain boundaries providing sites for segregation. Keeping this in mind, we can treat a composition to be single-phase if the main composition deviation is less than approximately 0.05. Based on results thus far, the Mn composition range 0.125 $\le$ $\alpha$ $\le$ 1.25 can be treated as single-phase in the (Co$_{2-\alpha}$Mn$_\alpha$)FeGe series.  

A table containing the measured compositions of main and secondary phases in (Co$_{2-\alpha}$Mn$_\alpha$)FeGe across the substitution range is provided in the supplementary information. Upon careful investigation, one finds the secondary phase to be rich in Ge but deficient in Fe compared to the main phase for lower Mn concentration, whereas opposite behaviour can be observed at very high value of Mn concentration. 

%%%%%%%%%%%%%%%%%%%%%%%
%%%%%%%%%%%%%%%%%%%
%%%%%%%%%%%%%%%%%

%%%%%%%%%%%%%%%%%%%%%%%
\begin{figure*}[ht] \centering
\includegraphics[width =1.75\columnwidth]{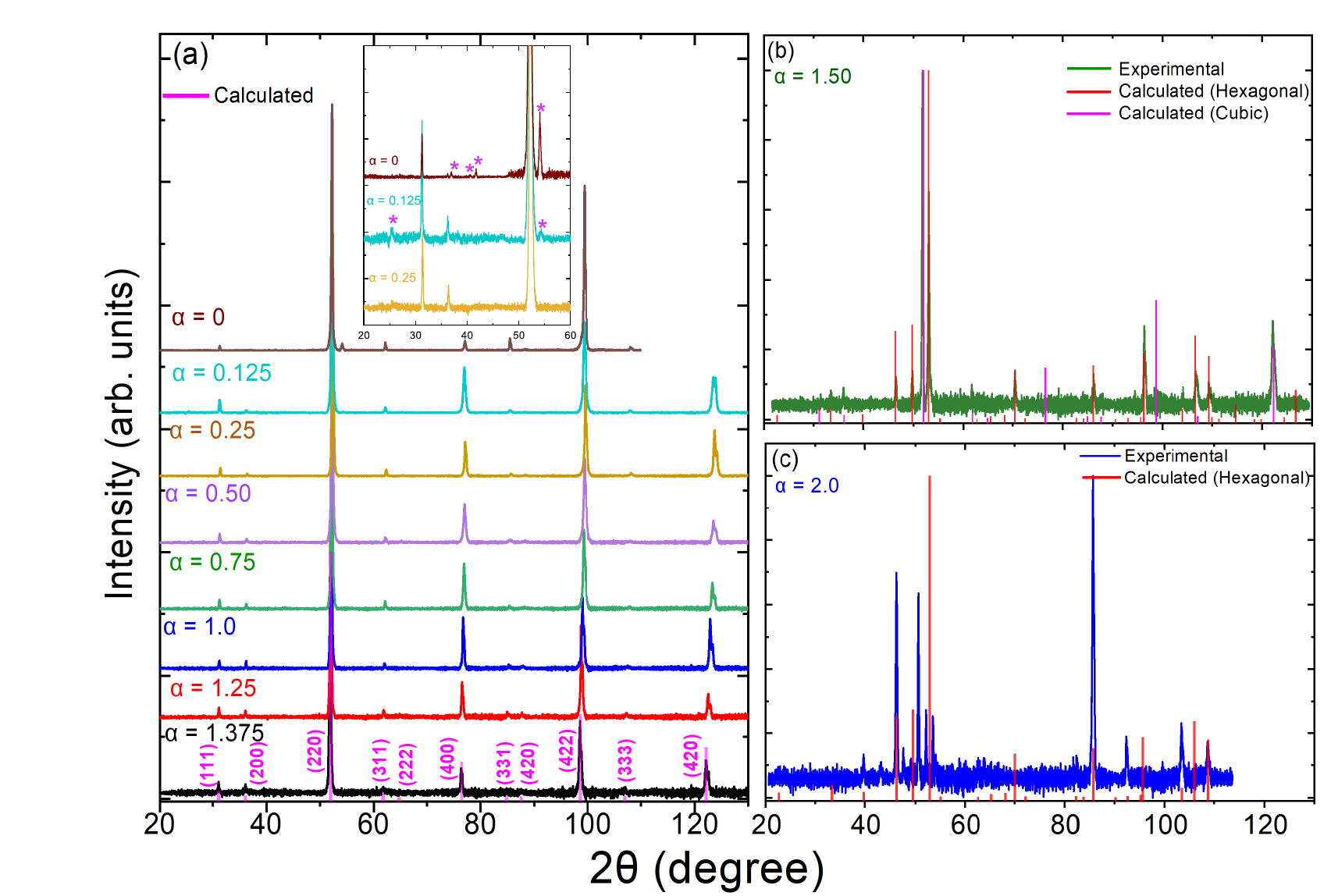}
\caption{(color online) XRD scans of (Co$_{2-\alpha}$Mn$_\alpha$)FeGe substitutional series heat treated at 900$\degree$C for 3 days. (a) Powder XRD patterns, which can be fit to an fcc structure, including the parent Co$_2$FeGe. (Inset) The low angle peaks are blown up to make any impurities peaks visible. (b) XRD pattern of Co$_{0.5}$Mn$_{1.5}$FeGe, which can be fitted to cubic plus hexagonal structures. (c) XRD pattern of multiphase Mn$_2$FeGe, which has one phase resembling hexagonal structure.}
\label{fig:xrd_Co2-aMna}
\end{figure*}
%%%%%%%%%%%%%%%%%%%%%%%%%%
%%%%%%%%%%%%

The XRD patterns of (Co$_{2-\alpha}$Mn$_\alpha$)FeGe alloy, taken on powder specimens at room temperature, are shown in figure \ref{fig:xrd_Co2-aMna}. Since many Heusler alloys are found to crystallise in a face centred cubic (fcc) structure, we begin by trying to index the observed peaks with an fcc unit cell. The calculated peaks based on an fcc unit cell (space group 216) are also shown in figure \ref{fig:xrd_Co2-aMna}(a). It is clear that the experimental patterns match very well with the calculated pattern, with the exception of a few extra peaks (denoted by *) in the cases of $\alpha$ = 0 and 0.125. These peaks should derive from the secondary phase (or significant segregation at the grain boundaries for $\alpha = 0.125$), as they cannot be fit even after assuming other crystal structures, such as tetragonal\cite{alijani2013tetragonal} and hexagonal\cite{keshavarz2019fe2mnge} structures seen in similar systems.\cite{Zhang2013structures} However, as the Mn concentration increases, the number of impurity peaks decreases, they become less intense, and finally disappear at $\alpha = 0.25$. What this simply means is that composition is trending towards single-phase composition as Mn concentration increases, in agreement with the micrographs and EDS analysis discussed above. The XRD patterns suggest single-phase behaviour up to $\alpha$ = 1.375. Hence, solely based on XRD pattern analysis the single-phase spectrum can be defined as 0.25 $\le$ $\alpha$ $\le$ 1.375. This is, however, in contrast to the range described in the previous section, where we defined 0.125 $\le$ $\alpha$ $\le$ 1.25 as single-phase range based on micrographs and EDS observations. The discrepancy is observed particularly at $\alpha$ = 1.375 and at $\alpha$ = 0.125. The $\alpha$ = 1.375 alloy appears phase pure in the XRD pattern despite having a small number of secondary phase regions verified by EDS. This is because the phase fraction needs to be at least of order $\sim$5 \% to be detected in XRD in our case, not to mention any amorphous material being harder to see by XRD, and for $\alpha\!=\!1.375$ the secondary phase fraction observed by microscopy is less than 2\%. On other hand, ignoring segregation at the grain boundary, the $\alpha\!=\!0.125$ composition seemed single-phase, but impurities peaks observed in XRD. This is why the study of phase stability based solely on XRD or solely on microscopy may not reflect the true phase structure of the alloy, and caution should be exercised when reviewing the literature -- ideally one should employ more direct microstructure investigation tools such optical and electron microscopes in conjunction with XRD to more accurately ascertain phase stability. (Of course, the microscopy techniques we have utilised have their limits as well, so those interested in the details of the grain boundaries and secondary phases will need to follow up with still better methods.) Now based on both the microscopy and XRD investigation, we find 0.125 $<$ $\alpha$ $<$ 1.375 to be the single-phase forming range for Co$_{2-\alpha}$Mn$_\alpha$FeGe alloys. 

A slightly different pattern was obtained in the cases of $\alpha$ = 1.50 and 2.0. First of all, there are too many peaks to be fit to a high symmetry fcc unit cell. The EDS analysis suggested two distinct phases in the sample, hence we tried to fit the pattern assuming mixture of two different crystal structures. Interestingly, all the observed peaks of $\alpha$ = 1.50 can be fit to a hexagonal structure (space group 194) plus a cubic structure (space group 216), as shown in figure \ref{fig:xrd_Co2-aMna}(b). The co-existence of cubic and hexagonal crystals is not surprising as EDS analysis showed two distinct phase compositions for $\alpha$ = 1.50. While pinpointing crystal structure of each phase compositions demands further investigation, given that $\alpha$ = 1.375 exhibits a cubic structure and main phase compositions at $\alpha\!=\!1.375$, $\alpha\!=\!1.50$ are the same within our measurement uncertainty, one can speculate the main phase observed for $\alpha$ = 1.50 is also the same cubic structure.  This is further corroborated by the fact that the experimental lattice parameter is found to be same in both cases, i.e., 5.782 \AA. Clearly there is competition between two different crystal structures for larger $\alpha$, which may be one of the reasons why adding too much Mn results in phase separation. In a similar fashion, many peaks observed in the pattern of $\alpha$ = 2.0 can be fit to a hexagonal structure, suggesting at least one of the phases exhibits a hexagonal structure. This is consistent with a recent report by Aryal {\em et al}\cite{aryal2021} who reported that bulk Mn$_2$FeGe indeed crystallises in a hexagonal DO$_{19}$ structure much like Fe$_2$MnGe\cite{keshavarz2019fe2mnge} and the parent compound Mn$_3$Ge.\cite{qian2014} Based on this, we conclude that at higher Mn concentration, the phase separation is due to the competition between two different crystal structures and hexagonal phase becomes dominant phase upon further increasing the Mn concentration. Clearly there is much more that can be done to investigate the phase competition in detail. 

%%%%%%%%%%%%%%%%%%%%%%%
\begin{figure}[ht]
\centering
\includegraphics[width =0.85\columnwidth]{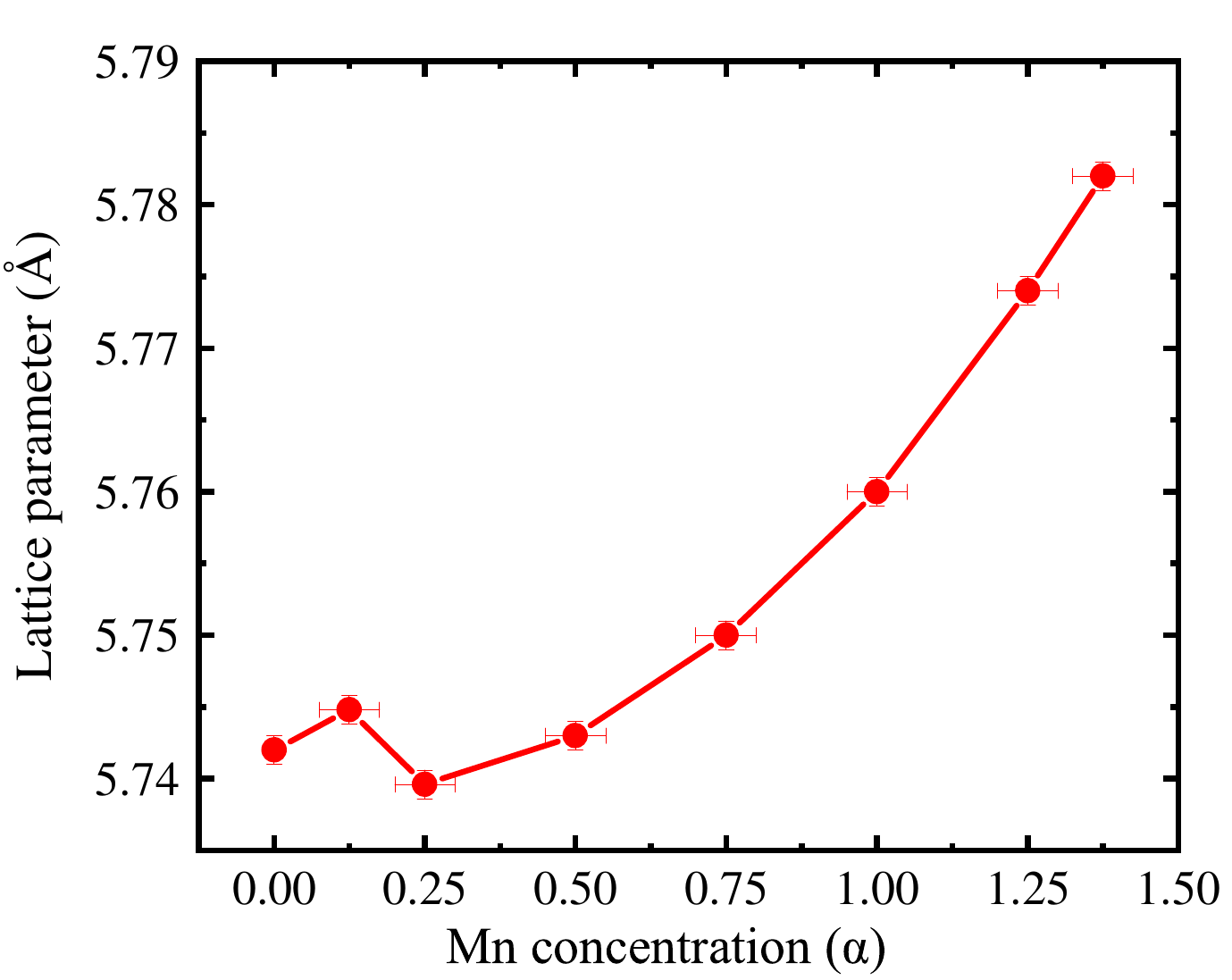}
\caption{(color online) Variation of experimental lattice parameter with Mn concentration in (Co$_{2-\alpha}$Mn$_\alpha$)FeGe . Vertical error bars are smaller than the points used.}
\label{fig:xrd_Co2-aMna_lattice}
\end{figure}
%%%%%%%%%%%%%%%%%%%%%%%%%%
To estimate the lattice parameter that is free of any instrumental uncertainty we utilised Cohen's method with Nelson-Riley error function, details can be found in Refs.\ \cite{cohen1935precision,nelson1945experimental}). The change in lattice parameter as a function of Mn substitution is shown in Fig. \ref{fig:xrd_Co2-aMna_lattice}. Since Mn (1.61 \AA) has a slightly larger atomic radius than Co (1.52 \AA) and Fe (1.56 \AA)\cite{atomradius}, an enhancement in lattice parameter is expected to accompany Mn substitution for Fe or Co. In Fig.\ \ref{fig:xrd_Co2-aMna_lattice}, a seemingly linear relationship between lattice parameter and Mn concentration can be observed, which is also supported by the slight shift of XRD peak positions toward lower angles (see Fig. \ref{fig:xrd_Co2-aMna}). The non-regular trend for low Mn concentration could be due the alloy being not phase pure, or atomic disorder. It is also possible that the non-linearity is due to experimental uncertainty we have missed, since the difference is less than 0.1 \%.  

%%%%%%%%%%%%%%%%%%%%%%%
\begin{figure}[ht]
\centering
\includegraphics[width =0.7\columnwidth]{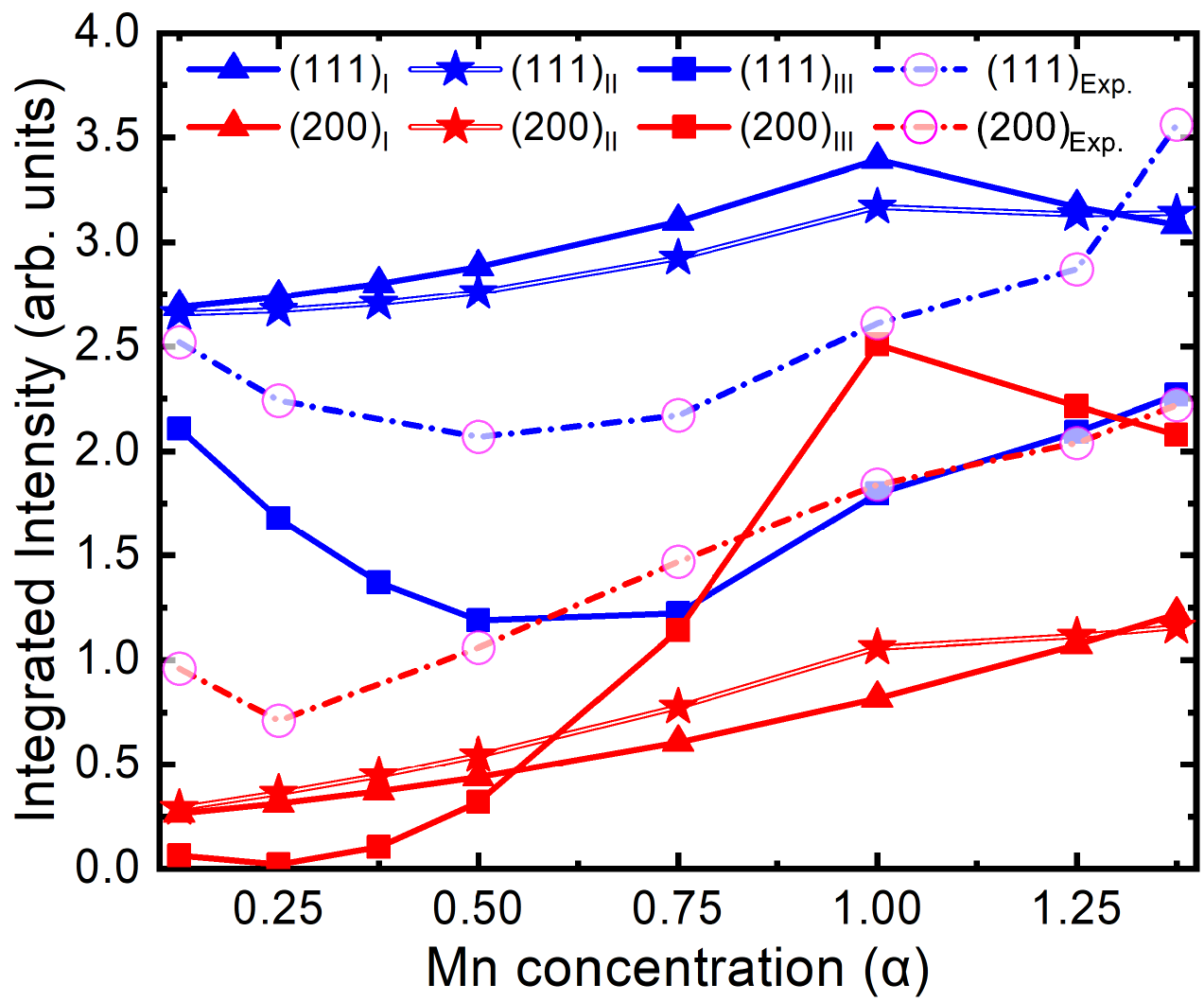}
\caption{(colour online) Variation of calculated and experimental low angle superlattice peaks intensity with the Mn concentration. The calculated values for structure type I, II and III are shown.}
\label{fig:xrd_Co2-aMna_intensity}
\end{figure}
%%%%%%%%%%%%%%%%%%%%%%%%%%

In order to gain some insight on the atomic order in these alloys, the low angle superlattice powder peak intensities are plotted against the Mn concentration in Fig. \ref{fig:xrd_Co2-aMna_intensity}. Here, the theoretical values were calculated for structure types I, II, and III following the atomic configuration in table \ref{tab:atom_config1}. None of these structures show an excellent match with experimental data, however. We also analysed the peak intensities of patterns collected on the bulk samples (\textit{i.e., not powdered}), and observed a significant degree of texturing (see supplementary \cite{supplementary2021}). In that light, the determination of detailed chemical order from XRD intensities alone is perhaps not well advised. Still, one can argue that type II and experimental data are closer in behaviour, which lends some support that type II is the favourable structure. A Rietveld refinement was also performed against these structures using the powder sample XRD patterns, and subsequently a very good refinement were achieved for the type II structure (the refinements are shown in the supplementary \cite{supplementary2021}). However, given the texturing issue we will refrain from concluding which structure is more favourable until considering additional evidence from magnetometry and theoretical calculations in the following sections. 
%%%%%%%%%%%%%%%%%%%%
\begin{figure}[ht]
\centering
\includegraphics[width = 0.8\columnwidth]{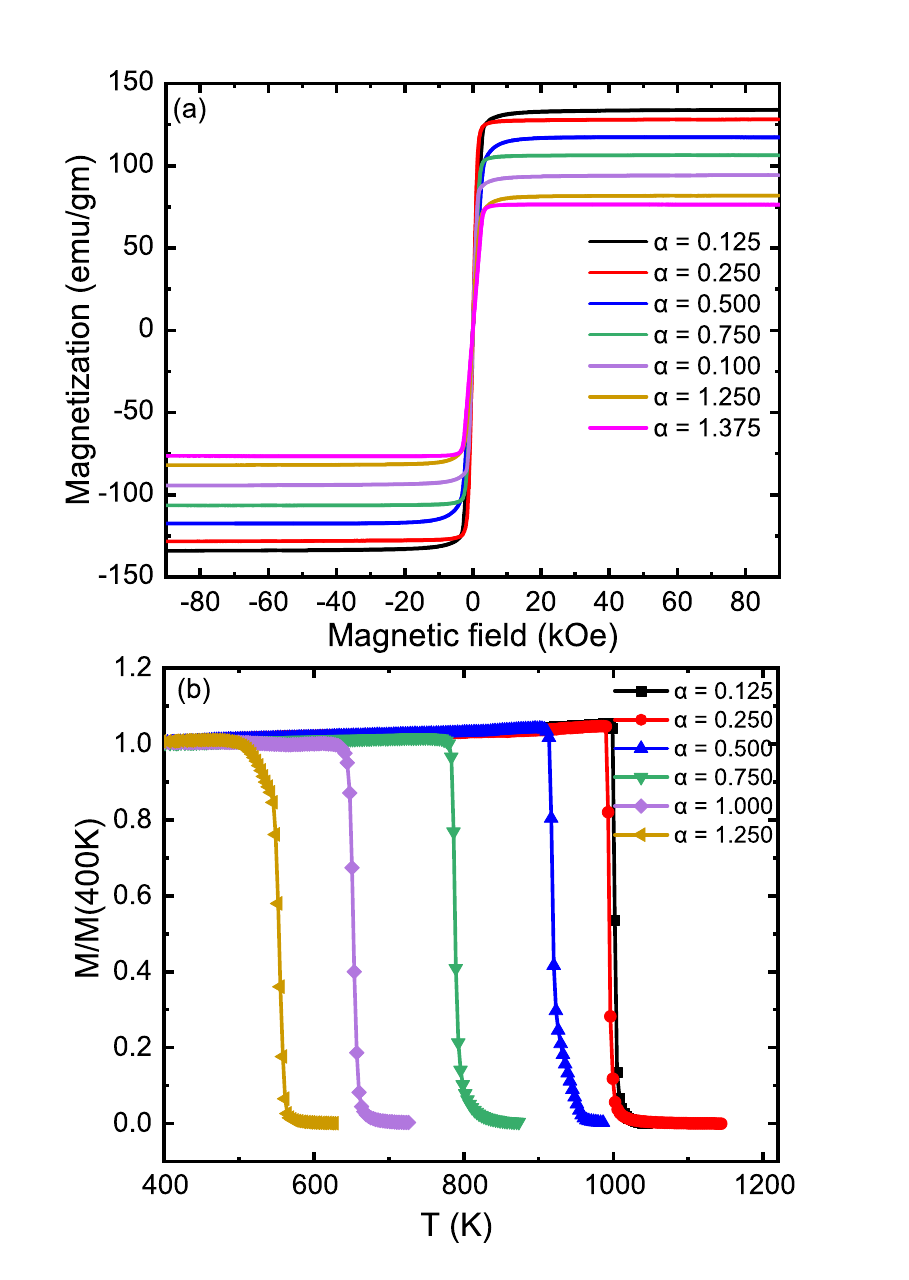}
\caption{(colour online) (a) The field-dependent magnetic measurements of (Co$_{2-\alpha}$Mn$_\alpha$)FeGe alloy, measured at T = 5 K. (b) The temperature dependent magnetic properties of the samples measured with an applied field of 100 Oe, exhibiting very high Curie temperatures.}
\label{fig:magnetic_Co2-aMna}
\end{figure}
%%%%%%%%%%%%%%%%%%%%%%

The magnetic measurements were performed only for the alloys which showed single-phase behaviour in microscopy and XRD analysis. The field- and temperature-dependent measurements are shown in figure \ref{fig:magnetic_Co2-aMna}. The magnetic hysteresis curves at T = 5 K (Fig. \ref{fig:magnetic_Co2-aMna}(a)) suggests the alloys are soft ferromagnets with very low coercivity. A decrease in the saturation magnetisation can be observed as the Mn concentration increases. Similarly, the temperature dependent magnetic measurement with an applied field of 100 Oe (Fig. \ref{fig:magnetic_Co2-aMna}(b)) shows that the alloys have very high Curie temperature (T$_\text{C}$), which also shows a downward trend as the Mn concentration increases. One can observe gradual increase in magnetisation close of transition temperature. This is likely a Hopkinson peak, arising from the very low measurement field used, and it is commonly observed in ferromagnetically ordered systems\cite{dhar1980peaks}.

%%%%%%%%%%%%%%%%%
\begin{figure}[ht]
\centering
\includegraphics[width = \columnwidth]{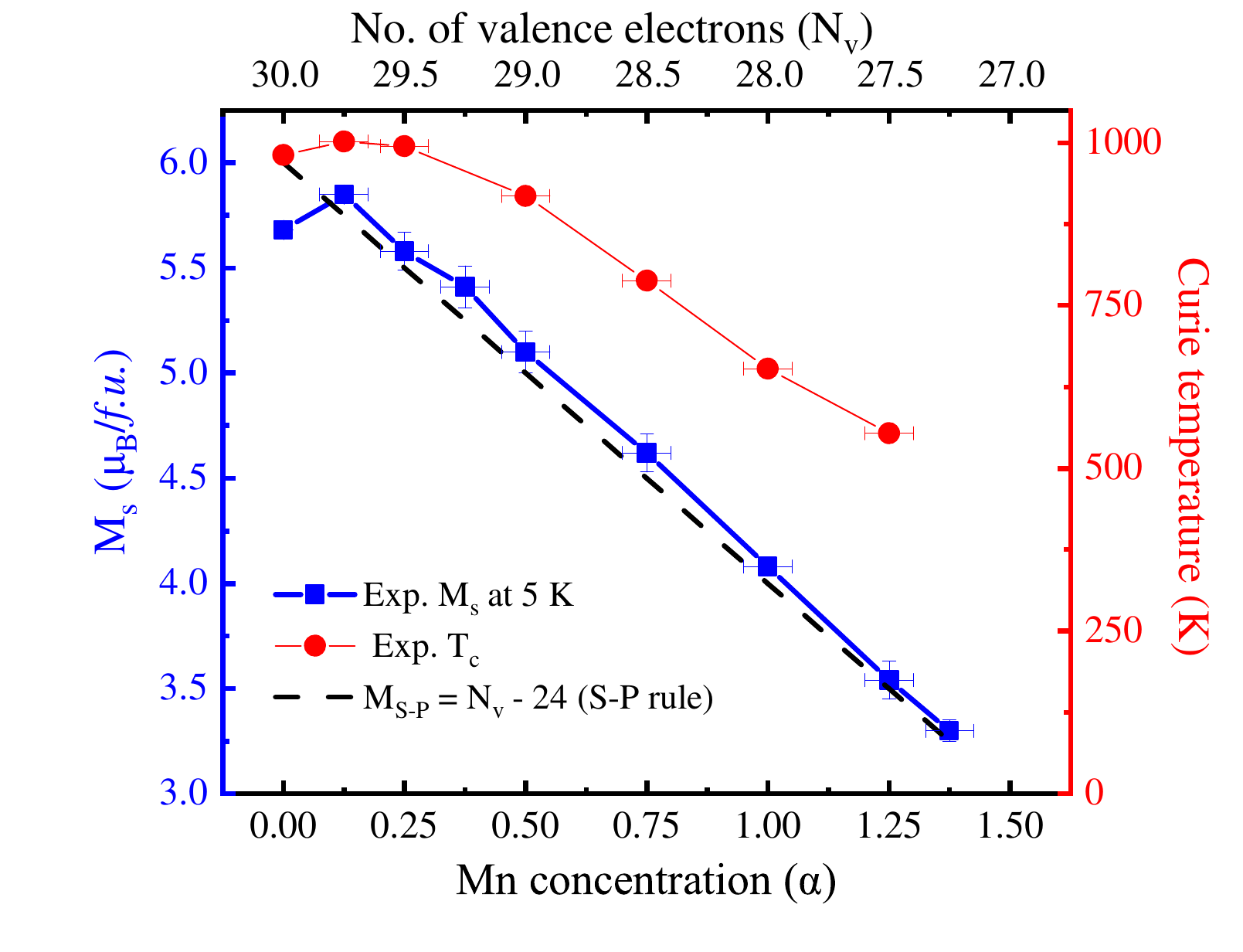}
\caption{(color online) (a) Variation of M$_\text{s}$ (measured at 5K) and T$_\text{C}$ with Mn concentration in (Co$_{2-\alpha}$Mn$_\alpha$)FeGe . Both exhibit more or less linear behaviour with N$_\text{v}$ matching reasonably well with the moment expected based on S-P rule.}
\label{fig:Ms}
\end{figure}
%%%%%%%%%%%%%%%%%%

Both the saturation magnetisation (M$_\text{s}$) at 5 K and Curie temperature (T$_\text{C}$) are extracted from the experimental curves, and are plotted relative to the Mn concentration (also to the number of valence electrons) in figure \ref{fig:Ms}. The M$_\text{s}$ exhibits a linear relation with Mn concentration, which in turn can be related to the number of valence electrons (N$_\text{v}$). This sort of behaviour, where magnetic moment varies linearly with N$_\text{v}$, is called Slater-Pauling behaviour and is one of the typical features of full Heusler alloys containing Co. In addition to the characteristic linear behaviour, the magnitude of the experimental M$_\text{s}$ at T = 5 K agrees well with the calculated Slater-Pauling moment (M$_{S-P}$). The M$_{S-P}$ is a theoretical moment obtained using the relationship M$_{S-P}$ = N$_\text{v}$ - 24, which stems from the fact that a gap occurs in the minority (spin-down) band after 12 electrons per unit cell. This means any excess or missing electrons will occupy the majority (spin-up) band and hence yielding a linear dependence of magnetic moment on the N$_\text{v}$. Since the Fermi level (E$_F$) will be pinned in the band gap in the minority channel, it is therefore expected that the compounds satisfying this relation should show half-metallic character.\cite{galanakis2002origin,galanakis2002slater} Thus, based on the quantitative agreement between the experimental and Slater-Pauling moment over a wide composition range, one can argue that the (Co$_{2-\alpha}$Mn$_\alpha$)FeGe ($0.125 < \alpha  < 1.375$) alloys are good candidate half-metals or near half-metals. We should note that a slightly higher moment has been measured experimentally compared to the Slater-Pauling value. This could be due to several factors -- an inherent systematical error, chemical disorder, or the presence of magnetic impurities undetectable in XRD or microscopic analysis. A slightly higher experimental moment has also been observed for several other Heusler alloys predicted to be half-metals\cite{shambhu2019tunable,alijani2011electronic,galanakis2016theory}. Still, the difference between the low temperature experimental M$_\text{s}$ and the expected Slater-Pauling value is at best barely outside our expected experimental uncertainty range (see error bars in Fig.\ \ref{fig:Ms}), and as such it probably does not merit further discussion. 

Another interesting feature of Co-based Heusler alloys is that T$_\text{C}$ also typically shows a linear variation with N$_\text{v}$. A roughly linear variation of T$_\text{C}$ has also been observed here in (Co$_{2-\alpha}$Mn$_\alpha$)FeGe alloys that are single-phase, with the value approaching 1000 K for lower $\alpha$. Even for the highest $\alpha$ values, T$_\text{C}$ is significantly above room temperature ({\it i.e.}, $>$ 500 K). High T$_\text{C}$ and M$_\text{s}$ are two critical features that materials should possess to withstand thermal effects for applications at room temperature and above. The (Co$_{2-\alpha}$Mn$_\alpha$)FeGe alloys studied here have both, and both are readily tuneable, which may render them suitable for practical applications.  

In figure \ref{fig:Ms}, one can observe an anomaly -- there is an enhancement of both M$_\text{s}$ and T$_\text{C}$ when increasing $\alpha$ = 0 to 0.125 even though N$_\text{v}$ trends downwards. It is worth positing an explanation for this anomalous behaviour. As M$_\text{s}$ and T$_\text{C}$ are directly related, explaining the M$_\text{s}$ trend should suffice to explain the variation of T$_\text{C}$ as well. First, the experimental moment of $\alpha$ = 0.125 agrees reasonably well with the M$_{S-P}$, but the moment of $\alpha$ = 0 is markedly less than the M$_{S-P}$ value. This is not so surprising since the $\alpha$ = 0 parent multi-phase whereas Co$_{1.875}$Mn$_{0.125}$FeGe ($\alpha$ = 0.125) is very nearly phase-pure -- there is only a bit of phase segregation at the grain boundaries. As the experimentally measured moment represents the average over moments of different phases that are present in the microstructure, the magnetic moment of a multi-phase alloy is very likely to be less than that of a single highly magnetic phase. Add to this the likelihood of highly disordered material with reduced magnetisation at the grain boundaries. The EDS analysis of Co$_2$FeGe alloy shows the main phase composition as Co$_{1.97}$Fe$_{1.06}$Ge$_{0.97}$, which is close to the target value and thus it should have a moment close to 6 $\mu_B$. However, the secondary phase composition (Co$_{2.14}$Fe$_{0.43}$Ge$_{1.43}$) is Fe-deficient but rich in non-magnetic Ge, so it is reasonable to expect that it is less magnetic and the cause for the overall volume experimental moment of Co$_2$FeGe to fall below the M$_{S-P}$ value calculated assuming a single phase. This is also consistent with the fact that the secondary phase volume fraction is essentially negligible at $\alpha$ = 0.125, so the moment is predominantly that of main phase, and hence closer to the theoretical M$_{S-P}$ value. While this reasoning seems plausible enough, it is based on a paucity of data, and therefore we will also consider the theoretical results below before drawing further conclusions.

%%%%%%%%%
\subsection{Findings from calculations}
\begin{figure}[ht]\centering
\includegraphics[width = 0.85\columnwidth]{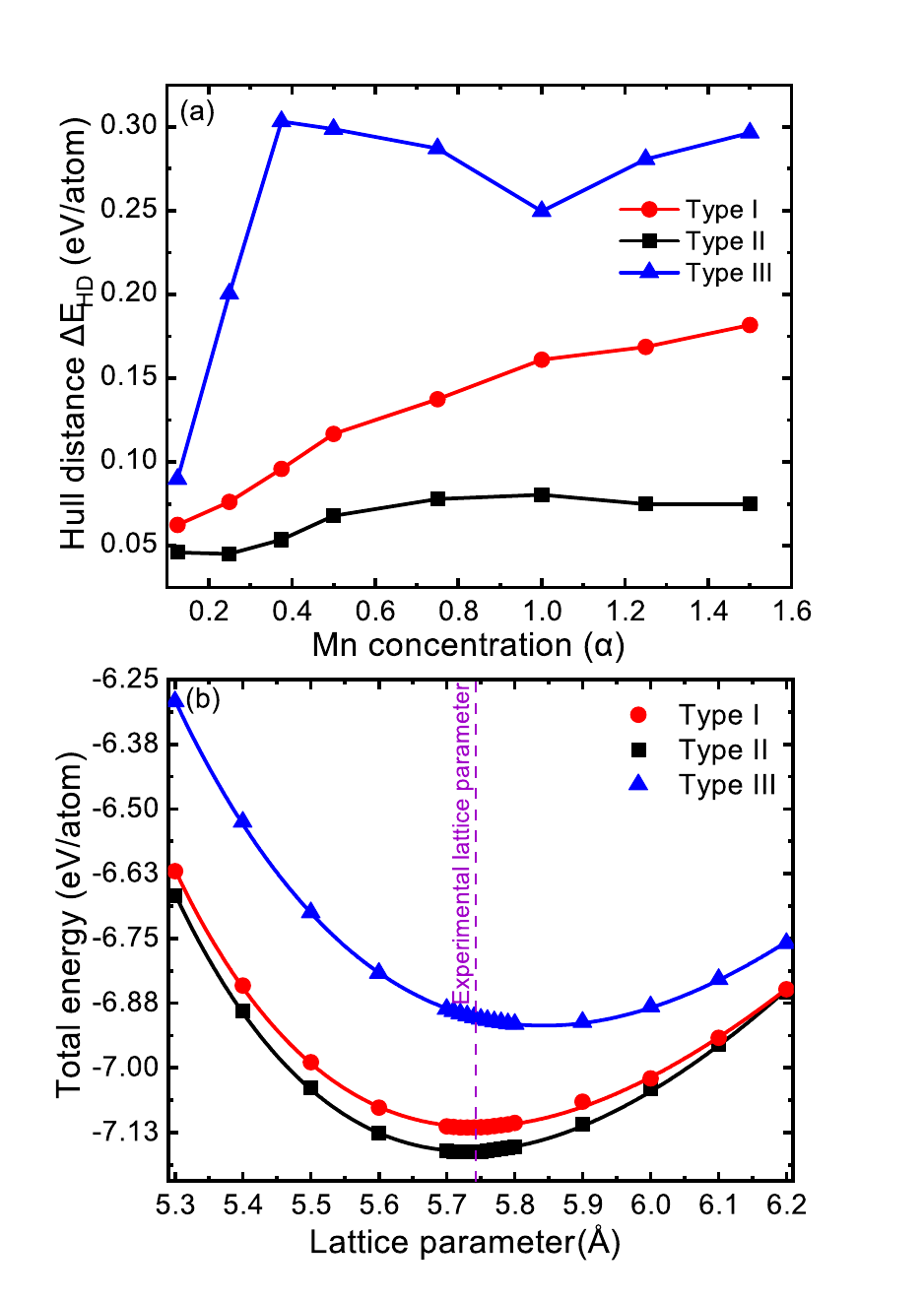}
\caption{(colour online) (a) The variation of hull distance ($\Delta E_{HD}$) as a function of Mn concentration. The type II has the minimum hull distance suggesting more stable phase than others. (b) Structure optimisation at $\alpha=0.50$ by calculating the total energy as a function of lattice parameter. This also suggests type II is energetically more favourable structure than other competing phases.}
\label{fig:hull_dis}
\end{figure}

At each Mn concentration, electronic structure calculations were performed for the three nonequivalent atomic configuration types, as reported in Table \ref{tab:atom_config1}. In order to obtain the total energy with greater accuracy, multiple volume relaxations were performed until the starting cell parameters were very close to the final cell parameters (i.e., the cell parameter does not change much after the relaxation). The magnetic moment, density of states, and the total energy were then calculated for the optimised structure, with no volume relaxation. The total energy was then utilised to calculate the hull distance ($\Delta E_{HD}$), for each configuration, following the equations (\ref{eq:form_energy}) and (\ref{eq:hull_dis}). The variation of $\Delta E_{HD}$ as a function of Mn concentration, for each configuration type, is shown in figure \ref{fig:hull_dis}(a). This clearly shows that $\Delta E_{HD}$ is lowest for configuration type II across the entire substitution range. Further, $\Delta E_{HD}$ increases rather slowly or remains constant in the case of type II as the Mn concentration increases, whereas that of type I and III increases significantly. Since $\Delta E_{HD}$ directly reflects the thermodynamic stability of a phase -- the greater the distance the less stable a phase is -- it can be argued that type II is more stable than the other configurations. This is consistent with the ``4-2 rule" discussed previously. As Mn is less electronegative (and also less valent) than Fe, placing Mn on on the B$_T$ site (i.e., sharing a sublattice with Ge) is more favourable than Mn occupying the vacant Co sites in sublattice A. See Fig. \ref{fig:unit_cell} for a visual interpretation of atomic arrangements.

In addition to optimisation by volume relaxation, structure optimisation was also performed by relaxing the structure for a set of fixed volumes, one such optimisation for $\alpha = 0.50$ is shown in figure \ref{fig:hull_dis}(b). From such optimisation, it is again found that type II yields the minimum energy for all unit cell volumes considered. Note that the energy difference between type I and II is small compared to that between II and III. This is because type I and II differ only in terms of site occupied by Fe and Mn. Due to the fact Fe and Mn are neighbours in the periodic table, one might initially suspect only a modest variation between these two structures. However, in light of the `4-2 rule'' discussion in the introduction, the double sublattice substitution of type II leads to a far more favourable electronic configuration for Fe in addition to reducing the valence electron count. At the energy minimum for $\alpha=0.50$, which occurs close to the experimental lattice parameter, the energy of type I is higher by 0.05 eV/atom than that of type II. We take the calculated energy difference between the structures in excess of about 0.05 eV/atom as a threshold to predict which phase is more likely, similar to our previous work\cite{shambhu2019tunable} and motivated by the findings of Ma \textit{et al.}\cite{ma2018computational}, which lends support to the hypothesis that type II is the more stable structure.
%%%%%%%%%%%%%%%%%%%%%%%
%%%%%%%%%%%%%%%%%%%

The observations above clearly suggest type II is the most stable atomic configuration type, however one should note here that $\Delta E_{HD}$ $\ne$ 0 even for this configuration. In principle, $\Delta E_{HD}$ should be exactly zero for a phase to be thermodynamically stable among many competing phases. For several reasons pointed out in section \ref{sec:level3}, it is possible for a phase to be stable in experiments despite lying above the convex hull (i.e., $\Delta E_{HD}$ $>$ 0). Perhaps the simplest is that the samples are annealed at $900^\circ$C, and given thermal energy of order $0.1\,$eV at that temperature, phases within $\sim\!0.1\,$eV/atom of the convex hull may be accessible. In recent studies, Ma \textit{et al.} have reported that many experimentally stable half Heusler compounds have $\Delta E_{HD}$ as much as 0.10 eV/atom \cite{ma2017computational}, whereas the full Heusler compounds have in the range 0.052 eV/atom \cite{ma2018computational}. Now keeping in mind of these possible values and since the $\Delta E_{HD}$ of type II is in the range 0.07 eV/atom, it is reasonable to believe that type II is thermodynamically stable. In contrast, the $\Delta E_{HD}$ of I and of III, both are markedly higher than 0.10 eV/atom even for moderate Mn concentration. This means these structures are thermodynamically unfavourable. Given how low the difference in energy is between structures I and II, and the fact that our experiments are not at $T=0$, it is perhaps reasonable to consider cases that are mixtures of types I and II, which amount to having both Fe and Mn on both sublattices. As we will see below, however, considering the magnetic properties of the structures exposes key difference between them and more strongly suggests structure type II is the favoured ordering.

\begin{figure}[ht]
\centering
\includegraphics[width = 0.75\columnwidth]{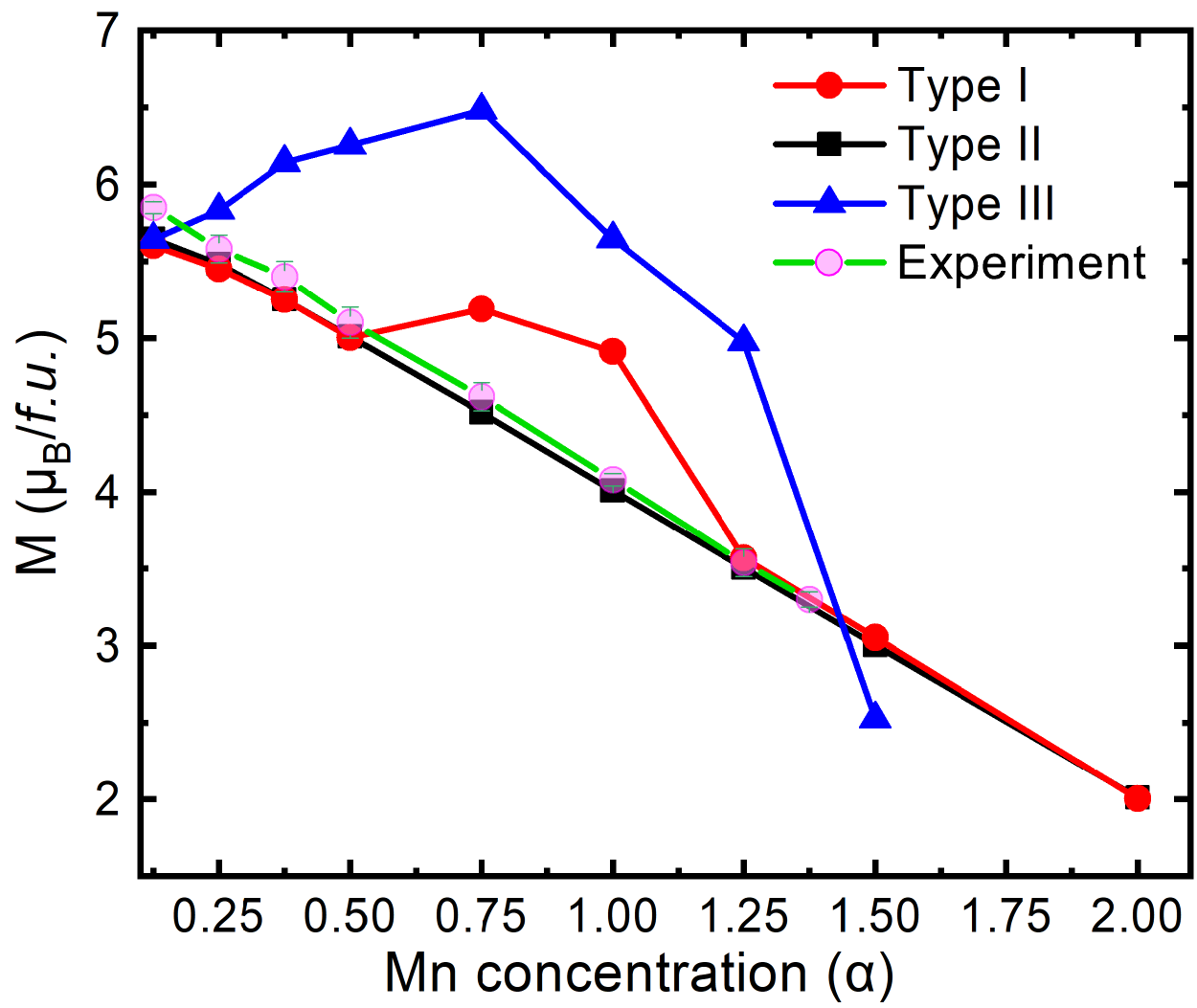}
\caption{(colour online) Variation of calculated total spin magnetic moment, and low temperature the experimental saturation moment with Mn concentration.}
\label{fig:mag_compar}
\end{figure}

\begin{figure}[ht]
\centering
\includegraphics[width = 0.75\columnwidth]{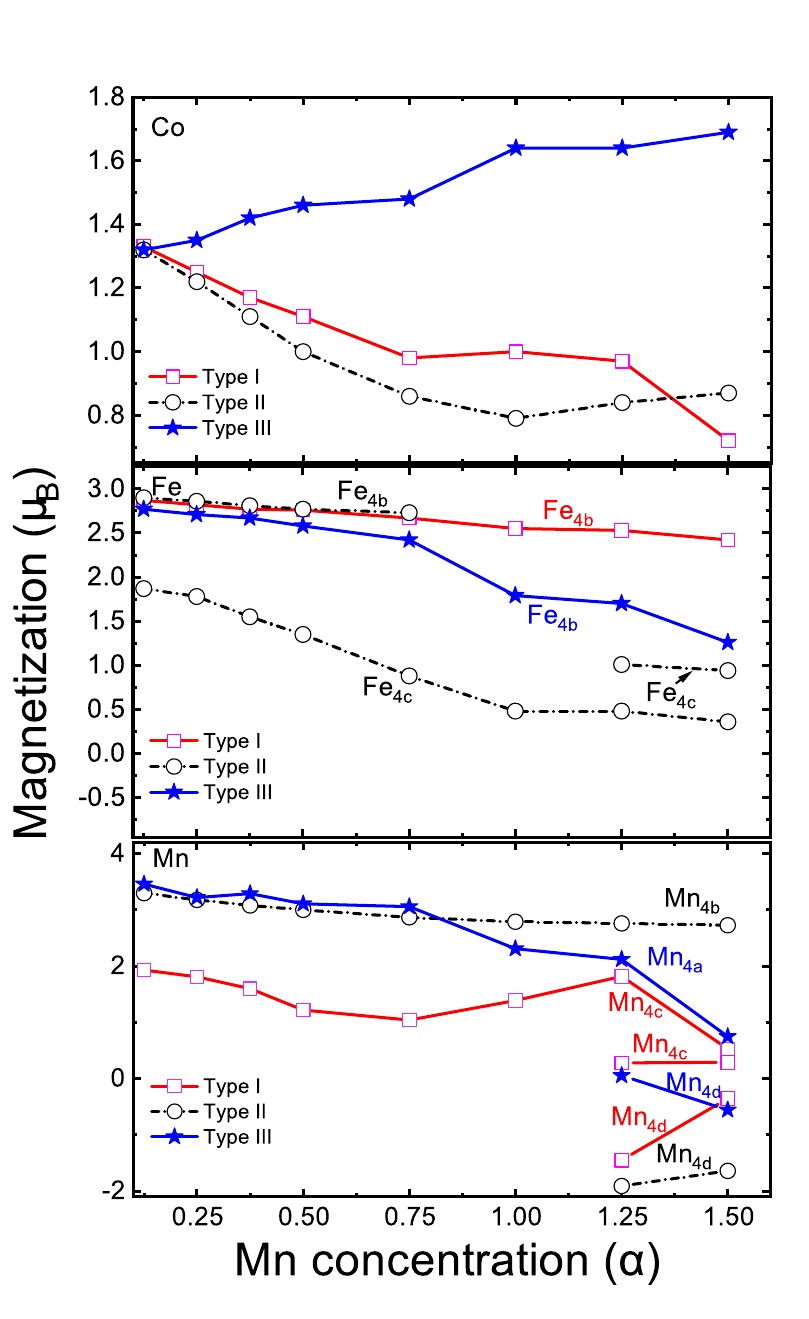}
\caption{(colour online) Variation of calculated atomic moments for three structure types, which shows clear distinction between the structures.}
\label{fig:atomic_moments}
\end{figure}

A comparison of calculated total moment with the experimental saturation moment near absolute zero (at 5 K in our case), shown in figure \ref{fig:mag_compar}, provides further insights and tighter constraints on the atomic configuration of the substituted alloys. The calculated magnetic moment of type III shows an irregular variation as the Mn concentration increases, and the moments are much higher than the experimental values -- well outside the plausible range of experimental uncertainty. This is because Mn and Fe contribute more to the total moment when they are placed in the sublattice B compared to when they are placed in the A sublattice with Co, see figure \ref{fig:atomic_moments}. Based on this, the type III structure can effectively be ruled out, which also agrees with the conclusion based on hull distance. From the 4-2 rule, one also expects $M$ to increase linearly with $\alpha$ the type 3 structure, contrary to experiemnt. For structures I and II, the moments are similar when Mn concentration is low, but as the Mn concentration increases these two moments diverge from each other and only the moments of type II seem to match fairly well with the experimental value over the entire single-phase range. Further, only type II shows the linear variation with Mn concentration observed in experiments and expected from the Slater-Pauling rule. In the light of this, we argue that on the whole only type II agrees reasonably well with the experimental results, suggesting this is the most stable configuration for the (Co$_{2-\alpha}$Mn$_\alpha$)FeGe alloy series. The detailed information obtained from our calculations on the atomic configurations, the total energy, total and atom specific magnetic moments, and the optimised lattice parameters are summarised in Table \ref{table:Co2-aMna_cal_res}. Note that for some of the structures, extra calculations were performed by swapping the atoms within the same sublattice (disordering), these are denoted with *. From those calculations, we noticed that the stable structure (i.e., type II) is not affected by such disordering as strongly as types I and type III (see table \ref{table:Co2-aMna_cal_res}).

Increasing the Mn concentration further than $\alpha$ = 1.0, it is observed that the total moment of type I structures decreases significantly, approaches the value of type II at $\alpha$ = 1.25, and remains nearly the same as type II up to $\alpha$ = 2.0. This behaviour clearly begs for an explanation. First, both configurations at $\alpha$ = 1.25 are identical to those at $\alpha$ = 1.0 (that is, 4 Fe atoms occupy B$_T$ site in type I, whereas this site is occupied by 4 Mn atoms in type II). The only difference is that the extra Mn atom at $\alpha$ = 1.25 directly occupies the vacant Co site (A sublattice, 4d site as shown in table \ref{table:Co2-aMna_cal_res}). Our calculations suggest that these Mn atoms on the A sublattice align antiferromagnetically with the other magnetic atoms. This is consistent with the 4-2 rule, which would suggest a 3$\uparrow$, 4$\downarrow$ configuration for Mn and a 5$\uparrow$, 4$\downarrow$ configuration for Co. Further, its influence on the atoms sharing the same sublattice (atoms on 4c site) is found to be different depending upon the atom types; the atomic moments have a great deal of variation if Mn occupies the 4c site (i.e., type I), compared to when 4c site is occupied by Fe atoms (i.e., type II). Notice the different moments for atoms of the same type at higher $\alpha$ in table \ref{table:Co2-aMna_cal_res} and in figure \ref{fig:atomic_moments}. Due this disproportionate influence, even though the individual atomic moments are remarkably different, the total moment is still found to be the same for types I and II.

While the overall picture clearly favours the type II structure, a more detailed experimental study using techniques sensitive to the local atomic ordering and especially the local magnetic state would be helpful. In the present case, $^{55}$Mn and $^{59}$Co NMR\cite{wurmehl2012nmr,wurmehl2013nmr,wurmehl2014nmr} and $^{57}$Fe M\"{o}ssbauer spectroscopy\cite{mende2021mos} would be particularly well suited, as would x-ray magnetic circular dichroism, and these techniques have been successfully employed on Heusler alloys previously. The possible antiferromagnetic alignment of Mn atoms and different moments for Fe and Mn atoms in different positions also clearly suggests neutron scattering. Hopefully one or more of these techniques can be applied in a future study. 

%%%%%%%%%%%%%%%%%%%%%%%%%%%
\begin{table*}
\scriptsize
\centering
\caption{DFT calculation results for possible atomic configurations. In the table 4a, 4b, 4c, and 4d represent the Wyckoff sites for the coordinates (0,0,0), (0.5,0.5,0.5), (0.25,0.25,0.25), and (0.75,0.75,0.75) respectively. E (in eV/atom) is total energy, m (in $\mu_B$) stands for magnetic moment, and $<a>$ (in \AA) represents the optimised lattice parameter (structures which exhibited structures other than cubic are represented by name, such as tetragonal (tetra.), orthorhombic (ortho.)). Note that in our model, 4c and 4d represents sublattice A, whereas 4a and 4b corresponds to sublattice B. While showing the atomic moments, if an atom exhibits two different values (depending on different site occupation), the moments are represented with the name of the site in the subscript.}
\clearrow
\begin{tabular}{c c c c c c c c}
\hline
\textbf{Configuration} & \textbf{4d} & \textbf{4c} & \textbf{4b} & \textbf{4a} & \textbf{E} & \textbf{m$_{tot}$} & \textbf{$<$a$>$} \\\hline
x = 0.125& & & & & & &  \\ 
I& 4Co& 3.5Co,0.5Mn& 4Fe& 4Ge& -6.96 & 5.60&5.746 \\[0.5ex]
\setrow{\bfseries}
II& 4Co& 3.5Co,0.5Fe& 3.5Fe,0.5Mn & 4Ge & -6.98 & 5.65&5.753 \\[0.5ex]
III& 4Co& 3.5Co,0.5Ge& 4Fe& 3.5Ge,0.5Mn& -6.93 & 5.64&5.755 \\[0.5ex] \hline
x = 0.250& & & &  & & &  \\ [0.5ex]
I& 4Co & 3Co,1Mn & 4Fe & 4Ge & -7.02 & 5.45& 5.741 \\[0.5ex]
\setrow{\bfseries}
II& 4Co & 3Co,1Fe & 3Fe,1Mn & 4Ge & -7.05 & 5.48 & 5.742 \\[0.5ex]
III& 4Co & 3Co,1Ge & 4Fe & 3Ge,1Mn & -6.89 & 5.83 & 5.802 \\[0.5ex]\hline
x = 0.375& & & & & & &  \\ [0.5ex]
I& 4Co & 2.5Co,1.5Mn & 4Fe & 4Ge & -7.07 & 5.251 & 5.736 \\[0.5ex]
\setrow{\bfseries}
II& 4Co & 2.5Co,1.5Fe & 2.5Fe,1.5Mn & 4Ge & -7.11 & 5.25 & 5.736 \\[0.5ex]
III& 4Co & 2.5Co,1.5Ge & 4Fe & 2.5Ge,1.5Mn & -6.86 & 6.14 & tetra. \\[0.5ex]\hline
x = 0.500& & & & & & &  \\ [0.5ex]
I& 4Co & 2Co,2Mn & 4Fe & 4Ge & -7.12 & 5.00 & 5.731 \\[0.5ex]
I*& 3Co,1Mn & 3Co,1Mn & 4Fe & 4Ge & -7.12 & 4.99 & 5.733 \\[0.5ex]
\setrow{\bfseries}
II& 4Co & 2Co,2Fe & 2Fe,2Mn & 4Ge & -7.17 & 5.01 & 5.728 \\[0.5ex]
II*& 3Co,1Fe & 3Co,1Fe & 2Fe,2Mn & 4Ge & -7.16 & 5.00 & 5.730 \\[0.5ex]
III& 3Co,1Ge & 3Co,1Ge & 4Fe & 2Ge,2Mn & -6.93 & 6.25 & 5.835 \\[0.5ex]
III*& 4Co & 2Co,2Ge & 4Fe & 2Ge,2Mn & -6.98 & 4.80 & Ortho. \\[0.5ex]\hline
x = 0.750& & & & & & &  \\ [0.5ex]
I& 4Co & 1Co,3Mn & 4Fe & 4Ge & -7.21 & 4.64 & 5.726 \\[0.5ex]
I*& 3Co,1Mn & 2Co,2Mn & 4Fe & 4Ge & -7.21 & 5.19 & 5.744 \\[0.5ex]
\setrow{\bfseries}
II& 4Co & 1Co,3Fe & 1Fe,3Mn & 4Ge & -7.27 & 4.51 & 5.716 \\[0.5ex]
II*& 3Co,1Fe & 2Co,2Fe & 1Fe,3Mn & 4Ge & -7.27 & 4.51 & 5.713 \\[0.5ex]
III& 4Co & 1Co,3Ge & 4Fe & 1Ge,3Mn & -7.06 & 6.48 & 5.851 \\[0.5ex]
III*& 3Co,1Ge & 2Co,2Ge & 4Fe & 1Ge,3Mn & -7.14 & 5.14 & tetra. \\[0.5ex]\hline
x = 1.000& & & & & & &  \\ [0.5ex]
I& 4Co & 4Mn & 4Fe & 4Ge & -7.29 & 4.91 & 5.739 \\[0.5ex]
\setrow{\bfseries}
II& 4Co & 4Fe & 4Mn & 4Ge & -7.37 & 4.01 & 5.711 \\[0.5ex]
III& 4Co & 4Ge & 4Fe & 4Mn & -7.21 & 5.65 & 5.768 \\[0.5ex]\hline
x = 1.250& & & & & & &  \\ [0.5ex]
I& 3Co,1Mn & 4Mn & 4Fe & 4Ge & -7.39 & 3.57 & 5.754 \\[0.5ex]
\setrow{\bfseries}
II& 3Co,1Mn & 4Fe & 4Mn & 4Ge & -7.49 & 3.51 & 5.716 \\[0.5ex]
II*& 3Co,1Fe & 3Fe,1Mn & 4Mn & 4Ge & -7.48 & 3.51 & 5.724 \\[0.5ex]
III& 3Co,1Mn & 4Ge & 4Fe & 4Mn & -7.28 & 4.97 & 5.766 \\[0.5ex]\hline
x = 1.500& & & & &  & &  \\ [0.5ex]
I& 2Co,2Mn & 4Mn & 4Fe & 4Ge & -7.39 & 3.053 & 5.711 \\[0.5ex]
\setrow{\bfseries}
II& 2Co,2Mn & 4Fe & 4Mn & 4Ge & -7.60 & 3.004& 5.728 \\[0.5ex]
II*& 2Co,2Fe & 2Fe,2Mn & 4Mn & 4Ge & -7.58 & 3.030 & 5.728 \\[0.5ex]
III& 2Co,2Mn & 4Ge & 4Fe & 4Mn & -7.37 & 2.523 & 5.728 \\[0.5ex]\hline
x = 2.00& & & & &  & &  \\ [0.5ex]
I& 4Mn & 4Mn & 4Fe & 4Ge & -7.70 & 2.004 & 5.710 \\[0.5ex]
\setrow{\bfseries}
II& 4Mn & 4Fe & 4Mn & 4Ge & -7.81 & 2.010 & 5.725 \\[0.5ex]
\end{tabular}
\label{table:Co2-aMna_cal_res}
\end{table*}
%%%%%%%%%%%%%%%%%%%%%%%%%%%%%%%%%%%%%%
%%%%%%%%%%%%%%%%%%%%%
%%%%%%%%%%%%%%%%%%%%%%%
%%%%%%%%%%
%%%%%%%%%%%%%%%%%%%%%
%%%%%%%%%%%%%%%%%%%%%%%
%%%%%%%%%%%%%%%%%%
\begin{figure*}[ht]
\centering
\includegraphics[width = 1.8\columnwidth]{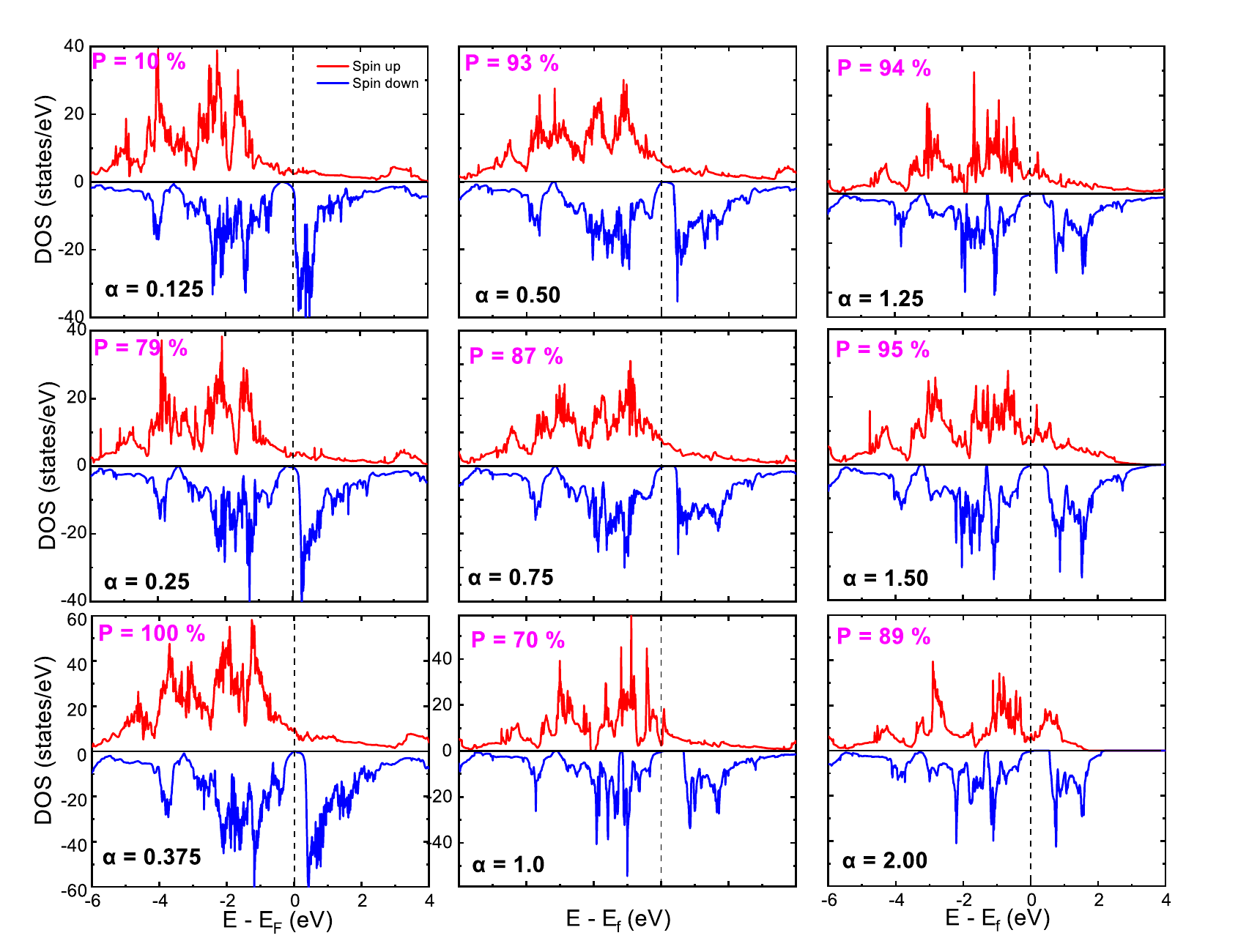}
\caption{(colour online) Spin-resolved total density of states plot for (Co$_{2-\alpha}$Mn$_\alpha$)FeGe (0.125 $\le$ $\alpha$ $\le$ 2) alloy series.}
\label{fig:Co2-aMna_dos}
\end{figure*}
%%%%%%%%%%%%

%%%%%%%%%%%%%%%%%
\begin{figure*}[ht]
\centering
\includegraphics[height = 1.2\columnwidth]{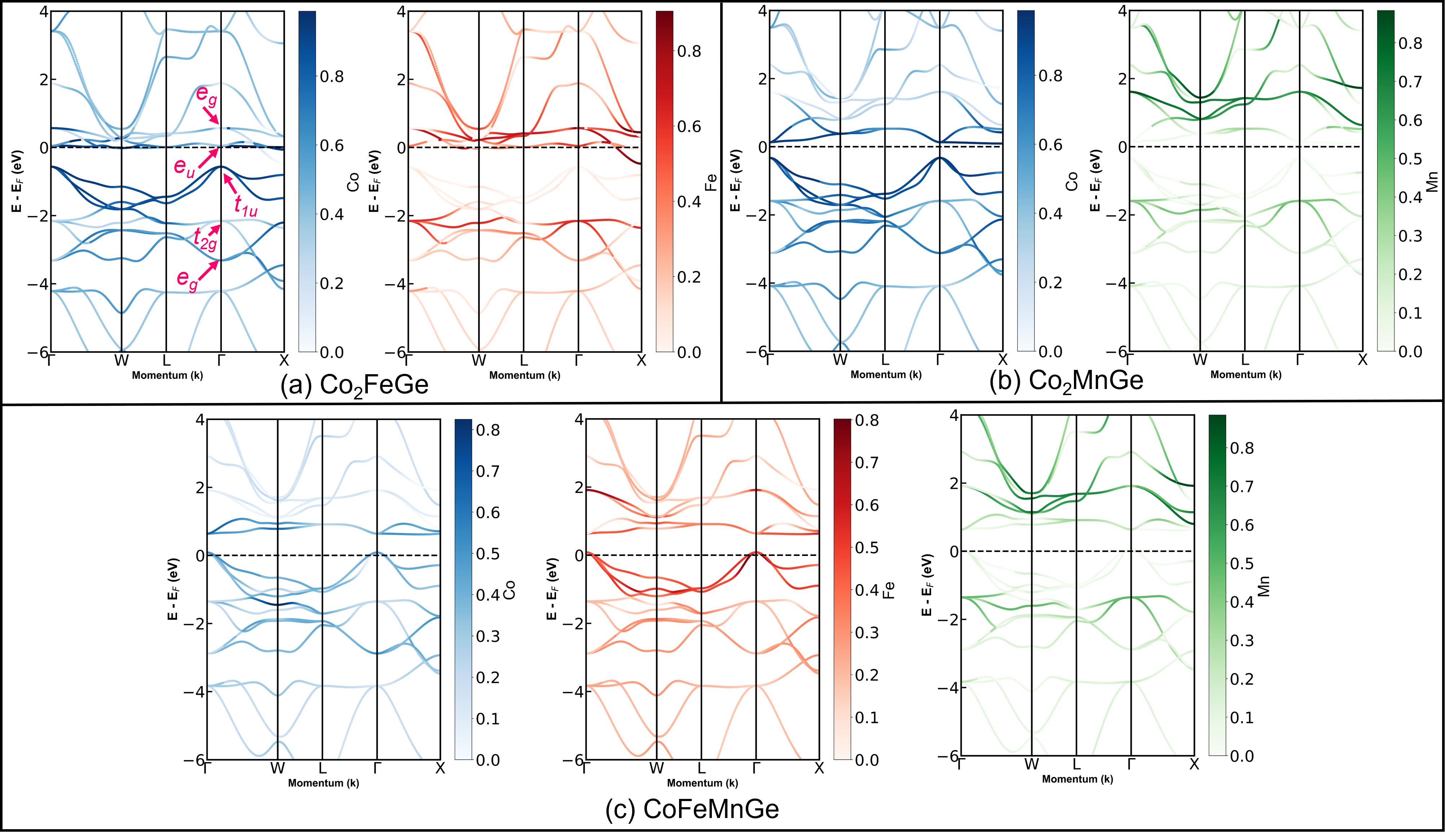}
\caption{(colour online) Minority channel band structure of (a) Co$_2$FeGe, (b) Co$_2$MnGe, and (c) CoFeMnGe. For each alloy, separate plot is shown for each transition element (Co in blue, Fe in red, and Mn in green), with the colour intensity that corresponds to weight of each atomic orbital to that band. The darker colour represents band is mainly contributed by that atomic orbital. See text for details.}
\label{fig:band_structure}
\end{figure*}
%%%%%%%%%%%%%%

So far, from the experimental and theoretical investigations, we have plausible evidence that type II, in which the substituted Mn atom indirectly substitutes the Fe atom instead of Co atom, is the stable atomic configuration. Hence, we use this configuration to further discuss the electronic properties of these alloys. The spin-resolved total density of states (DOS) plot of (Co$_{2-\alpha}$Mn$_\alpha$)FeGe at different $\alpha$ is shown in figure \ref{fig:Co2-aMna_dos}. For very low Mn concentration, $\alpha$ = 0.125, E$_F$ is located at the lower edge of the conduction band leading to finite minority states at the E$_F$, hence the alloy exhibits only 10 \% spin polarisation. The total calculated moment is found to be 5.65 $\mu_B/f.u.$, which is slightly less than the value expected from the S-P rule (i.e., 5.75 $\mu_B/f.u.$) and also less than the experimental value (i.e., 5.85 $\mu_B/f.u.$). Based on our previous experience\cite{shambhu2019tunable}, one could {\em potentially} have a discrepancy of $\pm$ 0.1 $\mu_B/f.u.$ between the experimental moment and the theoretical moment due to experimental uncertainties. However, an offset of $\pm$ 0.2 $\mu_B/f.u.$ in case of Co$_{1.875}$Mn$_{0.125}$FeGe is well beyond any anticipated experimental errors, and suggests on-site electron correlations (i.e., adding a $U$ parameter to GGA) may be required to find agreement between calculated and experimental moments. Note that similar correlations were necessary to match the experimental moment of Co$_2$FeSi\cite{wurmehl2005geometric}. It is worth mentioning here that an addition of similar correlation for Co$_2$FeGe also produces a moment of 6 $\mu_B$/f.u.\cite{uvarov2012electronic}. If this is true then it is clear that the experimentally measured moment (that is 5.68 $\mu_B$/f.u.) is not the true moment of Co$_2$FeGe; the moment is less due to the presence of a more weakly magnetic secondary phase. This provides further justification to our explanation of why there is upturn in both M$_\text{s}$ and T$_\text{C}$ in going from $\alpha$ = 0 to 0.125, as pointed out in the experimental section. 

At $\alpha$ = 0.25, it can be observed that the conduction band has pushed away from the E$_F$, see figure \ref{fig:Co2-aMna_dos}. However, there are still small number of minority states present at the E$_F$, thus this alloy exhibits a spin polarisation of 79 \%. The calculated total moment is found to be 5.48 $\mu_B$/f.u., which is only 0.02 $\mu_B$/f.u. less than the M$_{S-P}$ value. Upon further increasing Mn concentration, the conduction band has shifted further away from $E_F$, which pushes E$_F$ downwards such that it falls on the band gap. In fact, a perfect half-metallic behaviour is obtained at $\alpha$ = 0.375. Since there are no minority states, a 100 \% spin polarisation is achieved. The calculated moment is obtained to be 5.25 $\mu_B$/f.u., which is  equal to the S-P value.

With further increment of Mn, E$_F$ begins touching the valence band edge. This suggests there is not only the conduction band that is being shifted upwards but also the valence band. The half-metallic behaviour is lost and the spin-polarisation drops steadily up to $\alpha$ = 1.0. After crossing $\alpha$ = 1.0, there is again sudden increase in spin-polarisation. This  can be attributed to the higher number of states in the majority band. Thus, the spin-polarisation values in both sides of $\alpha$ = 1.0, that is, 0.75 and 1.25, are higher than that at $\alpha$ = 1.0. This suggests the spin-polarisation of CoMnFeGe, for which very high value of P has been measured in experiment, can be enhanced by slightly tweaking the stoichiometry of Co and Mn, either increasing Co by replacing Mn atoms or vice-versa. Indeed, this was one of the goals -- finding materials with improved properties -- of this work.

The investigation so far suggests the substitution of Mn for Co atoms in Co$_2$FeGe not only stabilises a single-phase compound, but also tunes the system to half-metallic character. Half-metallicity is realised as a consequence of band shifting (or in other words the increase in band energies due to substitution). Now it is worth explaining which bands are shifted and how that can lead to a half-metallic compound based on Co$_2$FeGe, which otherwise is a normal ferromagnet.

As noted above, Co$_2$FeGe doesn't exhibit half-metallic character. In ref.\ \cite{shambhu2019tunable}, we provided a simple explanation of why it is difficult to obtain half-metallicity for Co$_2$FeGe. Adding to that, a more compelling explanation can be found from the investigation of band structures, as shown in figure \ref{fig:band_structure}. The minority channel electronic band diagram for three alloys, viz. (a) Co$_2$FeGe, (b) Co$_2$MnGe, and (c) CoFeMnGe, are shown. The bands are depicted in terms of colour-map, where the colour intensity represents the extent of elemental contribution to that particular band. We are interested in the substitution between transition elements, so only the contribution of those elements are shown. Following the orbital hybridisation schemes suggested in ref.\cite{galanakis2002slater} for an X$_2$YZ type full Heusler alloy, the bands which are located just below and above E$_F$ (i.e., t$_{1u}$ and e$_u$ respectively) are contributed to mainly by atoms on the X site (the A sublattice), whereas the Y or B$_T$ atoms contribute mainly to the t$_{2g}$, e$_{1g}$, and e$_{g}$ bands. In fact, this is what one can observe in figure \ref{fig:band_structure}; the t$_{1u}$ and e$_u$ bands are strongly Co type in case of Co$_2$FeGe and Co$_2$MnGe, and also of Fe type in case of CoFeMnGe since Fe occupies one of the X sites for this alloy. However, in the band diagram of Co$_2$FeGe one can observe a band, which is predominantly localised on the Y site (i.e., Fe), drops below E$_F$ at the X point. As a result there are finite states at E$_F$, which provides a compelling argument why Co$_2$FeGe is not a half-metal. This seems qualitatively consistent with our speculations from the 4-2 rule: Fe in Co$_2$FeGe must adopt a 6 majority / 2 minority electron configuration, forcing electrons to occupy higher energy states. Similar behaviour was observed in related Heusler alloys with Fe occupying the Y site, such as Co$_2$FeSi, Co$_2$FeAl, Co$_2$FeSn, Fe$_2$CoSi, as pointed out by Faleev \textit{et al.}\cite{faleev2017unified} Hence, the general argument that it is unfavourable to obtain a half-metallic full Heusler compound with Fe occupying the B$_T$ site appears to hold, unless the Z atom is able to accommodate more electrons (e.g., Z=Al or Ga).

In the case of Co$_2$MnGe (Fig.~\ref{fig:band_structure}(b)), where Mn is located in the B$_T$ site instead of Fe, the band which {\em was} crossing E$_F$ has been pushed to higher energies and now lies above the E$_F$. This is because when a heavier atom (such as Fe) on a lattice site is exchanged for a lighter atom (such as Mn), the bands deriving from orbitals of atoms on that site move upward in energy. Since none of the bands cross the E$_F$ in case of Co$_2$MnGe, it exhibits half-metallic character. Further, keeping the B$_T$ site unchanged, if one replaces a heavier atom from sublattice A, such as going from Co$_2$MnGe to CoFeMnGe, one can again expect to see an upwards shifting of the bands that are localised in those sites (i.e., t$_{1u}$ and e$_u$). On comparing figs. \ref{fig:band_structure}(b) and (c), it is clearly observed that the t$_{1u}$ and e$_u$ bands have shifted upwards, and since as a result the t$_{1u}$ bands touch the Fermi level, half-metallic behaviour is lost again for CoFeMnGe. 

From the present investigation, it can be argued that the half-metallic character can be tuned in Co$_2$FeGe, if we substitute Fe by a less valence element. In case of (Co$_{2-\alpha}$Mn$_\alpha$)FeGe, where it appears as written that Mn is substituting for Co, the above theoretical and experimental results suggested that in fact Mn is more likely to occupy the Fe sites by displacing Fe towards the Co sites. This is why obtaining a half-metallic state is feasible, and we suggest may be realised at $\alpha = 0.375$. Since there is replacement of heavier atom by a lighter atom on two sites, that is replacement of Fe by Mn in B$_T$ site and replacement of Co by Fe in sublattice A, many bands are shifted upward in energy. The net result is a rather complicated shifting of bands that gives a narrow composition range for half-metallic behaviour, realised at $\alpha=0.375$ but lost by $\alpha=0.50$, and very narrow band gaps (see fig. \ref{fig:Co2-aMna_dos}).

%%%%%%%%%%%%%%%%%%%%%%%%%%%%%%%%%%
\section{\label{sec:level5}Summary for (Co$_{2-\alpha}$Mn$_\alpha$)FeGe}
The substitution of Mn for Co atoms in Co$_2$FeGe, viz.\ (Co$_{2-\alpha}$Mn$_\alpha$)FeGe series, can stabilise a single-phase compound over a wide range of Mn concentration, up to $\alpha$ $<$ 1.375. The single-phase compounds crystallise in fcc structure, with suggested chemical ordering that follows the 4-2 rule. The experimental magnetic moment shows a linear relationship with the number of valence electrons and agrees well with the Slater-Pauling rule,  suggesting possible half-metallic character in these alloys. The T$_\text{C}$ of the compounds is significantly higher than room temperature (as high as 1000 K), and it also exhibits a seemingly linear variation with the valence electron count. First principle calculations performed on several possible configurations suggested that the configuration type II, in which the substituted Mn atom occupies the Fe site by displacing it towards the vacant Co site, is the most stable configuration. Only in configuration type II do the theoretical and experimental low temperature saturation moments agree for all single-phase compounds. From the analysis of electronic structure, it is found that the introduction of Mn pushes the bands which are crossing the Fermi level in the minority channel towards higher energy. This opens up a minority gap and thus a half-metallic behaviour can be obtained at the Mn concentration $\alpha$ = 0.375. For other compounds, nearly half-metallic character with very high spin-polarisation is obtained. For instance, two compounds in the vicinity of CoMnFeGe, i.e., Co$_{1.25}$Mn$_{0.75}$FeGe and Co$_{0.75}$Mn$_{1.25}$FeGe, exhibit higher spin-polarisation than that of CoMnFeGe, suggesting one may enhance the spin-polarisation of CoMnFeGe by slightly tweaking the Co/Mn ratio. It is also found that competition between two crystal structures could be one of the reasons why the Mn rich alloys decompose into multiple phases. Hopefully, with the discovery of a few potentially technologically relevant materials and a careful and detailed investigation of a novel substitutional series over a very wide range of compositions, this paper can serve as motivation in the discovery of additional novel functional materials.

%%%%%%%%%%%%%%%%%
\begin{table*}[htb]\scriptsize
\caption{Possible atomic arrangements for X$_{2-\alpha}$X'$_\alpha$Y$_{1-\beta}$Y'$_\beta$Z substitutional series of X$_2$YZ Heusler alloy. Note that $\alpha, \beta \geq 0$, and we place the additional constraint that only one of $(\alpha,\beta)$ are nonzero -- when $\alpha > 0$, $\beta = 0$ and vice versa. Also note for $\alpha$ $>$ 1, the excess X' atoms occupy the vacant 4d site. The site interchanges 4a $\leftrightarrow$ 4b or 4c $\leftrightarrow$ 4d, and similarly the sublattice interchange (4a, 4b) $\leftrightarrow$ A(4c, 4d), generate equivalent structures.}
\renewcommand*{\arraystretch}{1.1}
\centering
\begin{tabular}{c c c c c}
\hline\hline
& \multicolumn{2}{c}{\bf Sublattice B} & \multicolumn{2}{c}{\bf Sublattice A}\\ \hline
\textbf{Type} & \multicolumn{1}{c}{\centering{\textbf{4a ($B_Z$)}}} & \multicolumn{1}{c}{\centering{\textbf{4b ($B_T$)}}} & \multicolumn{1}{c}{\centering{\textbf{4c}}} & \multicolumn{1}{c}{\centering{\textbf{4d}}}\\ \hline
 I & Z & (1-$\beta$)Y+$\beta$Y' & (1-$\alpha$)X+$\alpha$X'& X \\[1ex]
 II & Z & (1-$\alpha$-$\beta$)Y+$\alpha$X'+$\beta$X & (1-$\alpha$-$\beta$)X+$\alpha$Y+$\beta$Y' & X\\[1.5ex]
  III& (1-$\alpha$-$\beta$)Z+$\alpha$X'+$\beta$Y' & (1-$\beta$)Y+$\beta$Z & (1-$\alpha$)X+$\alpha$Z & X\\[1.5ex]
\hline\hline
\end{tabular}
\label{tab:atom_config2}
\end{table*}
%%%%%%%%%%%%%%%%%

%%%%%%%%%%%%%%%%%%%%%%%%%%%%%%%%%%%
%\section{\label{sec:level2}Experimental Details}
%\input{experimental}
%%%%%%%%%%%%%
%\section{\label{sec:level2}Computational Details}
%\input{computational}
%%%%%%%%%%%%%%%%%%%%%%%%%%%%%%%%%%
\section{\label{sec:level6}Result and Discussion: Co$_{2}$(Fe$_{1-\beta}$Mn$_{\beta}$)Ge and (Co$_{2-\alpha}$Fe$_{\alpha}$)MnGe }
%%%%%%%%%
\subsection{Co$_2$(Fe$_{1-\beta}$Mn$_\beta$)Ge}
\subsubsection{Finding from experiment}
\begin{figure}[htb]
\centering
\includegraphics[width =0.6\columnwidth]{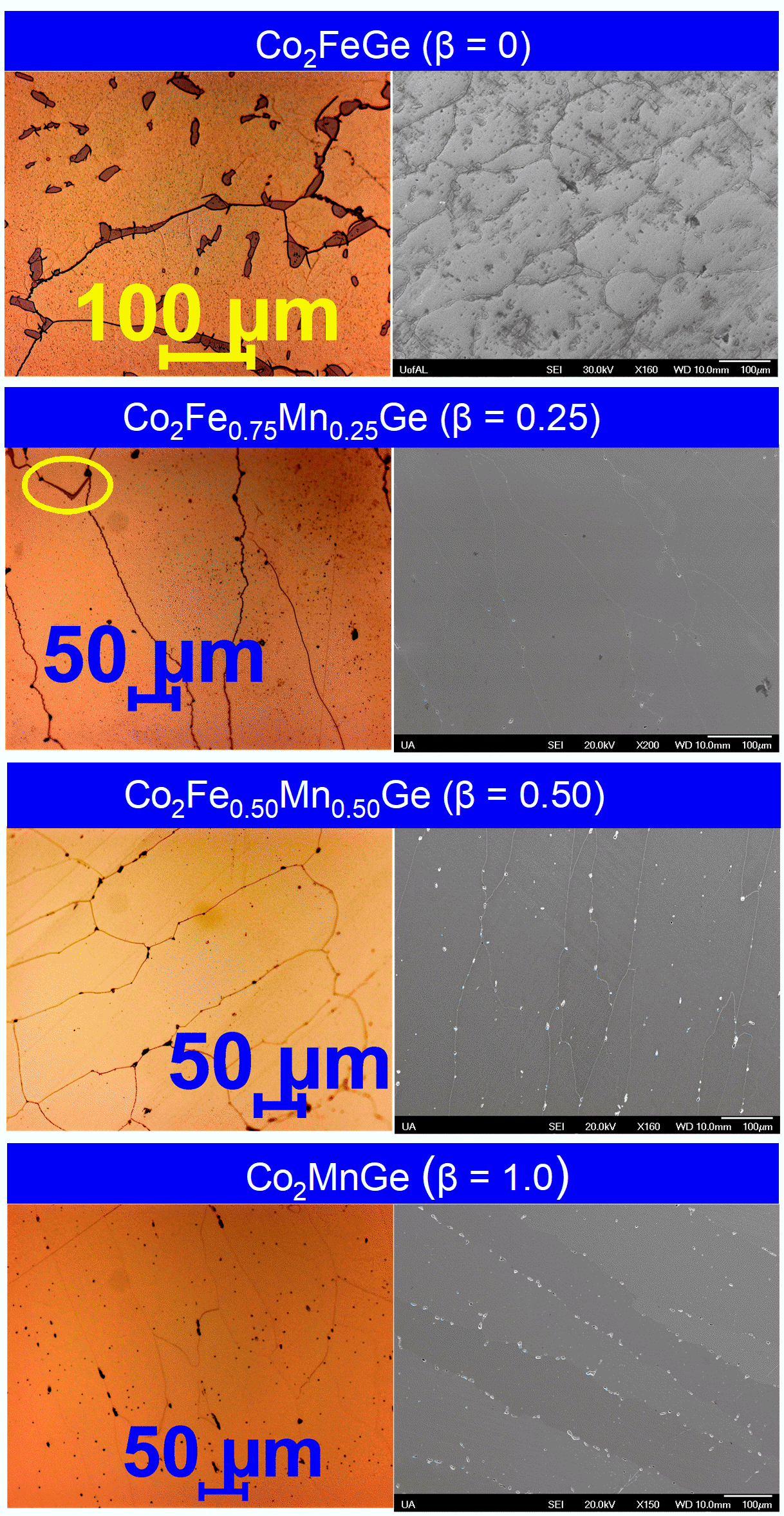}
\caption{(colour online) Optical (left) and electron (right) micrographs of Co$_2$(Fe$_{1-\beta}$Mn$_\beta$)Ge substitutional series heat treated at 900$\degree$C for 3 days. These micrographs reveal single-phase granular microstructure can be obtained after substituting Fe by Mn in parent Co$_2$FeGe alloy, which exhibits a multi-phase microstructure.}
\label{fig:optical_Co2Fe1-bMnb}
\end{figure}

For this series, four samples were prepared in steps of $\Delta \beta = 0.25$. The optical and electron micrographs are shown in figure \ref{fig:optical_Co2Fe1-bMnb}, which suggest formation of a single-phase granular microstructure after $\beta\!\geq\!0.25$, with grains varying in sizes from around 50 to 300 $\mu$m. Apart from grain boundaries, no other distinctive phases could be detected in the micrographs, suggesting substitution of Mn for Fe atoms in Co$_2$FeGe also stabilises a single-phase compound, similar to substitution of Mn for Co atoms.  
%%%%%%%%%%%%%%%%%%
%%%%%%%%%%%%%%%%%%
Careful observation reveals that at $\beta$ = 0.25 grain boundaries have somewhat different contrast, and are also thicker and darker in colour (notice the region inside yellow circle in \ref{fig:optical_Co2Fe1-bMnb}). This indicates the grain boundary composition could be different from the \textit{with-in} grain composition, which will be examined by EDS measurements in the next section.

\begin{table*}[htb]
\scriptsize
\caption{EDS determined grain, and grain boundary composition of Co$_2$(Fe$_{1-\beta}$Mn$_\beta$)Ge alloys. Note that the composition reported here were determined from the average of more 10 than measurements covering different regions across the sample. The reported composition may have an uncertainty (instrumental + random) of $\pm$ 5\%.}
\centering
\begin{tabular}{c c c}
\hline\hline
\textbf{Alloy} & \multicolumn{1}{p{3cm}}{\centering{\textbf{Grain \\  composition}}}& \multicolumn{1}{p{4cm}}{\centering{\textbf{Grain boundary \\ composition}}}\\
[0.5ex]
\hline
Co$_2$FeGe* & Co$_{1.97}$Fe$_{1.06}$Ge$_{0.97}$& Co$_{2.14}$Fe$_{0.43}$Ge$_{1.43}$  \\[1.5ex]
Co$_{2}$Fe$_{0.75}$Mn$_{0.25}$Ge & Co$_{1.97}$Fe$_{0.77}$Mn$_{0.27}$Ge$_{0.99}$ & Co$_{2.14}$Fe$_{0.38}$Mn$_{0.15}$Ge$_{1.33}$ \\[1.5ex]
Co$_{2}$Fe$_{0.50}$Mn$_{0.50}$Ge & Co$_{1.97}$Fe$_{0.52}$Mn$_{0.54}$Ge$_{0.98}$ & Co$_{1.97}$Fe$_{0.50}$Mn$_{0.54}$Ge$_{0.99}$ \\[1.5ex]
Co$_{2}$Fe$_{0.25}$Mn$_{0.75}$Ge & Co$_{1.97}$Fe$_{0.26}$Mn$_{0.80}$Ge$_{0.97}$ & Co$_{1.96}$Fe$_{0.23}$Mn$_{0.78}$Ge$_{1.03}$ \\[1.5ex]
Co$_{2}$MnGe &
Co$_{1.98}$Mn$_{1.04}$Ge$_{0.98}$ & Co$_{1.97}$Mn$_{0.98}$Ge$_{1.05}$ \\[1.5ex]
\hline\hline
\end{tabular}
\label{tab:Co2Fe1-bMnbEDS_compo}
\end{table*}
%%%%%%%%%%%%%%%

The EDS composition analysis, as shown in Table \ref{tab:Co2Fe1-bMnbEDS_compo}, indeed verifies that at $\beta$ = 0.25 the grain boundaries are rich in Ge but deficient in Fe compared to the target composition, markedly different from within the grains. The secondary phase observed in the parent Co$_2$FeGe also had a similar composition -- rich in Ge but deficient in Fe. This means after the Mn substitution, any remaining impurity phases are apparently localised at the grain boundaries, and they can be suppressed further by increasing the Mn concentration. For all other compositions ($\beta$ $>$ 0.25), both the compositions (\textit{\textit{i.e.}} within grains and at the grain boundaries) are fairly close to each other, and hence these alloys can be treated as single-phase alloys. This is also supported by the elemental maps. A map of $\beta$ = 0.50 is provided in the supplementary information, which shows homogeneous elemental distribution.\\
%%%%%%%%%%%%%%%%%%%%%%
\begin{figure}[htb]
\centering
\includegraphics[width =1.0\columnwidth]{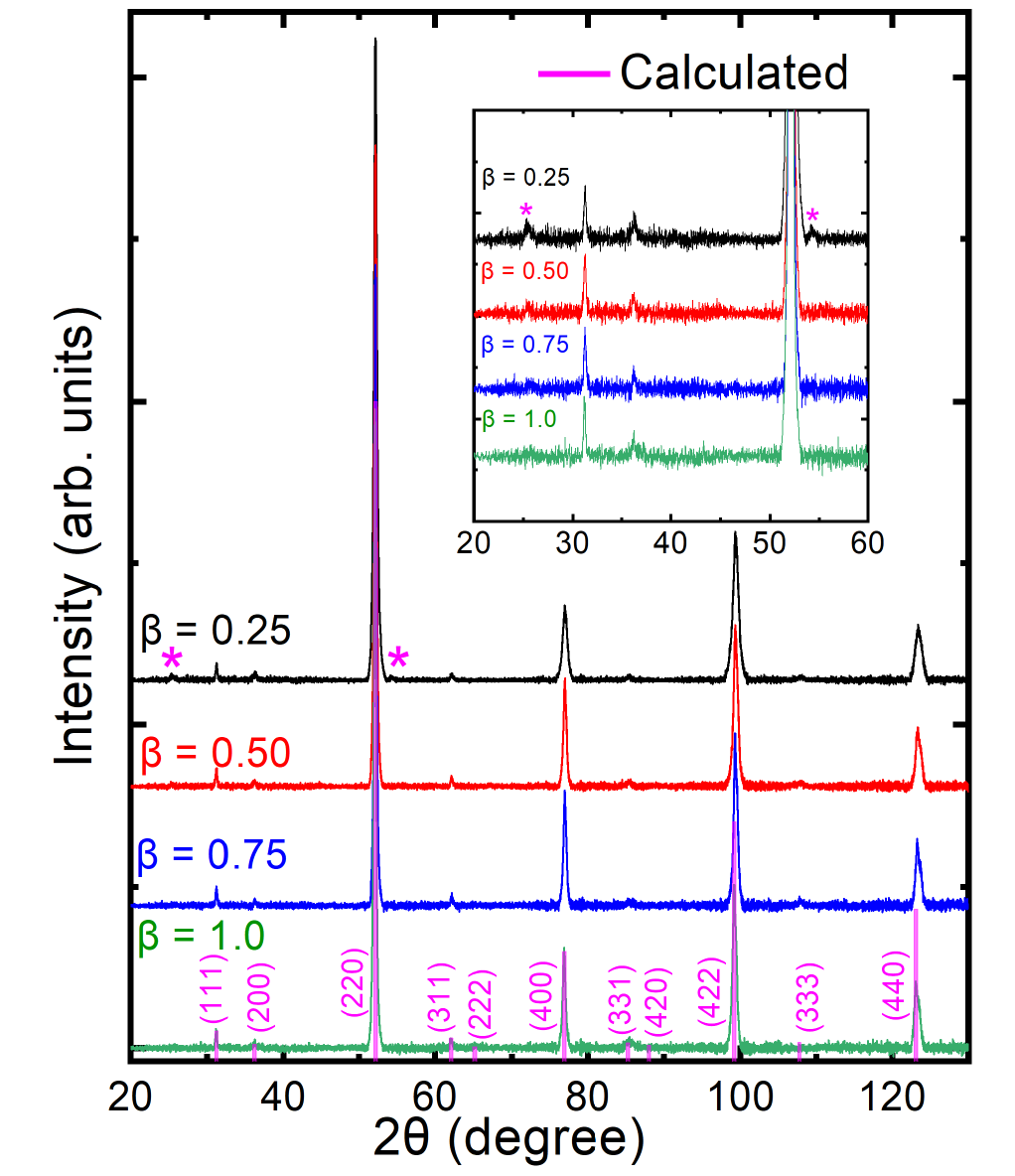}
\caption{(colour online) XRD scans of Co$_2$(Fe$_{1-\beta}$Mn$_\beta$)Ge alloy series. (Inset) The scans at lower angles are expanded to make impurity peaks clearly visible.}
\label{fig:xrd_Co2Fe1-bMnb}
\end{figure}
%%%%%%%%%%%%%%%%%%%%
%%%%%%%%%%%%%%%%%%%%%
\begin{figure}[htb]
\centering
\includegraphics[width =0.8\columnwidth]{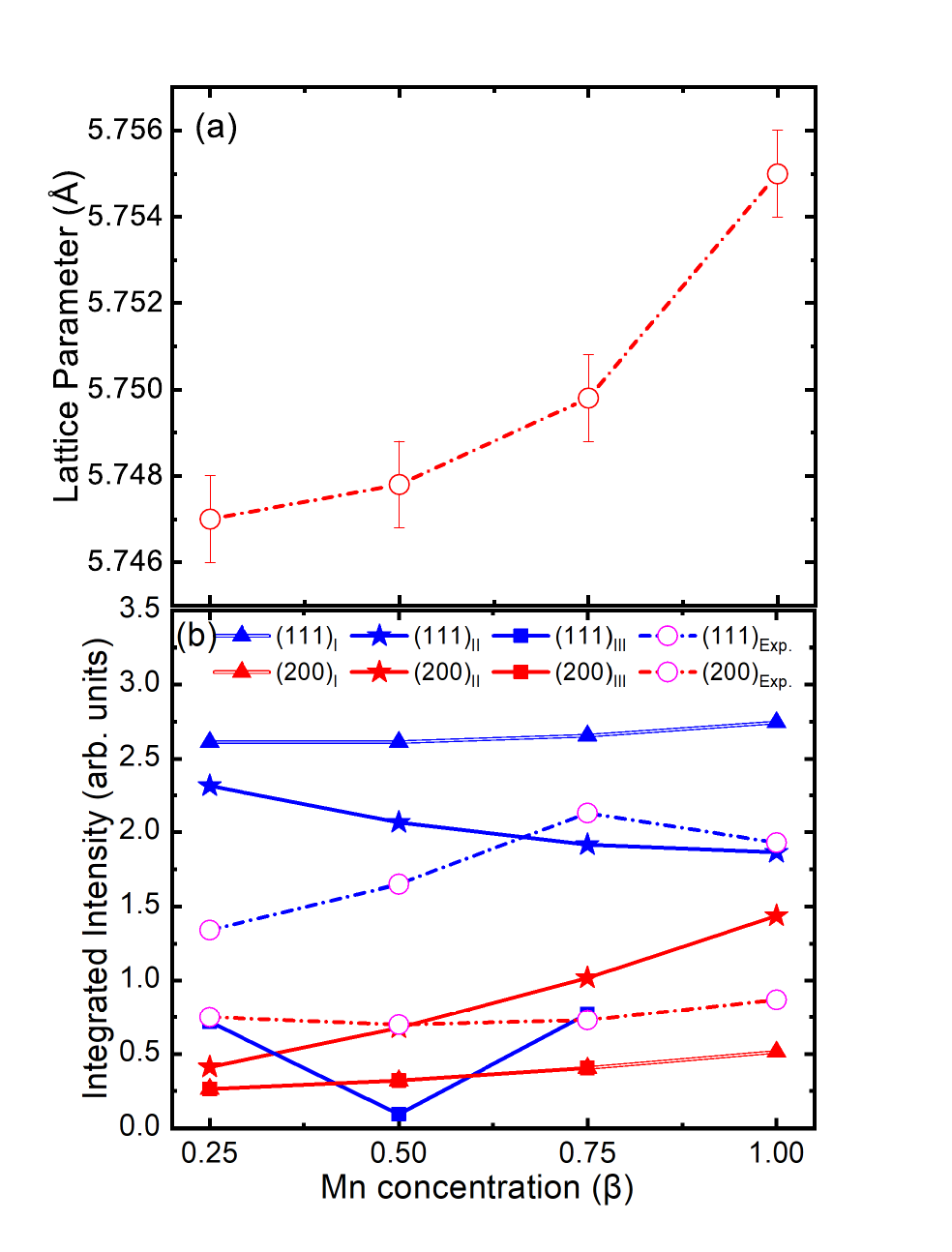}
\caption{(colour online) Variation of (a) lattice parameter and (b) experimental and calculated intensities of superlattice peaks. In figure subscript I, II, and III represents the calculated intensity assuming the atomic configuration types I, II, and III respectively as shown in table \ref{tab:atom_config2}, whereas Exp.\ stands for experimental intensity.}
\label{fig:xrd_CoFe1-bMnb_intensity}
\end{figure}
%%%%%%%%%%%%%%%%%%%%%%%%%%%%%
The {\em powder} XRD scans are shown in figure \ref{fig:xrd_Co2Fe1-bMnb}. In the inset, low angle peaks are blown up to make any less intense peaks clearly visible. We begin our analysis by fitting the observed pattern with an fcc structure, specifically space group 225. Analogous to Co$_{2-\alpha}$Mn$_{\alpha}$FeGe, for lower Mn concentration, there are a few extra peaks (denoted by *) which cannot be indexed to a cubic structure. Those extra peaks can plausibly be attributed to the grain boundaries, where EDS analysis verified that the grains and grain boundaries had clearly different compositions. As the Mn concentration is increased further (0.50, 0.75, 1.0), all of the impurity peaks disappear and the entire observed pattern can be fit to an fcc structure, suggesting formation of single-phase compounds. The XRD scans were also performed on the bulk samples (\textit{i.e.}, not powdered) of Co$_2$(Fe$_{1-\beta}$Mn$_\beta$)Ge, and are shown in the supplementary information \cite{supplementary2021}. They are consistent with the powder XRD patterns -- a few extra peaks for $\beta$ = 0.25, while for larger $\beta$ all peaks correspond to an fcc unit cell. The bulk XRD patterns also provide one additional bit information, the bulk samples have a (200) preferred orientation.

The variation of experimental lattice parameter is shown in figure \ref{fig:xrd_CoFe1-bMnb_intensity}(a). Lattice parameters were determined by Cohen’s method with a Nelson-Riley error function.\cite{nelson1945experimental,cohen1935precision} A slight enhancement in lattice parameter is detected after Mn substitution. This is an expected variation pertaining to the slightly larger atomic radius of Mn compared to Fe. We analysed the experimental and calculated peak intensities to get a better idea of atomic order in the unit cell, see figure \ref{fig:xrd_CoFe1-bMnb_intensity}(b). One can observe that the experimental intensity of the (111) peak increases as the Mn concentration increases. This trend matches with that of structure type I somewhat better than type II and III, which has the opposite trend. Further, the variation of experimental (200) peak also more closely resembles type I than type II. This would argue that type I structure is likely to be favourable for Co$_2$(Fe$_{1-\beta}$Mn$_\beta$)Ge series. This is in fact the same structure that one expects based on the Burch or 4-2 rules\cite{burch1974hyperfine,butler2011rational}. However, while suggestive, the XRD intensities alone are hardly enough to make a definitive conclusion. In figure \ref{fig:xrd_CoFe1-bMnb_intensity}(b), one can notice that the experimental intensity of (200) peak is higher than the calculated values, which suggests a degree of texturing even in the case of fine powder, and this was corroborated by XRD of bulk samples (see supplementary \cite{supplementary2021}). Particularly once the presence of texturing is established, one dare not draw too many conclusions from the XRD {\em intensities} -- not the site assignment details and especially not the degree of order, as the results would be questionable at best. We will therefore leave the question of site assignment aside until we discuss the magnetic measurements and theoretical results in order to form a more complete picture on which atomic configuration is most favourable. As noted earlier, local atomic probes like $^{59}$Co or $^{55}$Mn NMR or $^{57}$Fe M\"{o}ssbauer spectroscopy would be very valuable in this regard,\cite{wurmehl2012nmr,wurmehl2013nmr,wurmehl2014nmr,mende2021mos} as would element-specific moments from for example x-ray magnetic circular dichroism.

%%%%%%%%%%%%%%%%%%%%
\begin{figure}[htb]
\includegraphics[width =0.75\columnwidth]{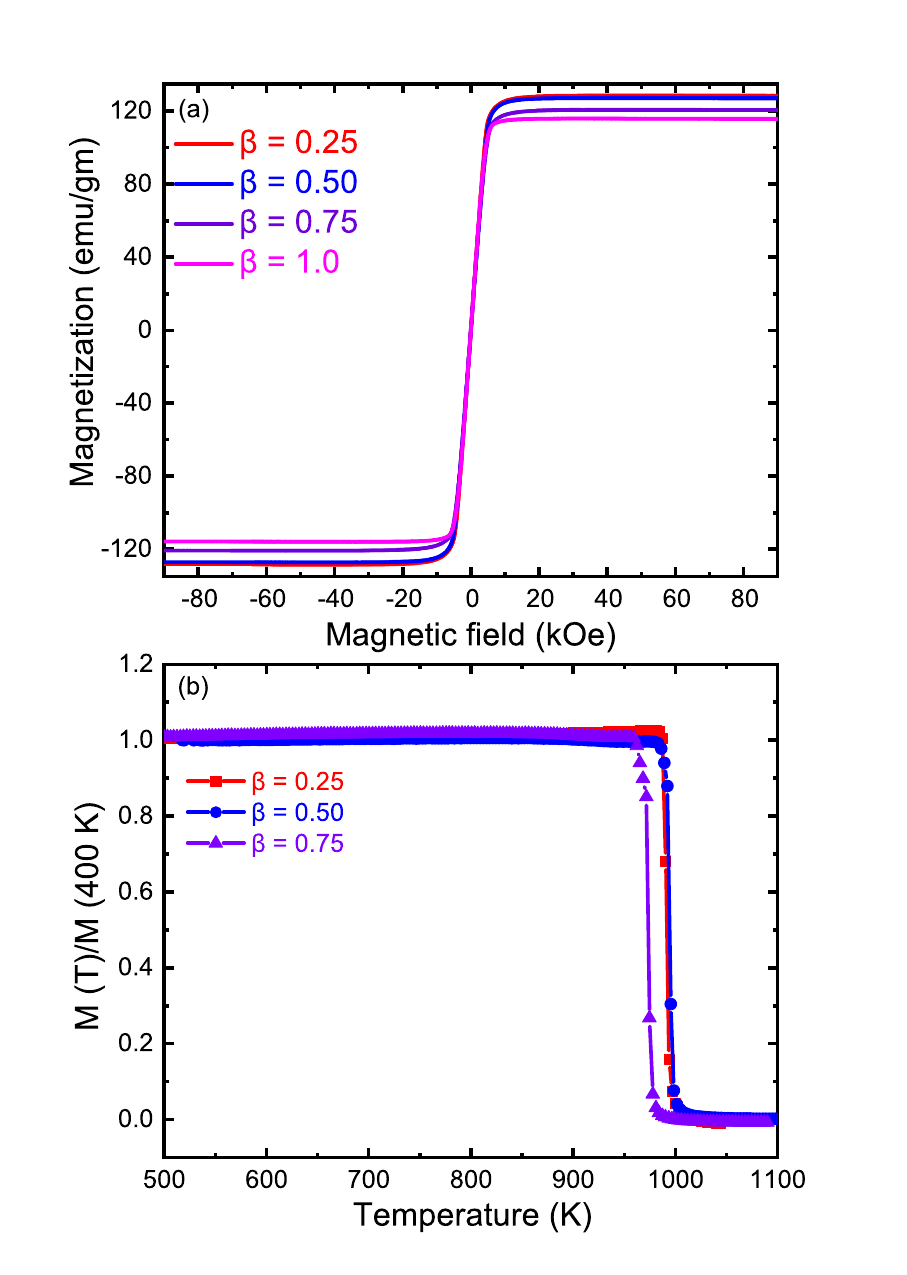}
\caption{(colour online) (a) M-H loops measured at T = 5 K for Co$_2$(Fe$_{1-\beta}$Mn$_\beta$)Ge alloy series. (b) Temperature variation of spontaneous magnetisation with an applied field of 100 Oe. The values are normalised to the value at 400 K.}
\label{fig:mh_Co2Fe1-bMnb}
\end{figure}
%%%%%%%%%%%%%%%%%%%%%%%%

\begin{figure}[htb]\centering
\includegraphics[width =0.85\columnwidth]{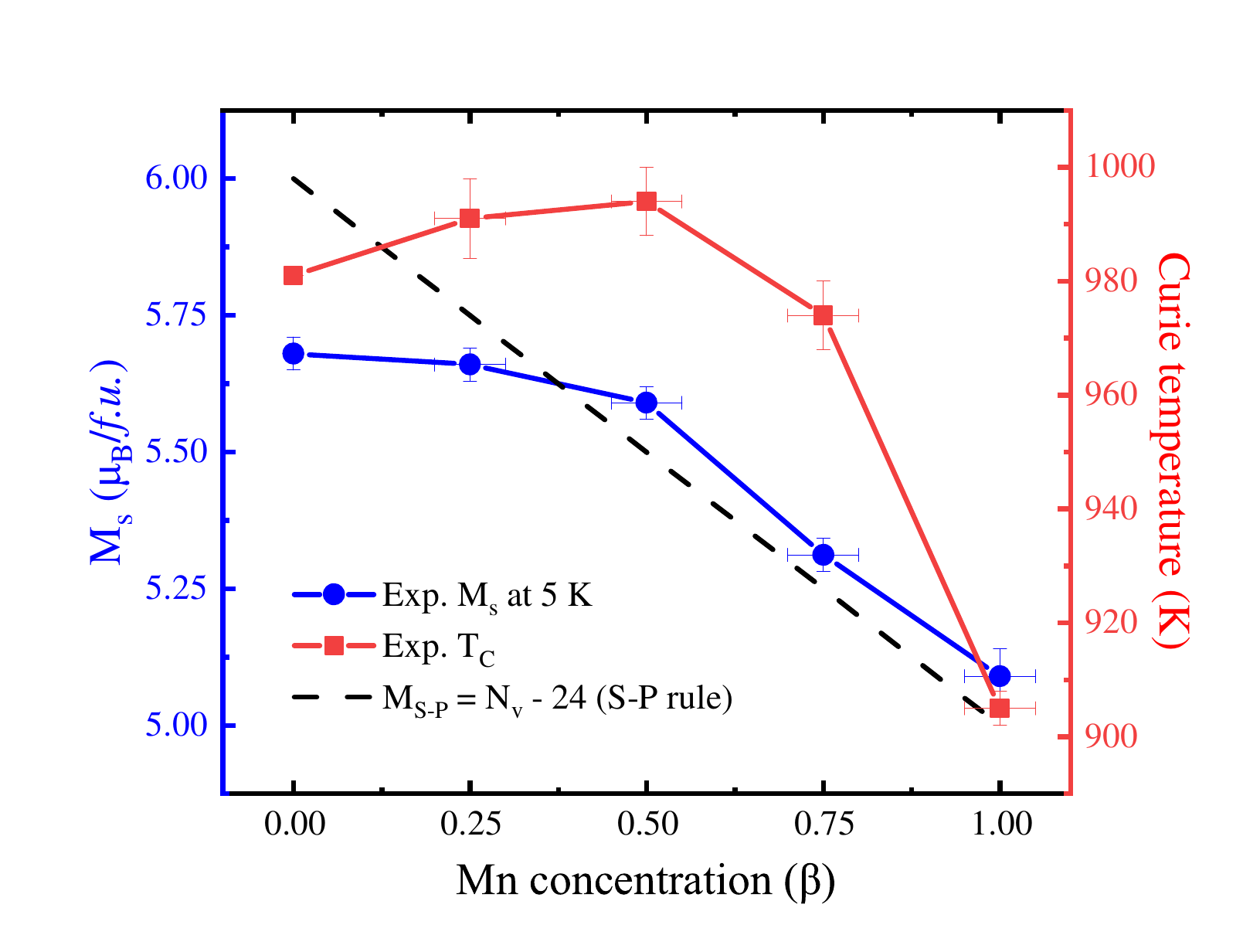}
\caption{(colour online) Variation of saturation magnetisation (blue) and Curie temperature (red) with Mn concentration $\beta$ in Co$_2$(Fe$_{1-\beta}$Mn$_\beta$)Ge. The solid line is guide to the eyes. The M$_\text{s}$ value expected from S-P rule is also shown. Note the moment measured here for Co$_2$MnGe matches with the value reported in Ref.\ \cite{buschow1983magneto}.}
\label{fig:mh_variation_Co2Fe1-bMnb}
\end{figure}

The magnetic isotherms (Fig. \ref{fig:mh_Co2Fe1-bMnb}(a)) suggest soft ferromagnetic behaviour for these alloys. The magnetisation decreases as the Mn concentration increases as one would expect. Similarly, the temperature dependent magnetic measurement (shown in Fig.~\ref{fig:mh_Co2Fe1-bMnb}(b)) reveals a very high T$_\text{C}$, close to 1000 K for low Mn concentrations. The variation of saturation magnetisation (M$_\text{s}$) measured at 5 K along with T$_\text{C}$ as a function of Mn concentration are plotted in Fig.~\ref{fig:mh_variation_Co2Fe1-bMnb}. Since the number of valence electrons N$_\text{v}$ decreases by 1 when replacing Fe by Mn in Co$_2$FeGe, the S-P rule predicts the moment decreases from 6 $\mu_B$ to 5 $\mu_B$ with a linear variation in between. There is significant discrepancy between the experimental and S-P moment for the parent alloy. As explained previously, this is due to the existence of secondary phases in the microstructure of Co$_2$FeGe. After replacing $\frac{1}{4}$ of Fe atoms by Mn ($\beta$ = 0.25) in Co$_2$FeGe, one can notice the experimental moment is still less than S-P moment, but the difference between them has narrowed. The experimental moment remains almost same, even though the N$_\text{v}$ decreases by 0.25. This is a direct consequence of the reduction of the secondary phase fraction -- the smaller the volume fraction of the secondary phase, the closer the experimental moment is to the expected value, as we also observed in case of (Co$_{2-\alpha}$Mn$_\alpha$)FeGe series. This makes sense when analysing the nature of variation in figure \ref{fig:mh_variation_Co2Fe1-bMnb}. Since the $\beta\!\ge\!0.50$ samples are single-phase (no extra peaks in the XRD and minimal segregation at the grain boundaries), the variation is almost linear for $\beta\!\ge\!0.50$. But the linearity is broken for $\beta\!<\!0.50$, because below that concentration the samples begin showing multi-phase behaviour. One can notice the same trend in T$_\text{C}$ as well. There is an enhancement in T$_\text{C}$ across the single-phase region as Mn concentration decreases, but as the samples begin to show impurity phases for low $\beta$, the enhancement ceases and a slight drop of T$_\text{C}$ is seen for the lowest $\beta$ samples.  Leaving the $\beta\!<\!0.50$ samples aside, the interesting and important discovery here is, for all other samples, the low temperature moment agrees reasonably well with the S-P moment, suggesting possible half-metallic behaviour\cite{galanakis2002slater} for higher Mn concentration alloys. 

%%%%%%%%%%%%%%%%%%%%%%%%%%
%%%%%%%%%%%%%%%%%%%

\subsubsection{Findings from calculations}
The three non-equivalent configurations that were considered for the calculations are as follows: type I, in which the substituted Mn atom directly occupies the vacant Fe position in B$_T$ site; type II, where the added Mn atom displaces Co atom from sublattice A towards vacant Fe position and itself resides in the Co position; and type III, where the Mn atom occupies the B$_Z$ site, by displacing the Ge atom towards vacant position in B$_T$ site. Type I is favoured from the 4-2 rule since it has Mn in a ferromagnetic configuration with Co, and Co is in a reasonable 5$\uparrow$, 4$\downarrow$ state. Type II has Co in the unreasonable 7$\uparrow$, 2$\downarrow$ state, and type III would require some Ge to be in an wholly unrealistic 0$\uparrow$, 4$\downarrow$ state to satisfy the 4-2 rule. This suggests that if we wish to have a half-metallic state, we will need to have the type I structure as the most stable. 

Since type I and type III structures involve substitution on the same sublattice, we expect small differences in the calculated parameters for these structures. Further, at $\beta\!=\!1.0$, both structures are equivalent -- both contain 8 Co atoms on sublattice A, and 4 Mn and 4 Ge atoms on sublattice B. In addition to the aforementioned three structures, we also considered the possibility of several disordered structures for each type. These disordered structures are denoted by asterisk (*) in Table \ref{table:Co2Fe1-bMnb_cal_all}.

%%%%%%%%%%%%%%%%%
\begin{figure}[htb]
\includegraphics[width = 0.9\columnwidth]{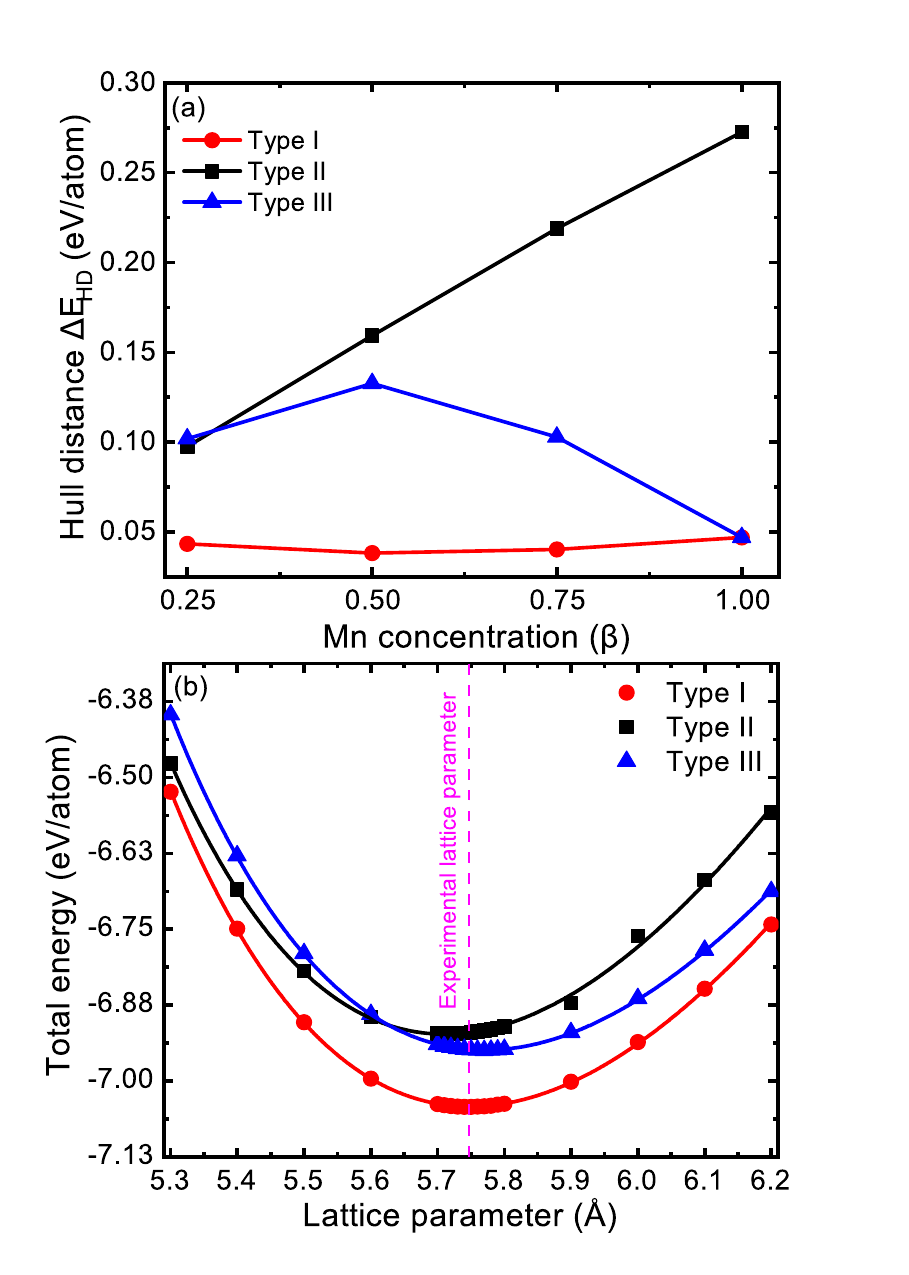}
\caption{(colour online) (a) The variation of hull distance ($\Delta E_{HD}$) as a function of Mn concentration $\beta$ in Co$_2$(Fe$_{1-\beta}$Mn$_\beta$)Ge. The type I has the minimum hull distance suggesting it is the more stable phase. (b) Structure optimisation by calculating the total energy as a function of lattice parameter. The optimisation is for $\beta\!=\!0.50$, which also suggests type I is the energetically more favourable structure.}
\label{fig:Co2Fe1-bMnbGe_hull_dis}
\end{figure}
%%%%%%%%%%%%%%%%%%%%
Starting with the hull distance variation, as shown in figure \ref{fig:Co2Fe1-bMnbGe_hull_dis}(a), it can be clearly observed that type I structure has the smallest hull distance. Also, $\Delta E_{HD}$ is below 0.05 eV/atom over the entire substitution range for the type I structure, which is the threshold value that we set for phase stability. For other (non-equivalent) structures $\Delta E_{HD}$ is at least 0.05 eV/atom higher than type I. Based on this, it is plausible to argue that type I structure is the most stable structure. The type I structure being more stable is also supported by the structure relaxation calculations for a set of fixed unit cell volumes, shown in figure \ref{fig:Co2Fe1-bMnbGe_hull_dis}(b). Type I exhibits the lowest total energy on varying the lattice parameter by $\pm$ 10 \% of the experimental value. Furthermore, close to the experimental lattice parameter, the total energy of other structures is greater than that of type I by more than 0.05 eV/atom. 

%%%%%%%%%%%%%%%%%
\begin{figure}[htb]
\centering
\includegraphics[width = 0.8\columnwidth]{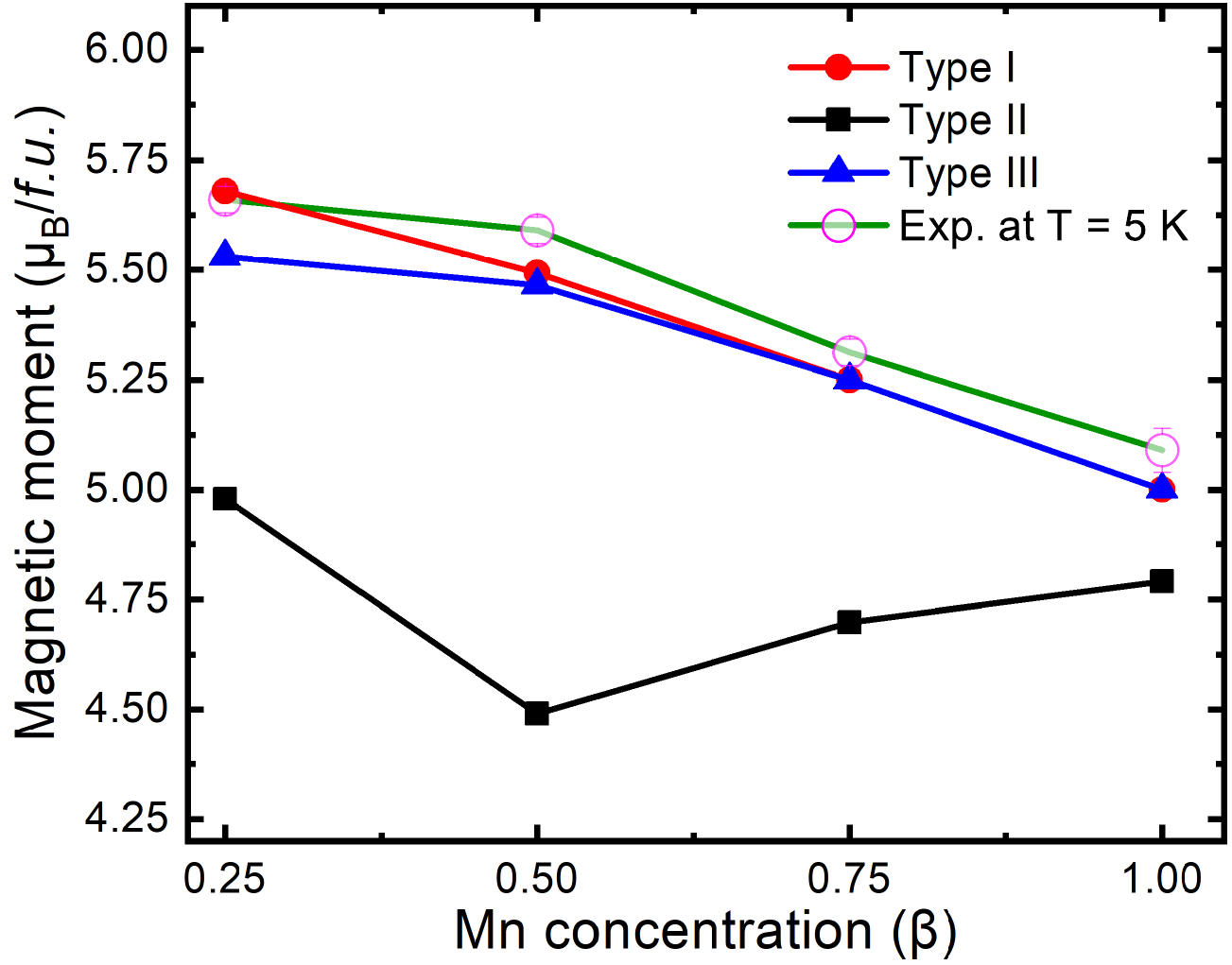}
\caption{(colour online) Comparison of calculated magnetic moments for different structures and the experimental moment as a function of Mn concentration $\beta$ in Co$_2$(Fe$_{1-\beta}$Mn$_\beta$)Ge.}
\label{fig:Co2Fe1-bMnbGe_cal_mh}
\end{figure}
%%%%%%%%%%%%%%%%%%%%
%%%%%%%%%%%%%%%%%
\begin{figure}[htb]
\includegraphics[width = 0.85\columnwidth]{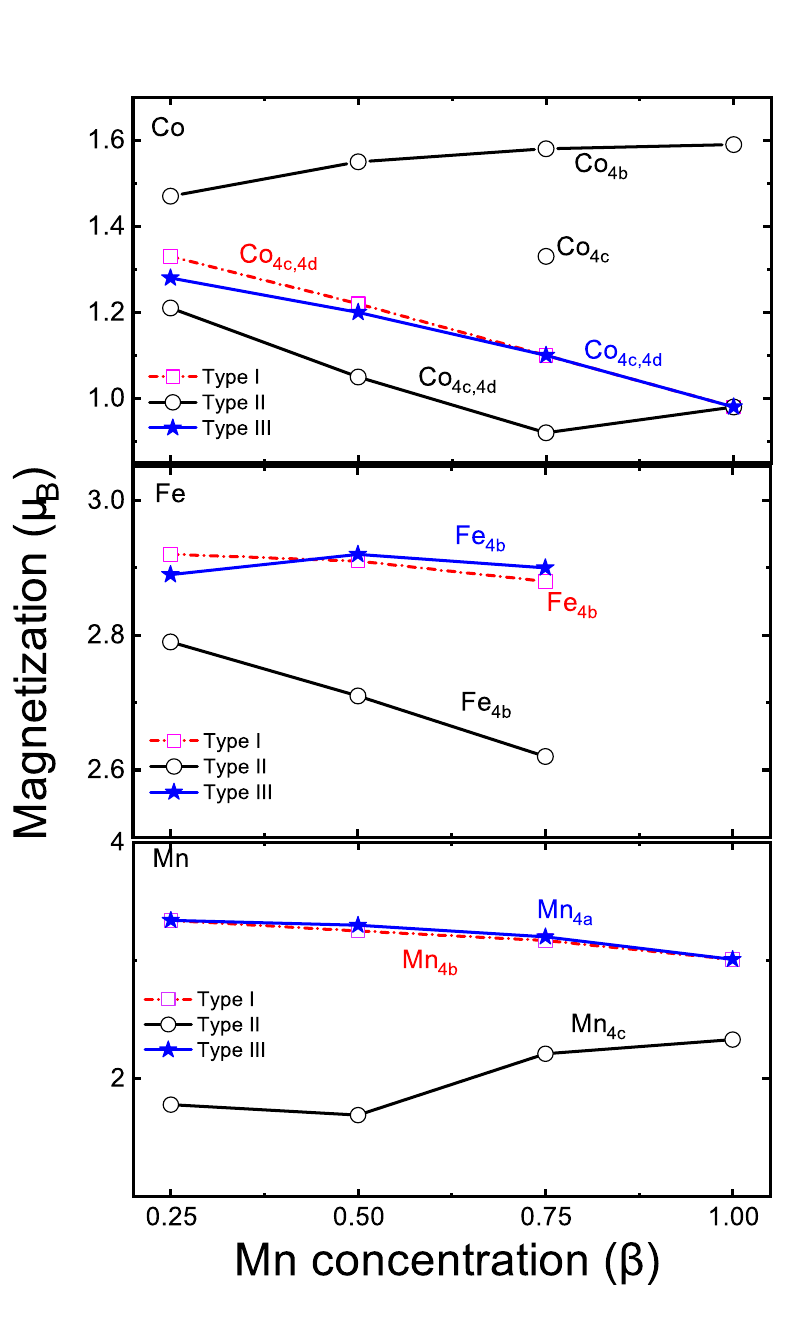}
\caption{(colour online) Atom specific moment variation in Co$_2$(Fe$_{1-\beta}$Mn$_\beta$)Ge. The subscript (\textit{\textit{i.e.},} 4a, 4b, 4c, and 4d) refers to the Wyckoff site that the atom resides. One can observe that the type I structure shows systematic variation as compared to II and III.}
\label{fig:Co2Fe1-bMn-atomic_m}
\end{figure}
%%%%%%%%%%%%%%%%%%%%
%%%%%%%%%%%%%%%%%%%
\begin{table*}[htb]
\centering
\scriptsize
\caption{DFT results for several possible atomic configurations of Co$_{2}$(Fe$_{1-\beta}$Mn$_\beta$)Ge. $E$ (in eV/atom) is total energy, $m$ (in $\mu_B$) stands for magnetic moment, and $<\!a\!>$ (in \AA) represents the optimised lattice parameter. The configuration which is most stable is in bold, and the configurations with asterisks (*) represent disordered structures that can be generated for a particular structure type.}

\begin{tabular}{c c c c c c c c}
\hline\hline
\textbf{Configuration} & \textbf{4d} & \textbf{4c} & \textbf{4b} & \textbf{4a} & \textbf{E} & \textbf{m$_{tot}$} & \textbf{$<$a$>$} \\\hline
x = 0.25& & & & &  & &  \\ 
\setrow{\bfseries}I& 4Co& 4Co& 3Fe,1Mn& 4Ge& -6.97 & 5.68&5.752 \\[0.5ex]
I*& 4Co& 3Co,1Fe& 2Fe,1Mn,1Co& 4Ge& -6.95 & 5.02&5.728 \\[0.5ex]
II& 4Co& 3Co,1Mn& 3Fe,1Co & 4Ge & -6.92 & 4.98&5.729 \\[0.5ex]
III& 4Co& 4Co& 3Fe,1Ge& 3Ge,1Mn& -6.92 & 5.53&Ortho. \\[0.5ex]
III*& 4Co& 3Co,1Ge& 3Fe,1Mn& 3Ge,1Co& -6.80 &5.142&Tetra. \\[0.5ex] \hline
x = 0.50& & & & & & &  \\ [0.5ex]
\setrow{\bfseries}I& 4Co & 4Co & 2Fe,2Mn & 4Ge & -7.04 & 5.49 & 5.748 \\[0.5ex]
I*& 4Co & 2Co,2Fe & 2Co,2Mn & 4Ge & -6.96 & 4.50 &Tetra. \\[0.5ex]
II& 4Co & 2Co,2Mn & 2Fe,2Co & 4Ge & -6.92 & 4.49 & Ortho. \\[0.5ex]
III& 4Co & 4Co & 2Fe,2Ge & 2Ge,2Mn & -6.95 & 5.46 & Ortho. \\[0.5ex]
III*& 4Co & 2Co,2Ge & 2Fe,2Mn & 2Ge,2Co & -6.85 & 5.90 & Ortho. \\[0.5ex]\hline
x = 0.75& & & & & & &  \\ [0.5ex]
\setrow{\bfseries}I& 4Co & 4Co & 1Fe,3Mn & 4Ge & -7.11 & 5.25 & 5.743 \\[0.5ex]
I*& 4Co & 3Co,1Fe & 3Mn,1Co & 4Ge & -7.06 & 4.68 & 5.725 \\[0.5ex]
II& 4Co & 1Co,3Mn & 1Fe,3Co & 4Ge & -6.93 & 4.70 & 5.734 \\[0.5ex]
III& 4Co & 4Co & 1Fe,3Ge & 1Ge,3Mn & -7.05 & 5.25 & Ortho. \\[0.5ex]
III*& 4Co & 1Co,3Ge & 1Fe,3Mn & 1Ge,3Co & -6.94 & 4.69 & Ortho. \\[0.5ex]\hline
x = 1.0& & & & &  & &  \\ [0.5ex]
\setrow{\bfseries}I& 4Co & 4Co & 4Mn & 4Ge & -7.17 & 5.00 & 5.738 \\[0.5ex]
II& 4Co & 4Mn & 4Co & 4Ge & -6.94 & 4.79 & 5.743 \\[0.5ex]
III& 4Co & 4Co & 4Ge & 4Mn & -7.17 & 5.00 & 5.738 \\[0.5ex]
III*& 4Co & 4Co & 3Mn,1Ge & 3Ge,1Mn & -7.06 & 5.00 & 5.762 \\[0.5ex]
III**& 4Co & 3Co,1Ge & 4Mn & 3Ge,1Co & -6.99 & 5.55 & 5.801 \\[0.5ex]\hline
\end{tabular}
\label{table:Co2Fe1-bMnb_cal_all}
\end{table*}
%%%%%%%%%%%%%%%%%%%%%%
The variation of magnetic moments calculated for different structures along with the experimental moment at 5 K are shown in figure \ref{fig:Co2Fe1-bMnbGe_cal_mh}. The moments calculated for structure I agree best with the experimental moment at each Mn concentration, though moments for structure III also agree well. Further, considering the calculated site-specific moments (shown in figure \ref{fig:Co2Fe1-bMn-atomic_m}), types I and III show the most systematic variation. When combining the XRD intensity variation (Fig. \ref{fig:xrd_CoFe1-bMnb_intensity}), the $\Delta E_{HD}$, the total energy, and the magnetic moment variation, it seems safe to argue Co$_2$(Fe$_{1-\beta}$Mn$_\beta$)Ge favours the type I structure. This is in contrast to the (Co$_{2-\alpha}$Mn$_\alpha$)FeGe series, where the type II structure was observed to be most stable. Although these two series favour two different structure types, one thing that is common to both structures is that the atoms occupy sites based on the 4-2 and Burch rules. The Burch rule says that the less valence transition metal element prefers to reside in the same sublattice as the main group element, while the 4-2 rule dictates that the electron count in the minority band alternates between 4 and 2 moving from site to site, which when coupled with which electron configurations are plausible dictates site preferences. Still, given that the difference between the hull distances for structures I and III less than $0.15\,$eV/atom and the calculated properties are very similar, one cannot rule out disorder between these two cases, which simply means some amount of mixing of Ge and Mn atoms on the 4a/4b Wycoff sites. 

%%%%%%%%%%%%%%%%%
\begin{figure}[htb]
\centering
\includegraphics[width = 1.1\columnwidth]{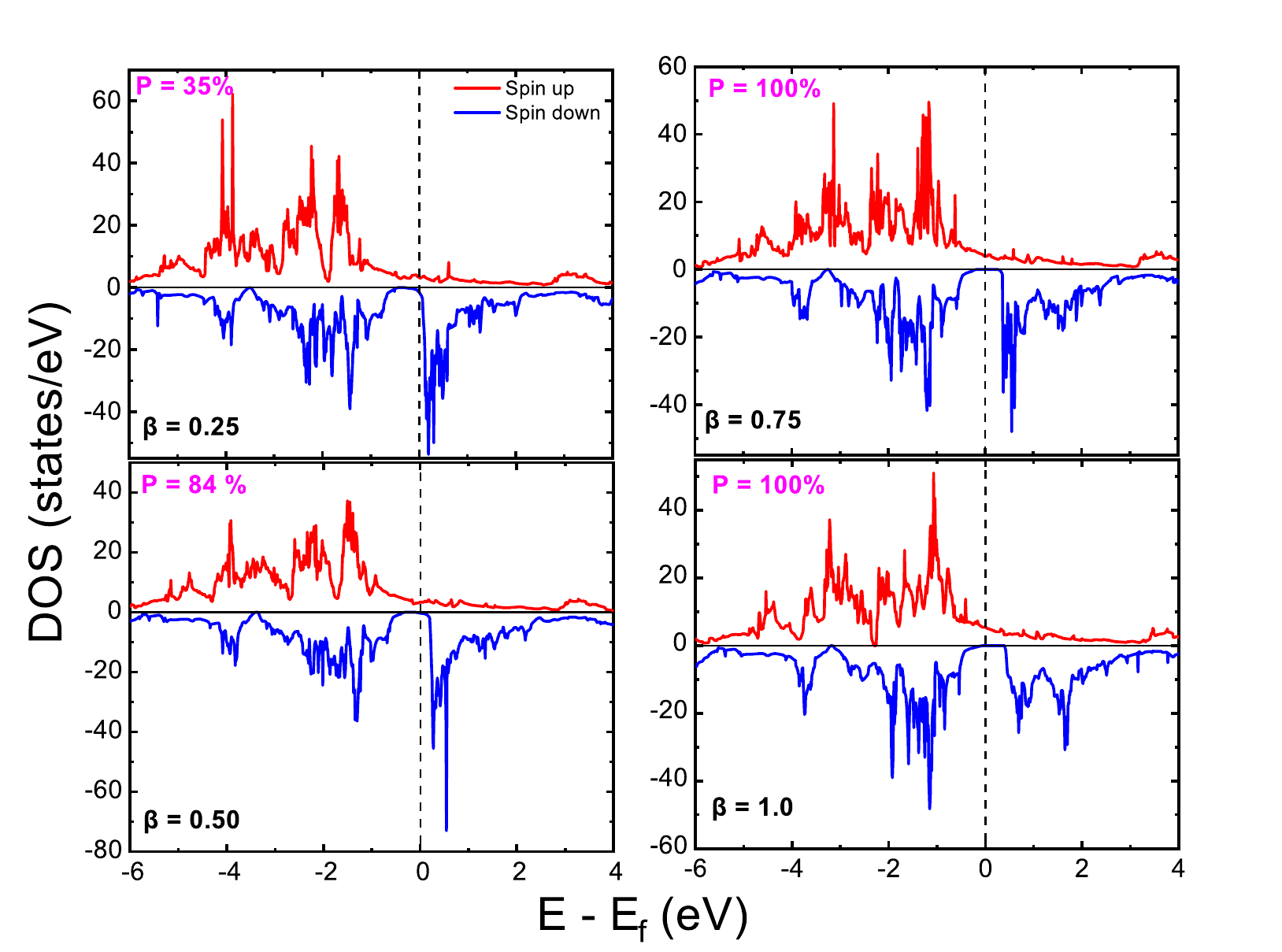}
\caption{(colour online) Spin resolved total density of states plot for Co$_2$(Fe$_{1-\beta}$Mn$_\beta$)Ge alloy series. Half-metallic behavior can be obtained for $\beta$ $\ge$ 0.75.}
\label{fig:Co2Fe1-bMnbGe_dos}
\end{figure}
%%%%%%%%%%%%%%%%
Electronic structure calculations for Co$_2$(Fe$_{1-\beta}$Mn$_\beta$)Ge were performed based on the type I structure. In this structure, the substituted Mn atoms directly occupy the B$_T$ site, so this type of atomic configuration is favourable to tune the system toward half-metallic character as explained earlier. There is an upward shift of conduction band edge as a result of Mn substitution, which can be observed in the spin-resolved total density of plots shown in figure \ref{fig:Co2Fe1-bMnbGe_dos}. But, unlike the (Co$_{2-\alpha}$Mn$_\alpha$)FeGe series, in the case of Co$_2$(Fe$_{1-\beta}$Mn$_\beta$)Ge, where there is only a single substitution, only the bands contributed by atoms of the B sublattice gain energy (the t$_{2g}$ and e$_g$ bands are affected more than the t$_{1u}$ and e$_u$ bands, noted previously) and are therefore shifted upwards. As a result, the band gap widens as the Mn concentration increases. As shown in figure \ref{fig:Co2Fe1-bMnbGe_dos}, the band gap increases from 0.030 eV at $\beta$ = 0.25 to 0.4194 eV at $\beta$ = 1.0. Because of this, the Fermi level has moved downward, resulting in an enhancement of spin polarisation. At $\beta$ = 0.75, the Fermi level is located within the band gap, and hence a half-metallic behaviour is achieved with 100 \% spin polarisation. Upon further increasing Mn, the half-metallic behaviour is retained, even at $\beta$ = 1.0, but the Fermi level approaches closer to the valance band edge. 

The Co$_2$MnGe alloy (\textit{\textit{i.e.}} $\beta$ = 1.0) is one of the most extensively studied alloys in the family of Heusler compounds because of its two very important characteristics: potential half-metallic character and its ultra-low damping parameter. Our experimental as well as theoretical investigations also provide plausible evidence that Co$_2$MnGe is likely half-metal. However, a close observation of DOS plot of Co$_2$MnGe reveals that while E$_F$ is located at the band gap, it is only 0.02 eV away from the top edge of the valence band. This very small gap between the valence band and E$_F$ means the half-metallicity in Co$_2$MnGe is not so robust, particularly when one considers thermal energy. A direct way of mitigating this issue is to move E$_F$ away from valence band to the middle of the gap. Our observations show that this is achievable by partially occupying the Mn sites with a heavier atom, such as Fe. This is, in fact, what one notices in case of Co$_2$Fe$_{0.25}$Mn$_{0.75}$Ge, as shown in Fig. \ref{fig:Co2Fe1-bMnbGe_dos}, where E$_F$ is located almost in the middle of the gap, 0.16 eV higher than the top of the valence band and 0.26 eV lower than the bottom of the conduction band. The half-metallic character can be argued to be more robust in Co$_2$Fe$_{0.25}$Mn$_{0.75}$Ge compared to Co$_2$MnGe. Further, the number of majority states at E$_F$ of Co$_2$Fe$_{0.25}$Mn$_{0.75}$Ge is less than the number of states at E$_F$ of Co$_2$MnGe. Since the Gilbert damping parameter depends directly on the number of states the E$_F$, it is also plausible to expect that Co$_2$Fe$_{0.25}$Mn$_{0.75}$Ge exhibits an even lower damping parameter than Co$_2$MnGe \cite{liu2009origin,guillemard2020engineering}.  %% CITATION for damping ~ states at Ef

%%%%%%%%%%%%%%%%%%%%%%%%%%%%
\subsubsection{Conclusions for Co$_{2}$(Fe$_{1-\beta}$Mn$_\beta$)Ge}
Similar to the substitution of Co by Mn atoms, the substitution of Fe by Mn atoms can also stabilise a single-phase compound based on Co$_{2}$FeGe. Slightly more Mn is needed to make a phase-pure sample in the latter case, however. For single-phase samples, M$_\text{s}$ and T$_\text{C}$ are high, and exhibit linear variation with Mn concentration. The M$_\text{s}$ also agrees with the moment expected from Slater-Pauling rule. The computational results suggest type I as the most stable structure, which is also backed by the experimental results. The electronic structure calculation suggests the Fermi level can be tuned, with half-metallic character emerging at $\beta$ = 0.75. It is also suggested that the half-metallicity of Co$_2$Fe$_{0.25}$Mn$_{0.75}$Ge should be even more robust than that of Co$_{2}$MnGe, with the possibility of obtaining an even lower Gilbert damping parameter.  

%%%%%%%%%%%%
\subsection{(Co$_{2-\alpha}$Fe$_\alpha$)MnGe}
\subsubsection{Findings from experiments}
%%%%%%%%%%%%%%%%%%%%
\begin{figure}[htb]
\centering
\includegraphics[width =0.6\columnwidth]{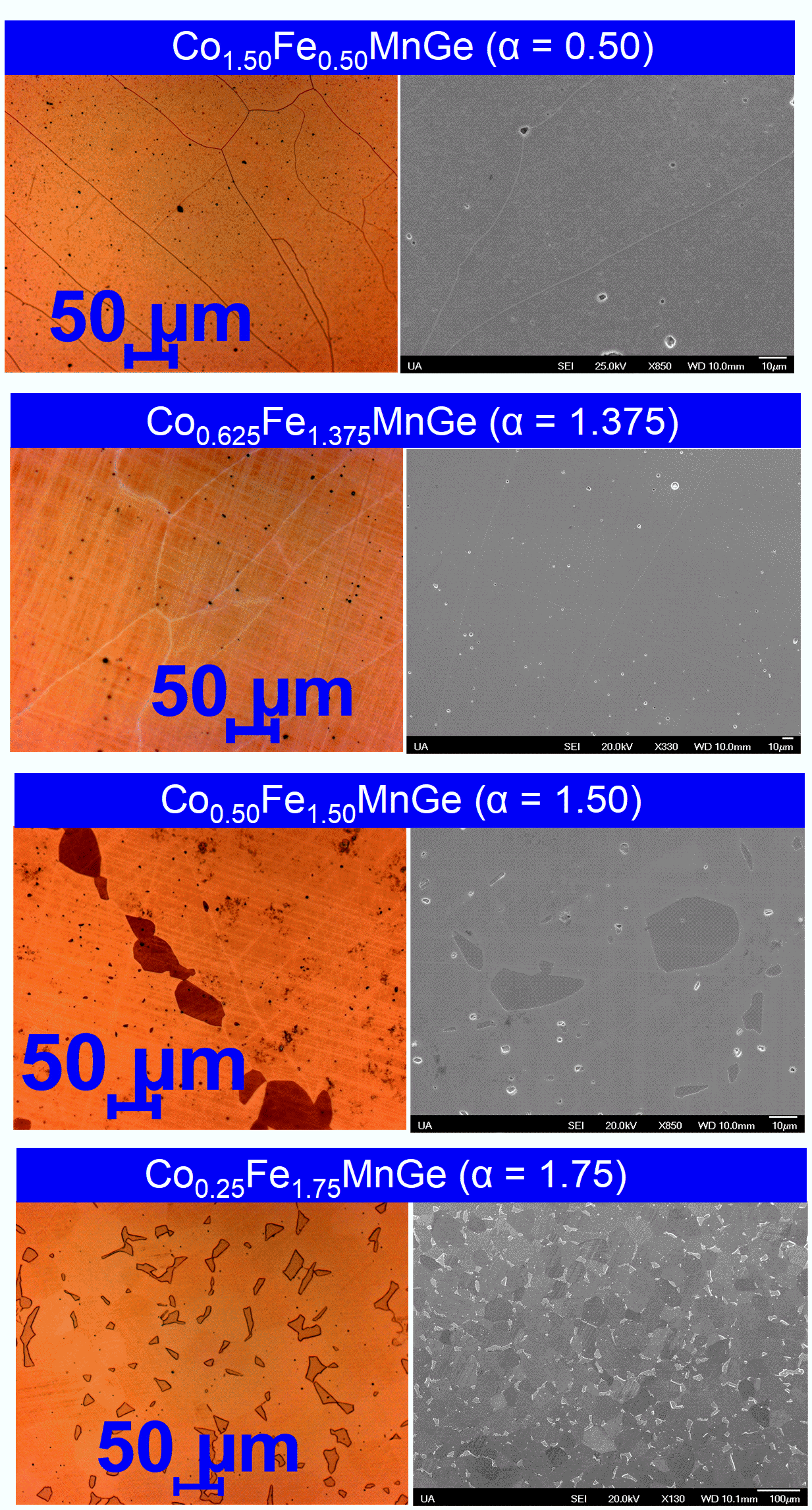}
\caption{(colour online) Optical (left) and electron (right) micrographs of Co$_{2-\alpha}$Fe$_{\alpha}$MnGe substitutional series at various substitution value. Since the parent Co$_2$MnGe alloy is single-phase, the phase-stability holds upto $\alpha$ = 1.375 when substituting Co by Fe. For $\alpha$ $\ge$ 1.50, multi-phase microstructure is observed as evident by the occurrence of different contrast regions.}
\label{fig:optical_Co2-aFeaMnGe}
\end{figure}
%%%%%%%%%%%%%%%%%%

Bulk samples of (Co$_{2-\alpha}$Fe$_{\alpha}$)MnGe were prepared up to $\alpha$ = 1.75. The Fe$_2$MnGe (\textit{\textit{i.e.}}, $\alpha$ = 2.0) alloy was not prepared in this work, as it has already been shown to exhibit single-phase microstructure crystallising in a hexagonal DO$_{19}$ structure\cite{keshavarz2019fe2mnge}. Our investigation has already covered $\alpha$ = 0 (Co$_2$MnGe) above and $\alpha$ = 1.0 (CoFeMnGe), both are single-phase alloys and have same cubic crystal structure. It therefore seems likely that the intermediate compositions (\textit{i.e.} $\alpha$ = 0.25, 0.50, 0.75) are also single-phase and cubic. The optical as well as electron micrographs indeed exhibit no occurrence of regions of different contrast, thereby suggesting formation of single-phase compound. The microstructure for $\alpha$ = 0.50 is shown in Fig. \ref{fig:optical_Co2-aFeaMnGe}. The microstructures of other intermediate compositions are provided in the supplementary information. However, going from $\alpha\!=\!1.0$ to $\alpha\!=\!2.0$ (\textit{i.e.}, from CoFeMnGe to Fe$_2$MnGe), though both endpoints exhibit single-phase microstructure the crystal structure changes from cubic to hexagonal. The competition between the two crystal structures should lead to mixed phase behaviour for intermediate compositions for $1\!<\alpha<2$. Indeed, the micrographs show single-phase like microstructure up to $\alpha$ = 1.375, beyond which the sample exhibits multi-phase microstructure (see Fig. \ref{fig:optical_Co2-aFeaMnGe}). The phase separation and the possible role of the competition between two different crystal structures will be explored in more detail in the XRD section.

%%%%%%%%%%%%%%%%%%%%%%%%
\begin{table*}[htb]
\scriptsize
\caption{EDS determined grain and grain boundary compositions for (Co$_{2-\alpha}$Fe$_\alpha$)MnGe alloys. Note in case of multi-phase samples (denoted by *), the grain corresponds to the main phase, whereas the grain boundary composition represents that of secondary phase (a region of different contrast). The reported compositions have an uncertainty of $\pm$ 5\%.}

\centering
\begin{tabular}{c c c}
\hline\hline
\textbf{Alloy} & \multicolumn{1}{p{3cm}}{\centering{\textbf{Grain or \\ main phase composition}}}& \multicolumn{1}{p{4cm}}{\centering{\textbf{Grain boundary or \\ secondary phase composition}}}\\
[0.5ex]
\hline
Co$_{1.75}$Fe$_{0.25}$MnGe & Co$_{1.73}$Fe$_{0.27}$Mn$_{1.04}$Ge$_{0.96}$& Co$_{1.72}$Fe$_{0.26}$Mn$_{1.05}$Ge$_{0.97}$ \\[1.5ex]
Co$_{1.50}$Fe$_{0.50}$MnGe & Co$_{1.43}$Fe$_{0.53}$Mn$_{1.08}$Ge$_{0.96}$& Co$_{1.36}$Fe$_{0.50}$Mn$_{1.11}$Ge$_{1.03}$  \\[1.5ex]
Co$_{1.25}$Fe$_{0.75}$MnGe & Co$_{1.22}$Fe$_{0.74}$Mn$_{1.03}$Ge$_{1.01}$& Co$_{1.19}$Fe$_{0.71}$Mn$_{1.05}$Ge$_{1.05}$  \\[1.5ex]
Co$_{0.80}$Fe$_{1.20}$MnGe & Co$_{0.81}$Fe$_{1.21}$Mn$_{0.98}$Ge$_{0.99}$& Co$_{0.83}$Fe$_{1.13}$Mn$_{1.02}$Ge$_{1.02}$  \\[1.5ex]
Co$_{0.625}$Fe$_{1.375}$MnGe & Co$_{0.630}$Fe$_{1.369}$Mn$_{0.99}$Ge$_{1.00}$& Co$_{0.654}$Fe$_{1.278}$Mn$_{1.04}$Ge$_{1.03}$  \\[1.5ex]
Co$_{0.50}$Fe$_{1.50}$MnGe* & Co$_{0.50}$Fe$_{1.47}$Mn$_{1.03}$Ge$_{0.99}$& Co$_{0.28}$Fe$_{1.64}$Mn$_{1.09}$Ge$_{0.99}$  \\[1.5ex]
Co$_{0.25}$Fe$_{1.75}$MnGe* & Co$_{0.24}$Fe$_{1.78}$Mn$_{1.00}$Ge$_{0.98}$& Co$_{0.47}$Fe$_{1.02}$Mn$_{1.17}$Ge$_{1.34}$  \\[1.5ex]
\hline\hline
\end{tabular}
\label{tab:EDS_compo_Co2-aFeaMnGe}
\end{table*}
%%%%%%%%%%%%%%%%%%%%%%%%%%%%

%%%%%%%%%%%%%%%%%%%%%%%
The alloy compositions, as determined by EDS, are shown in Table \ref{tab:EDS_compo_Co2-aFeaMnGe}. Both the grain and the grain boundary compositions are very close to the target compositions up to $\alpha$ = 1.375, which suggests formation of a single-phase compound, in agreement with the micrograph observations. In case of $\alpha$ = 0.50 and 0.75, one may, however, argue that measured composition does not meet our criterion (within 5 \% of target) to make them a single-phase. The discrepancy in those two alloys have trivial explanations, however. Prior to arc melting, we put in weights for each element that differs slightly from final target weight, particularly for Mn (usually 5 \% excess), to compensate for volatility and ensure that the final composition is in line with what we desire. In those two cases, our initial prediction of the necessary weights was not quite right, but the resulting measured composition and atomic ratios were within what we deemed an acceptable range. (Similarly, for $\alpha$ = 1.20, our initial intention was to make $\alpha$ = 1.25, but mistakenly we put in the wrong weights for Co and Fe, so we adjusted the Mn and Ge weights to target a composition of $\alpha$ = 1.20.) In short, verifying the {\em actual} final composition and its uniformity is crucial for a variety of reasons.

In the case of $\alpha$ = 1.50, two different contrast can be observed in the optical micrograph, and EDS verifies that those regions indeed have different elemental compositions. The secondary phase is poor in Co and slightly rich in Fe. But, since the secondary phase volume fraction is much smaller, the main phase composition is still close to the target composition. Similar behaviour can be observed at $\alpha$ = 1.75, where there are two distinct phases. Interestingly, the secondary phase of the $\alpha$ = 1.50 sample is almost identical to the main phase at $\alpha$ = 1.75, as though we are seeing a competition between those two phases unfold. Following this line of thought, the main phase for $\alpha=1.75$ should become more dominant as $\alpha$ increases further, and it should be possible to obtain singe-phase microstructure upon further increasing the Fe concentration. This assertion seems plausible when one notes that Fe$_2$MnGe exhibits single-phase microstructure and crystallises in hexagonal DO$_{19}$ structure, a different structure than of CoFeMnGe ($\alpha$ = 1.0). One then hazards a guess that one is seeing a Co-doped hexagonal Fe$_2$MnGe phase emerge as the main phase at the expense of the cubic phase that dominates for $\alpha\leq 1.5$. We will investigate this idea further in examining the XRD data below.

%%%%%%%%%%%%%%%%%%%%%%%%%%
\begin{figure*}[h]\centering
\includegraphics[width =1.75\columnwidth]{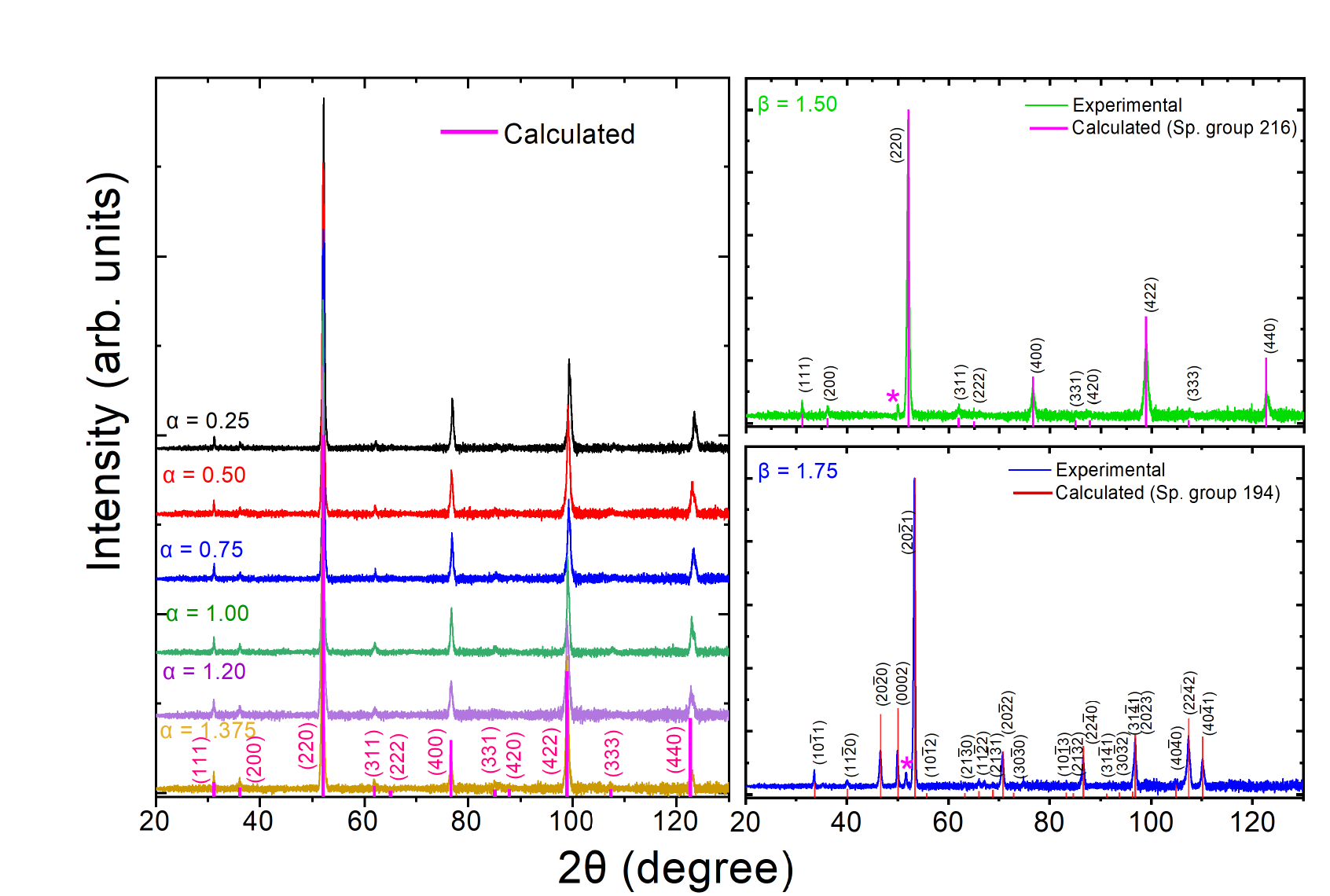}
\caption{(colour online) XRD scans of (Co$_{2-\alpha}$Fe$_{\alpha}$)MnGe alloy series. (Inset) The scans at lower angles are expanded up to make any impurities peaks clearly visible.}
\label{fig:xrd_Co2-aFeaMnGe}
\end{figure*}
%%%%%%%%%%%%%%%%%%%%%%%%%%%%%%%
The powder XRD patterns, shown in figure \ref{fig:xrd_Co2-aFeaMnGe}, also suggest samples up to $\alpha$ = 1.375 are single-phase. These patterns can be very well indexed to an fcc unit cell, with reasonably good agreement between the experimental and theoretical intensity, the latter being calculated based on space group 216. For $\alpha$ = 1.50, the peaks still can be fit to a cubic structure except the peak at 2$\theta$ = 50.02$\degree$. The XRD scan on a bulk sample with $\alpha$ = 1.50 (shown in the supplementary information) further reveals additional peaks at 33.55$\degree$, and at 53.30$\degree$. Interestingly, all of these extra peaks can be indexed to the DO$_{19}$ hexagonal structure adopted by Fe$_2$MnGe as: 33.55$\degree$ (10$\overline{1}$1), 50.02 $\degree$ (0002), and 53.30$\degree$ (20$\overline{2}$1). In the light of this, the $\alpha$ = 1.50 alloy can be deemed as a mixture of two structures, viz. cubic and hexagonal (DO$_{19}$). Since almost all peaks corresponding to cubic structure are visible but only a few related to DO$_{19}$, the main phase can be thought of as cubic whereas the secondary phase as hexagonal. The details of the structural phase transition between the two and its impact on electronic and magnetic properties would be a useful future study; previously we have investigated a similar competition in Fe$_{3-x}$V$_x$Ge.\cite{mahat2020}

\begin{figure}[h]
\centering
\includegraphics[width =0.75\columnwidth]{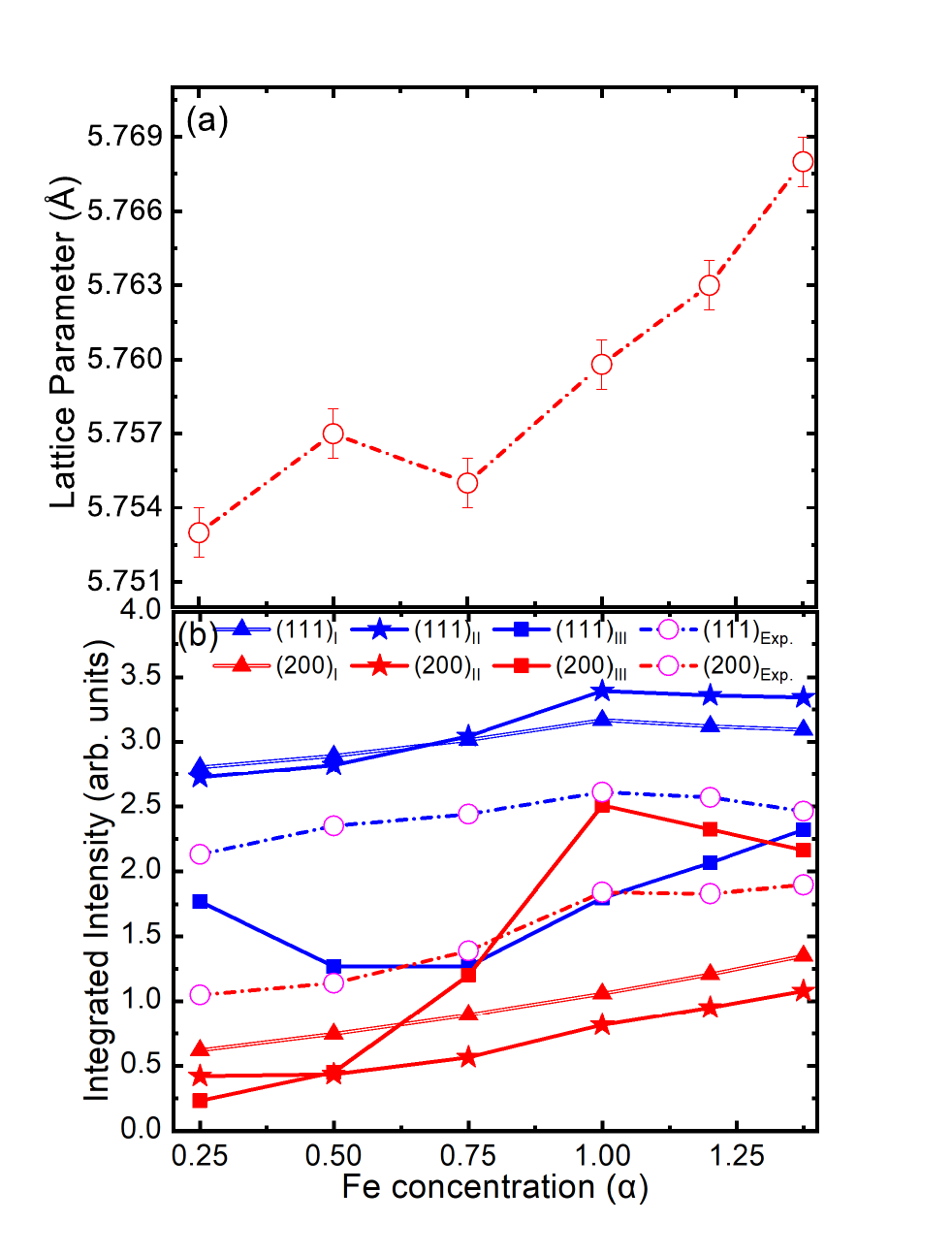}
\caption{(colour online) Variation of (a) lattice parameter and (b) superlattice peak intensities with Fe concentration $\alpha$ in (Co$_{2-\alpha}$Fe$_\alpha$)MnGe. The intensities calculated assuming type I configuration has closer semblance of experimental intensity.}
\label{fig:Co2-aFeaMnGe_lattice_variation}
\end{figure}

In contrast to $\alpha$ = 1.50, the XRD pattern of $\alpha$ = 1.75 shows a majority of peaks that correspond to a DO$_{19}$ structure, with the exception of the peak at 51.62$\degree$, which can be indexed as (220) peak of the cubic structure. This means the main phase at $\alpha$ = 1.75 crystallines in a hexagonal structure with a secondary cubic phase. Note in the EDS analysis, we realised the secondary phase of $\alpha$ = 1.50 and the main phase of $\alpha$ = 1.75 have identical compositions and argued that they could have same crystal structure. The XRD analysis seems to verify this argument and thus we contend that the hexagonal structure dominates upon increasing the Fe concentration en route to the Fe$_2$MnGe endpoint. 

The variation of lattice parameter is shown in figure \ref{fig:Co2-aFeaMnGe_lattice_variation}(a). There is a slight extension in lattice parameter as an effect of substituting Fe (atomic radius 156 pm) for Co (atomic radius 152 pm). The comparison of experimental and calculated superlattice peak intensity is shown in figure \ref{fig:Co2-aFeaMnGe_lattice_variation}(b). The integrated intensities show an upward trend as Fe concentration increases with a slight downturn after $\alpha > 1$. The calculated intensities for structure I and II (recall the site assignments in Table~\ref{tab:atom_config2}) also exhibit a similar trend, with type I somewhat closer to the experimental curve. This lends some support that type I is preferable. As  with the previous series, the experimental (111) peak has lower intensity, whereas (200) has higher intensity than the calculated values. This can be attributed to the preferred texturing of samples (which can be clearly observed on the bulk XRD patterns (see supplementary information \cite{supplementary2021}). As a result, we will again reconsider the site assignments after examining the magnetic and theoretical information.

%%%%%%%%%%%%%%%%%%%%%%%%
\begin{figure}[h]
\centering
\includegraphics[width =0.75\columnwidth]{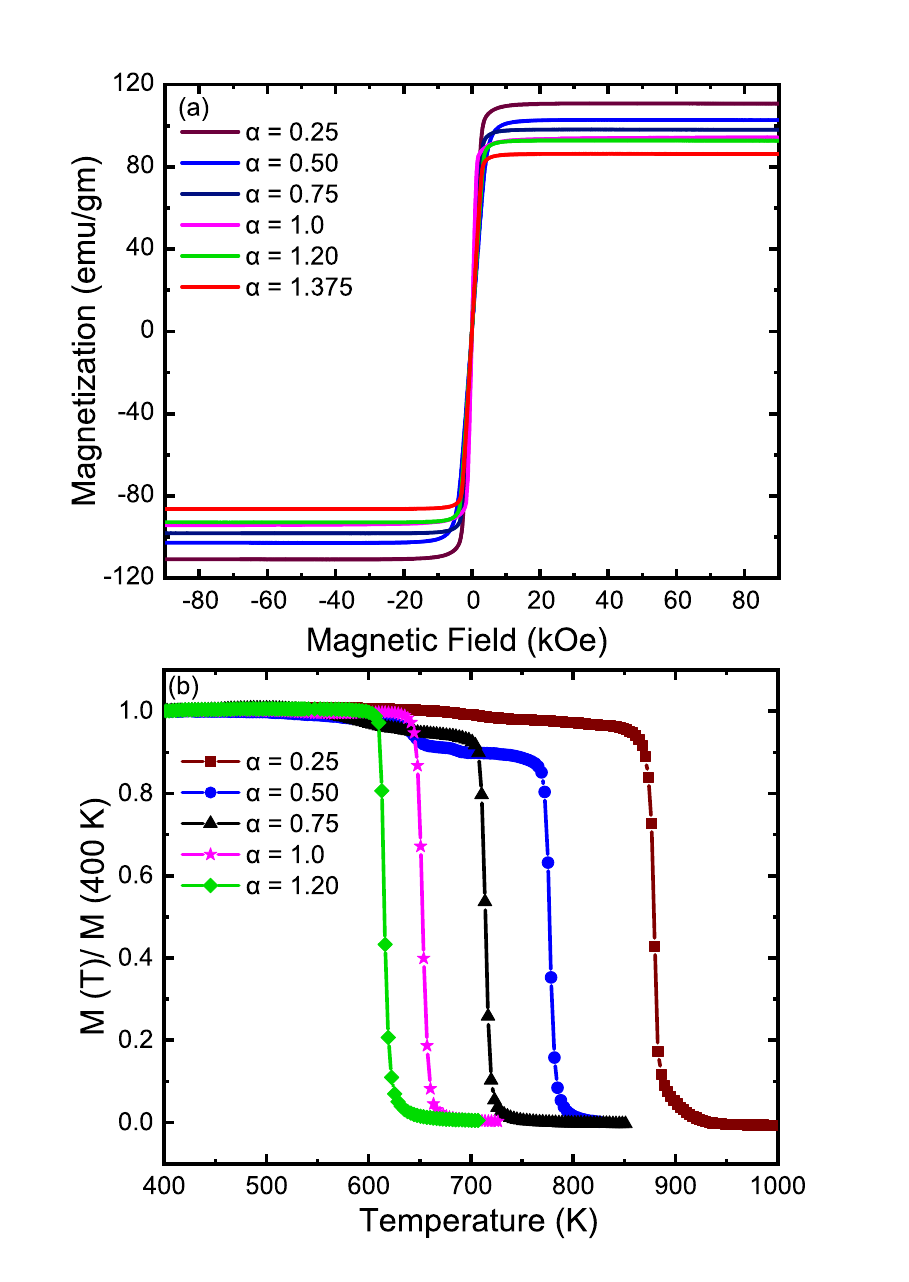}
\caption{(colour online) (a) M-H loops measured at T = 5 K for (Co$_{2-a}$Fe$_{\alpha}$)MnGe alloy series. (b) Temperature dependent variation of spontaneous magnetisation, which suggests drop in T$_\text{C}$ as the Fe concentration increases.}
\label{fig:mh_Co2-aFeaMnGe}
\end{figure}
%%%%%%%%%%%%%%%%%%%%%%

\begin{figure}[h]
\centering
\includegraphics[width =0.9\columnwidth]{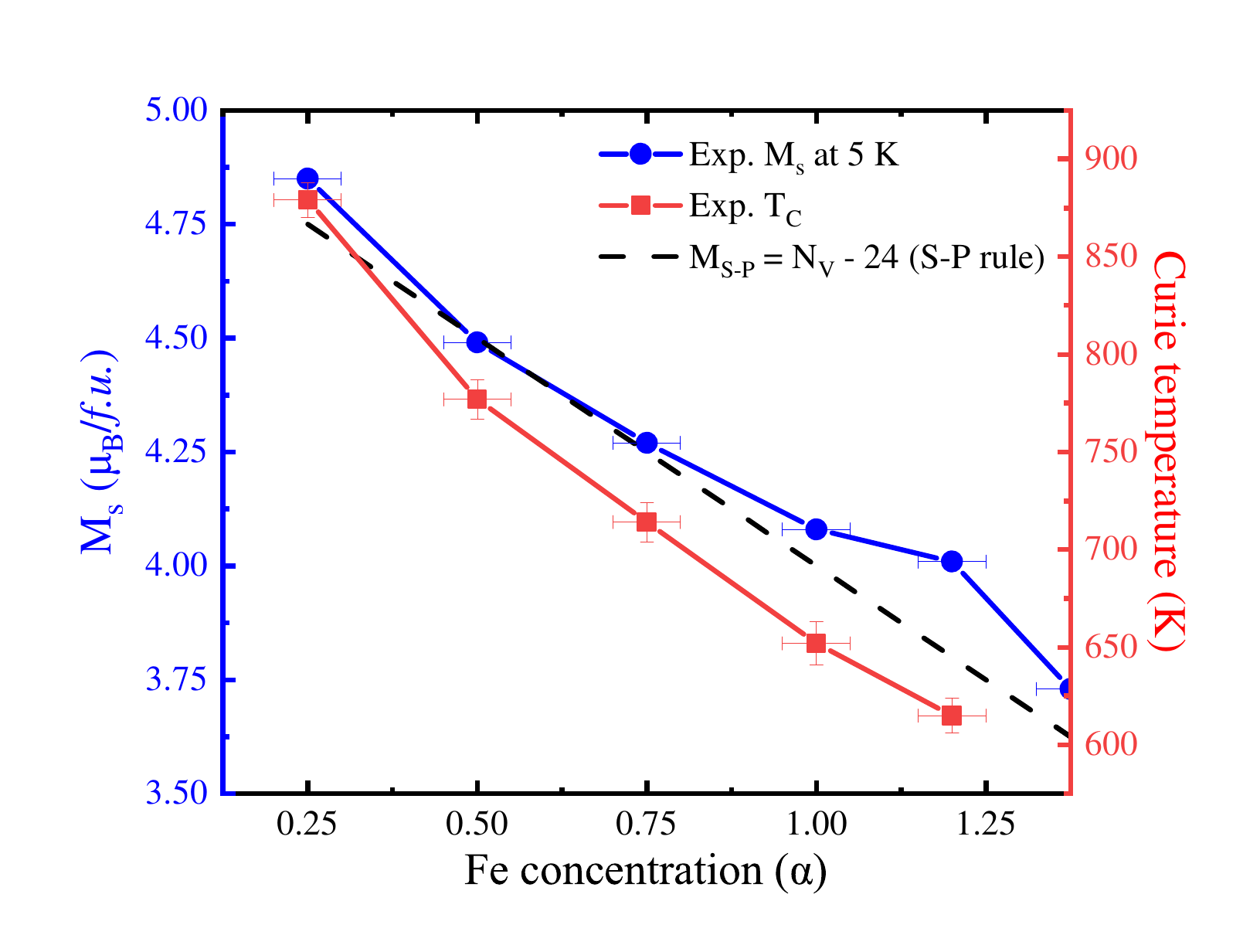}
\caption{(colour online) The variation of M$_\text{s}$ and T$_\text{C}$ with Fe concentration $\alpha$ in (Co$_{2-\alpha}$Fe$_\alpha$)MnGe. A roughly linear decrease in both the parameters can be observed with the increase in Fe concentration.}
\label{fig:mh_variation_Co2-aFeaMnGe}
\end{figure}

The alloys are soft ferromagnets, as indicated by magnetic hysteresis loops shown in figure \ref{fig:mh_Co2-aFeaMnGe}(a). The magnetisation versus temperature characteristics in \ref{fig:mh_Co2-aFeaMnGe} (b) indicate that the these alloys have very high T$_\text{C}$. Since N$_\text{v}$ decreases as Fe concentration increases,  M$_\text{s}$ and T$_\text{C}$ also decrease as the Fe concentration increases (see Fig. \ref{fig:mh_variation_Co2-aFeaMnGe}). It is again interesting to note M$_\text{s}$ and T$_\text{C}$ exhibit similar trends with N$_\text{v}$, implying a common origin, and the measured M$_\text{s}$ at 5 K agrees rather well with the Slater-Pauling value. However, there is not quite the strict linear variation we had witnessed with the other series. This is perhaps due to compositions being not exactly on target, particularly in the case of $\alpha$ = 0.50, 0.75, and 1.20, which we already pointed out in the EDS analysis section above. 

%%%%%%%%%%%%%%%%%%%%
%%%%%%%%%%%%%%%%%%%

\subsubsection{Findings from calculations}

%%%%%%%%%%%%%%%%%%
\begin{figure}[h]\centering
\includegraphics[width = 0.75\columnwidth]{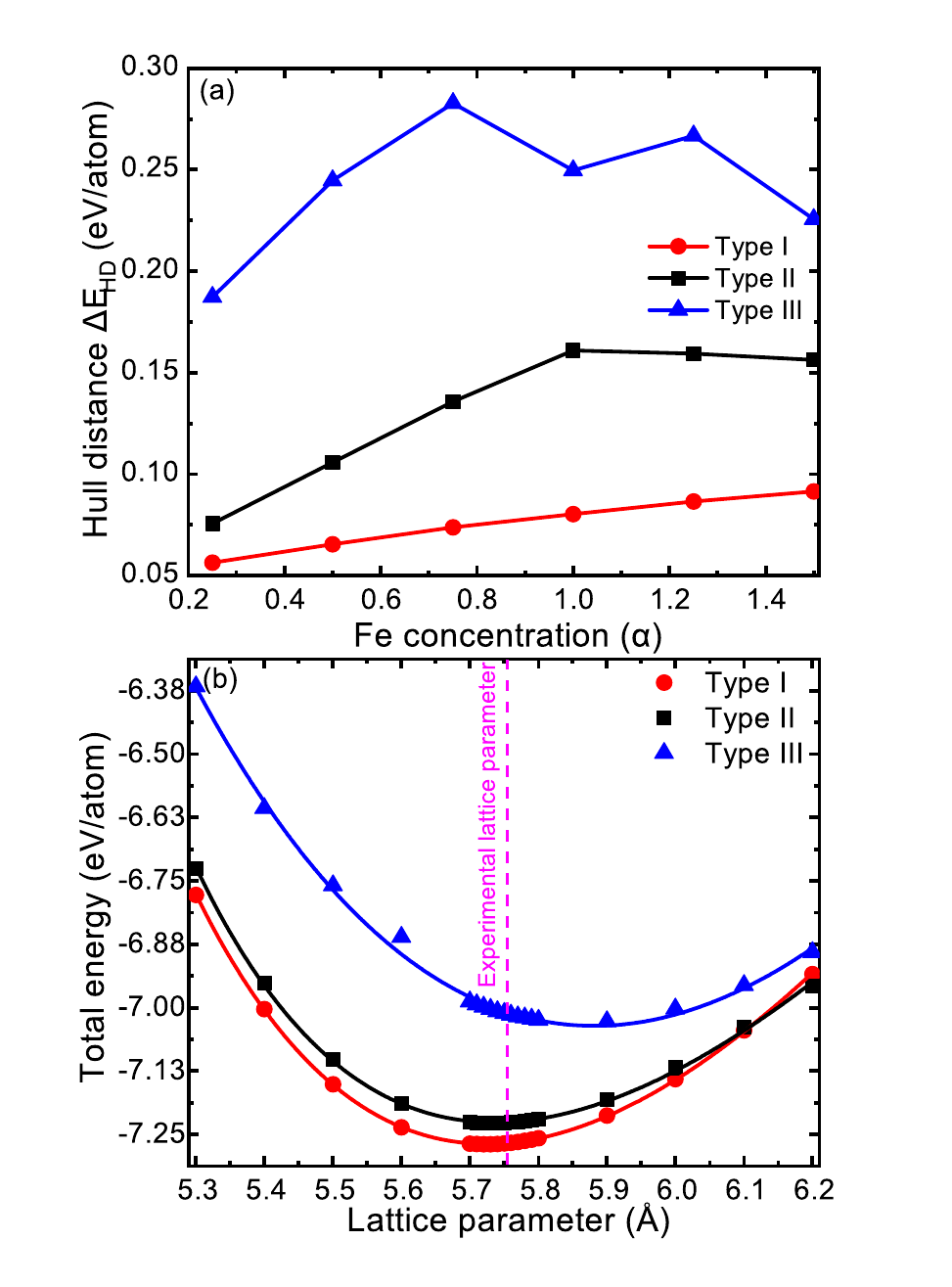}
\caption{(colour online) (a) The variation of hull distance ($\Delta E_{HD}$) as a function of Fe concentration $\alpha$ in (Co$_{2-\alpha}$Fe$_\alpha$)MnGe. The type I structure has the minimum hull distance, suggesting it is more stable phase than others. (b) Structure optimisation by calculating the total energy as a function of lattice parameter. The optimisation is for $\alpha$ = 0.50, which also suggests type I is energetically the most favourable structure of the three considered.}
\label{fig:hull_dis2}
\end{figure}
%%%%%%%%%%%%%%%%%

The three main configurations in this case can be summarised as: type I, where a substituted Fe atom directly occupies a vacant Co site in sublattice A; type II, where a substituted Fe atom displaces a Mn atom from the B$_T$ site towards the vacant Co site and itself occupies B$_T$ site; and type III, where a substituted Fe atom occupies the B$_Z$ site by displacing a Ge atom toward a vacant Co site. The $\Delta E_{HD}$ is calculated as a function of Fe concentration and the corresponding plot is shown in figure \ref{fig:hull_dis2}(a). Over the entire substitution range, the type I structure has the lowest $\Delta E_{HD}$. It appears that $\Delta E_{HD}$ of type I is slightly higher than 0.05 eV/atom, the threshold value for the prediction of phase-stability. Even so, when compared with other configurations studied -- where, for the most of substitution range, the $\Delta E_{HD}$ is noticeably greater than 0.10 eV/atom -- type I is lower by at least by 0.05 eV/atom. One should also keep in mind that there is possible uncertainty in the DFT calculated total energies, and also in the convex hull energy that we use from the OQMD database. We therefore consider it plausible to argue that the type I structure is the most favourable configuration for (Co$_{2-\alpha}$Fe$_\alpha$)MnGe series. This argument of type I being most favourable is further supported by the plot of total energy for different lattice volumes (see Fig. \ref{fig:hull_dis2}(b)) at Fe concentration of $\alpha$ = 0.50, which shows that the type I structure has the lowest total energy.

%%%%%%%%%%%%%%%%%%%%%
\begin{figure}[h]\centering
\includegraphics[width = 0.75\columnwidth]{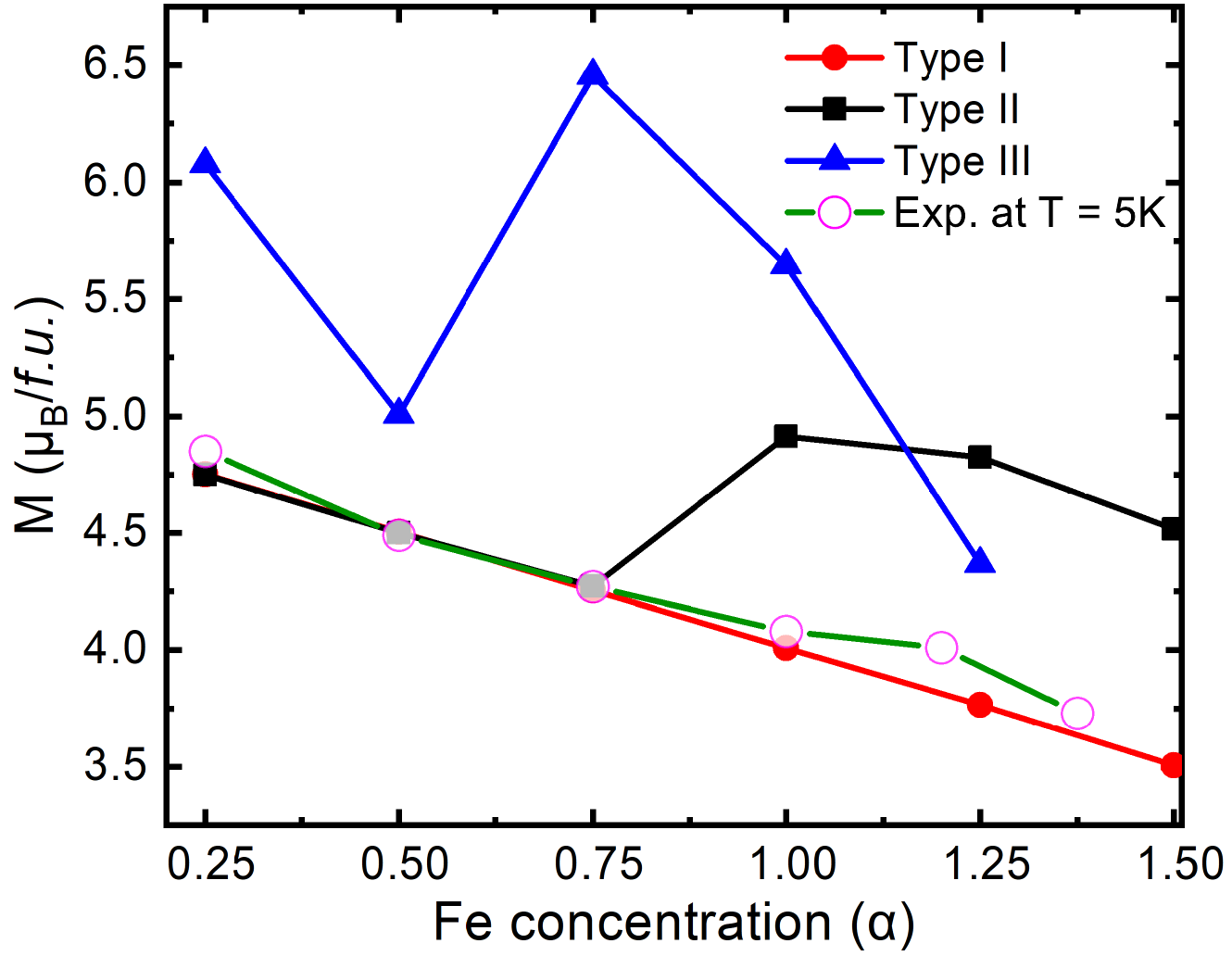}
\caption{(colour online) The variation of calculated and experimental magnetic moment with Fe concentration $\alpha$ in (Co$_{2-\alpha}$Fe$_\alpha$)MnGe, which suggest the values of type I structure matches reasonably well with the experimental values.}
\label{fig:Co2-aFeaMnGe_cal_mh}
\end{figure}
%%%%%%%%%%%%%%%%%

In figure \ref{fig:Co2-aFeaMnGe_cal_mh}, the calculated moment for three structure types are compared with the experimental moment at T = 5 K. The moment variation as a function of Fe concentration shows only the calculated moment for type I matches with the experimental value over the entire single-phase spectrum. The calculated moment of type II seems to agree with the experimental moment up to $\alpha$ = 0.75, but after $\alpha$ = 0.75 the two moments differ markedly. Further, the linear variation, which means linear dependence of M$_\text{s}$ with N$_\text{v}$, is obtained only for the type I structure. This provides more persuasive support that (Co$_{2-\alpha}$Fe$_\alpha$)MnGe series favours type I structure.  

%%%%%%%%%%%%%%%%%%%%%
\begin{figure}[h]
\centering
\includegraphics[width = 0.8\columnwidth]{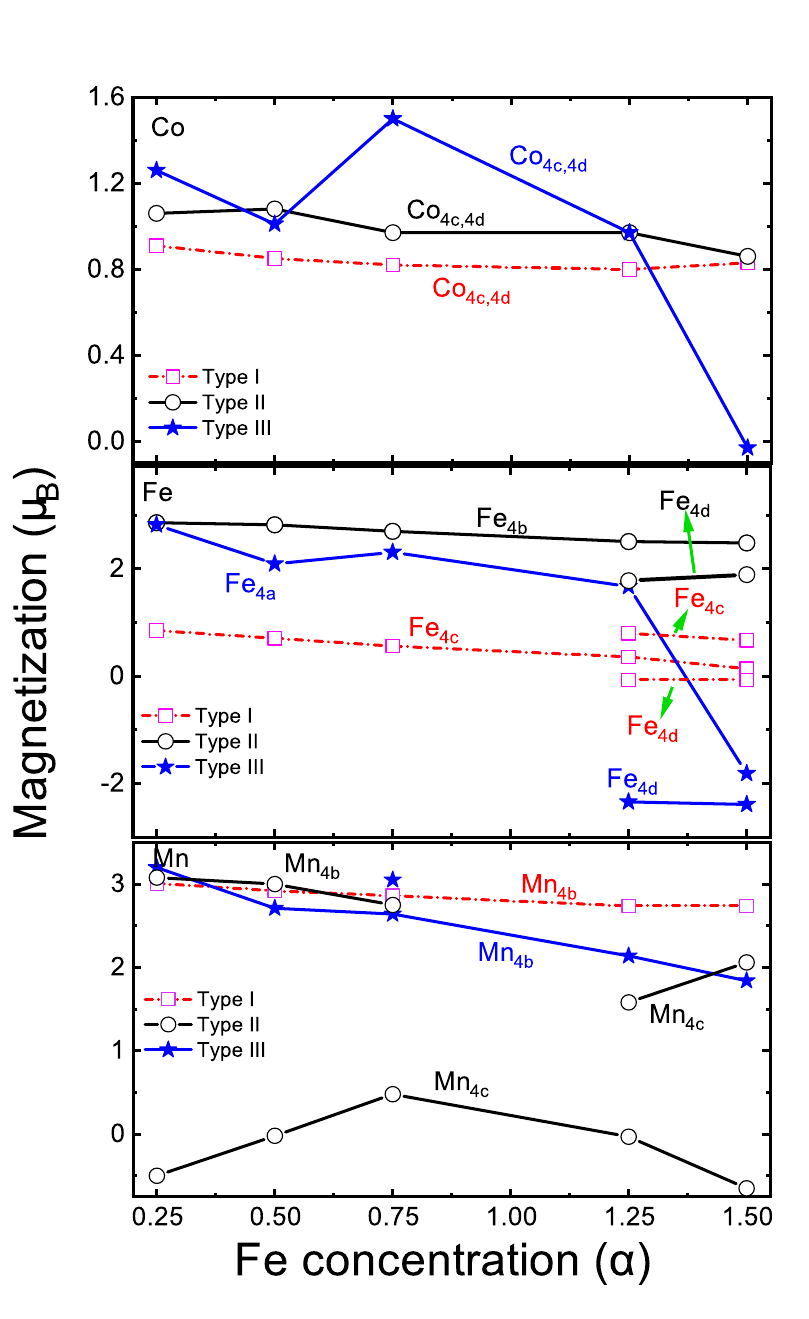}
\caption{(colour online) The variation of calculated atomic magnetic moments in (Co$_{2-\alpha}$Fe$_\alpha$)MnGe. The type I structure exhibits a linear moment variation, in contrast to the others. At some compositions, atoms exhibited different moments while occupying the same Wyckoff site.}
\label{fig:Co2-aFeaMnGe_atomic_m}
\end{figure}
%%%%%%%%%%%%%%%%%

Now, when compared with the (Co$_{2-\alpha}$Mn$_\alpha$)FeGe series, where we suggested that the type II structure is the most stable, the (Co$_{2-\alpha}$Fe$_\alpha$)MnGe series favours the type I structure. These two series are similar in that both involve replacement of Co from sublattice A. However, they differ in which atom is replacing the Co atom and which is already occupying the B$_T$ site. From the observation of these two series, together with the Co$_2$Fe$_{1-\beta}$Mn$_{\beta}$Ge, one thing that becomes clear is, {\em no matter how the substitution is made, the favoured structure is the one that obeys the 4-2 rule with each element in a plausible electron configuration.} Saying that the less valence transition element always prefers to share the same sublattice with the main group element gives the same result, but in our opinion without a satisfactory physical motivation. 

%% CITE BURCH

%%%%%%%%%%%%%%%%%%%%%%%%
\begin{table*}[h]
\centering
\scriptsize
\caption{DFT results for some possible atomic configurations of (Co$_{2-\alpha}$Fe$_\alpha$)MnGe. In the table 4a, 4b, 4c, and 4d represents the Wyckoff sites for the coordinates (0,0,0), (0.5,0.5,0.5), (0.25,0.25,0.25), and (0.75,0.75,0.75) respectively. E (in eV/atom) is total energy, m (in $\mu_B$) stands for magnetic moment, and $<a>$ (in \AA) represents the optimised lattice parameter. Note that in our model, 4c and 4d corresponds to sublattice A, whereas 4a and 4b corresponds to sublattice B. And if an element exhibits different moment values, the values are shown with the Wyckoff site (which the element occupies) in the subscript.}

\begin{tabular}{c c c c c c c c}%
 \hline\hline
\textbf{Configuration} & \textbf{4d} & \textbf{4c} & \textbf{4b} & \textbf{4a} & \textbf{E} & \textbf{m$_{tot}$} & \textbf{$<$a$>$} \\\hline
x = 0.25& & & & & & &  \\ 
\setrow{\bfseries}I& 4Co& 3Co,1Fe& 4Mn& 4Ge& -7.22 & 4.75&5.730 \\[0.5ex]
II& 4Co& 3Co,1Mn& 3Mn,1Fe & 4Ge & -7.20 & 4.75&5.740 \\[0.5ex]
III& 4Co& 3Co,1Ge& 4Mn& 3Ge,1Fe& -7.09 & 6.08&5.819 \\[0.5ex]\hline
x = 0.50& & & & & & &  \\ [0.5ex]
\setrow{\bfseries}I& 4Co & 2Co,2Fe & 4Mn & 4Ge & -7.27 & 4.50 & 5.723 \\[0.5ex]
II& 4Co & 2Co,2Mn & 2Mn,2Fe & 4Ge & -7.23 & 4.50 & 5.733 \\[0.5ex]
III& 4Co & 2Co,2Ge & 4Mn & 2Ge,2Fe & -7.09 & 5.01 & Ortho. \\[0.5ex]\hline
x = 0.75& & & & & & &  \\ [0.5ex]
\setrow{\bfseries}I& 4Co & 1Co,3Fe & 4Mn & 4Ge & -7.32 & 4.26 & 5.717 \\[0.5ex]
II& 4Co & 1Co,3Mn & 1Mn,3Fe & 4Ge & -7.26 & 4.27 & 5.723 \\[0.5ex]
III& 4Co & 1Co,3Ge & 4Mn & 1Ge,3Fe & -7.11 & 6.45 & 5.844 \\[0.5ex]\hline
x = 1.25& & & & & & &  \\ [0.5ex]
\setrow{\bfseries}I& 3Co,1Fe & 4Fe & 4Mn & 4Ge & -7.43 & 3.76 & 5.707 \\[0.5ex]
II& 3Co,1Fe & 4Mn & 4Fe & 4Ge & -7.36 & 4.83 & 5.751 \\[0.5ex]
III& 3Co,1Fe & 4Ge & 4Mn & 4Fe& -7.25 & 4.37 & 5.762 \\[0.5ex]\hline
x = 1.50& & & & & & &  \\ [0.5ex]
\setrow{\bfseries}I& 2Co,2Fe & 4Fe & 4Mn & 4Ge & -7.48 & 3.51 & 5.704 \\[0.5ex]
II& 2Co,2Fe & 4Mn & 4Fe & 4Ge & -7.42 & 4.51 & Tetra. \\[0.5ex]
III& 2Co,2Fe & 4Ge & 4Mn & 4Fe& -7.35 & -1.18 & 5.740 \\[0.5ex]
\end{tabular}
\label{table:Co2-aFeaMnGe_cal_table}
\end{table*}

\begin{figure}[h]
\centering
\includegraphics[width = 1.1\columnwidth]{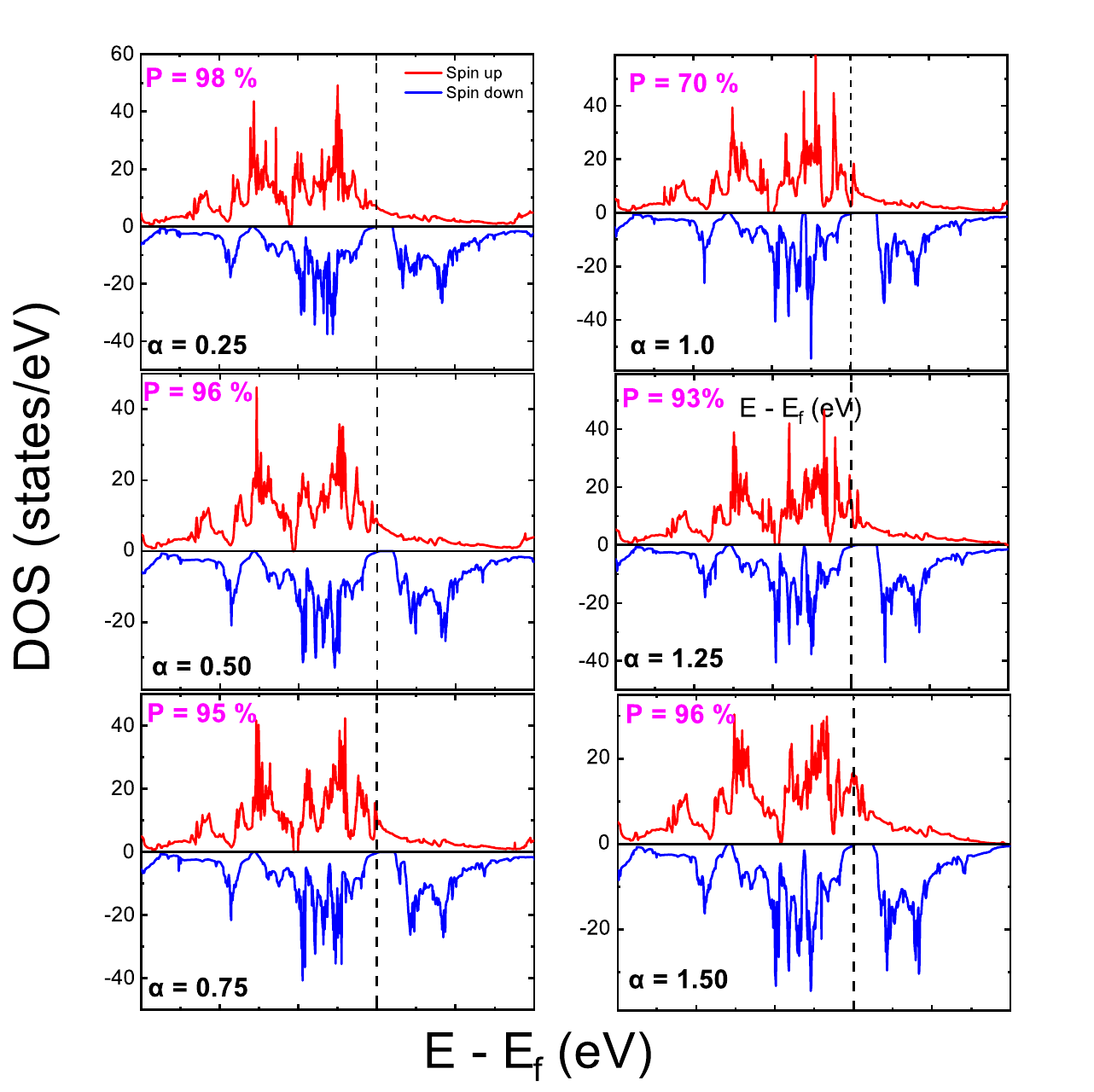}
\caption{(colour online) Spin resolved total density of states plot for (Co$_{2-\alpha}$Fe$_\alpha$)MnGe alloy series. All the alloys exhibit very high value of spin polarisation.}
\label{fig:Co2-aFeaMnGe_cal_dos}
\end{figure}

%%%%%%%%%%%%%%%%%%%%%%%%%
The total density of states (DOS) plots at different Fe concentration level are shown in figure \ref{fig:Co2-aFeaMnGe_cal_dos}. Unlike the (Co$_{2-\alpha}$Mn$_\alpha$)FeGe series with its parent Co$_2$FeGe, the parent alloy Co$_2$MnGe of the (Co$_{2-\alpha}$Fe$_\alpha$)MnGe series is already half-metallic. As such, E$_F$ is located deep in the gap in the minority channel, slightly closer to the valence band (see Fig. \ref{fig:Co2Fe1-bMnbGe_dos}). After Fe is introduced at the expense of Co, the bands in both the majority and minority channel are shifted higher in energy. As a result, the valence band start crossing the E$_F$ in the minority channel and the half-metallic character is destroyed. Upon further increasing the Fe concentration, the number of minority states at E$_F$ grows more and more. But, at the same time there is also increase of states in the majority band. The net effect is that spin polarisation doesn't drop significantly. The calculated results suggest the alloys retain very high spin polarisation, more than 95 \%, up to $\alpha$ = 0.75. There is, however, sudden drop in spin polarisation at $\alpha$ = 1.0. This drop is mainly because E$_F$ falls in the local minimum in the majority band. The minimum is very narrow, hence, by slightly changing the composition, it is possible to tune the Fermi level towards energies with higher densities of states, which in turn increases the spin polarisation. This is exactly what one can see at $\alpha$ =  1.25, where the spin polarisation climbs back to $\sim$ 93 \%, even though there is continuous increase in the number of minority states. Notice that on the both sides of $\alpha$ = 1.0, that is $\alpha$ = 0.75 and 1.25, the spin polarisation is very high. This suggests one can also tune material with higher spin polarisation than that of CoFeMnGe by slightly tweaking the Co/Fe ratio, similar to Co/Mn ratio discussed earlier.

%%%%%%%%%%%%%%%%%%%
%%%%%%%%%%%%%%%%%%%%

\subsubsection{Conclusions for (Co$_{2-\alpha}$Fe$_\alpha$)MnGe}
The substitution of Co by Fe atoms in Co$_2$MnGe doesn't destroy it's single-phase microstructure, and single-phase behaviour persists as far as $\alpha$ = 1.375, somewhat farther than in (Co$_{2-\alpha}$Mn$_\alpha$)FeGe (single-phase up to $\alpha$ = 1.25). The EDS and XRD analyses suggest the phase separation at high Fe could be due to a competition between cubic and hexagonal structures, the latter being dominant for higher Fe concentration and already known to be a stable structure for Fe$_2$MnGe. The theoretical and experimental results together point to type I being favourable structure for (Co$_{2-\alpha}$Fe$_\alpha$)MnGe, in agreement with the 4-2 rule. The electronic calculations suggest the samples are nearly half-metallic with very high spin polarisation (more than 90\%) and it could be possible to enhance the spin polarisation of CoFeMnGe by slightly altering the Co/Fe ratio.  

%%%%%%%%%%%%%

%%%%%%%%%%%%%%%%%%%%%%%%%%%%%%%%%%
\section{\label{sec:level7}Summary}
Here we summarise the results of all three series and discuss how substituting different sublattices impacts various alloy properties, such as phase-stability, structural, magnetic and the electronic properties. All alloys that were prepared for this work are presented in table \ref{table:all_series} along with their phase-stability, experimental M$_\text{s}$ and T$_\text{C}$, and calculated spin polarisation.

%%%%%%%%%%%%%%%%%%%
\begin{table*}[h]
\scriptsize
\caption{A brief summary of alloy properties with reference to N$_\text{v}$. The phase-stability, experimental saturation magnetisation (M$_\text{s}$), experimental Curie temperature (T$_\text{C}$), and calculated spin polarisation (P) are shown. In the table, the left side includes the alloys that are obtained from (Co$_{2-\alpha}$Mn$_\alpha$)FeGe series, whereas the alloys on the right hand side are the result of Co$_2$(Fe$_{1-\beta}$Mn$_\beta$)Ge followed by (Co$_{2-\alpha}$Fe$_\alpha$)MnGe series. The alloy properties are found to be dependent on the N$_\text{v}$ regardless of substitution schemes. The data for Fe$_2$MnGe are from ref.\ \cite{keshavarz2019fe2mnge}. Note Ayan {\em et al.}\cite{aryal2021} found Mn$_2$FeGe to single-phase hexagonal under different annealing conditions than used here.}
\begin{tabular}{>{\rowmac}c>{\rowmac}c>{\rowmac}c>{\rowmac}r>{\rowmac}c>{\rowmac}c>{\rowmac}c>{\rowmac}c>{\rowmac}c>{\rowmac}c>{\rowmac}%
  c<{\clearrow}}
  \hline\hline
\textbf{N$_\text{v}$} & \textbf{Alloy} & \multicolumn{1}{p{1cm}}{\centering{\textbf{Single-phase}}} & \multicolumn{1}{p{0.1cm}}{\centering{\textbf{M$_\text{s}$($\mu_B$)}}} & \textbf{T$_\text{C}$(K)} & \textbf{P (\%)} & \textbf{Alloy} &\multicolumn{1}{p{1cm}}{\centering{\textbf{Single-phase}}} & \textbf{M$_\text{s}$($\mu_B$)} &\textbf{T$_\text{C}$(K)} &\textbf{P (\%)} \\\hline
30& Co$_2$FeGe& No& 5.68(2) & 981 & -52 & ---&---&---&---&---\\[0.5ex]
29.75& Co$_{1.875}$Mn$_{0.125}$FeGe& No& 5.85(4) & 1002(3) & 10 &Co$_2$Fe$_{0.75}$Mn$_{0.25}$Ge &No&5.66(3)&991(7)&35\\[0.5ex]
29.50& Co$_{1.75}$Mn$_{0.25}$FeGe& Yes& 5.58(9) & 995(4) & 79 &Co$_2$Fe$_{0.50}$Mn$_{0.50}$Ge &Yes&5.59(3)&994(6)&84\\[0.5ex]
29.25& Co$_{1.625}$Mn$_{0.375}$FeGe& Yes& 5.4(1) & --- & 100 &Co$_2$Fe$_{0.25}$Mn$_{0.75}$Ge &Yes&5.31(3)&974(6)&100\\[0.5ex]
29.0& Co$_{1.50}$Mn$_{0.50}$FeGe& Yes& 5.1(1) & 918(4) & 93 &Co$_2$MnGe &Yes&5.09(5)&905(3)&100\\[0.5ex]
28.75& ---& ---& --- & --- & --- &Co$_{1.75}$Fe$_{0.25}$MnGe &Yes&4.85(5)&879(9)&98\\[0.5ex]
28.50& Co$_{1.25}$Mn$_{0.75}$FeGe& Yes& 4.62(9) & 788(4) & 87 &Co$_{1.50}$Fe$_{0.50}$MnGe &Yes&4.49&777(10)&96\\[0.5ex]
28.25& ---& ---& --- & --- & --- &Co$_{1.25}$Fe$_{0.75}$MnGe &Yes&4.27&714(9)&95\\[0.5ex]
28.0& CoMnFeGe& Yes& 4.08(4) & 653(5) & 70 &CoFeMnGe &Yes&4.08(4)&653(5)&70\\[0.5ex]
27.80& ---& ---& --- & --- & --- &Co$_{0.80}$Fe$_{1.20}$MnGe &Yes&4.01(4)&615(9)&---\\[0.5ex]
27.625& ---& ---& --- & --- & --- &Co$_{0.625}$Fe$_{1.375}$MnGe &Yes&3.73(4)&---&---\\[0.5ex]
27.50& Co$_{0.75}$Mn$_{1.25}$FeGe& Yes& 3.54(9) & 554(6) & 94 &Co$_{0.50}$Fe$_{1.50}$MnGe &No&---&---&96\\[0.5ex]
27.25& Co$_{0.625}$Mn$_{1.375}$FeGe& No& 3.30(5) & --- & --- &Co$_{0.25}$Fe$_{1.75}$MnGe &No&---&---&---\\[0.5ex]
27.0& Co$_{0.50}$Mn$_{1.50}$FeGe& No& --- & --- & 95 &Fe$_{2}$MnGe &Yes (hexagonal)&5.02&$\sim\!500$&---\\[0.5ex]
26.0&Mn$_{2}$FeGe& No& --- & --- & 89 &--- &---&---&---&---\\[0.5ex]
\end{tabular}
\label{table:all_series}
\end{table*}
%%%%%%%%%%%%%%%%
\textbf{Phase-stability}: Stabilising single-phase compounds based on Co$_2$FeGe (which itself is multi-phased) was one of the primary goals of this work. In this regard it was observed that substitution of Mn for either Co (\textit{i.e.}, (Co$_{2-\alpha}$Mn$_\alpha$)FeGe) or Fe (\textit{i.e.}, Co$_2$(Fe$_{1-\beta}$Mn$_\beta$)Ge) successfully stabilises a single-phase compound. However, the minimum amount of Mn that is needed to obtain phase-pure material is found to differ based on the substitution scheme; our investigation suggest $\alpha$ $>$ 0.125 and $\beta$ $>$ 0.25 as lower Mn concentration limits for (Co$_{2-\alpha}$Mn$_\alpha$)FeGe and Co$_2$(Fe$_{1-\beta}$Mn$_\beta$)Ge respectively. Interestingly, at those threshold values, the total number of valence electrons (N$_\text{v}$) of the alloy is found to be the same, \textit{i.e.}, 29.75 (see table \ref{table:all_series}). We therefore suggest that N$_\text{v}$ has to be less than 29.75 to stabilise a phase pure compound by substituting Mn in Co$_2$FeGe. This is consistent with the fact that the 30-electron parent Co$_2$FeGe is not single phase. The fact that Co$_2$FeSi {\em is} a stable single-phase material suggests a somewhat greater ability for Si to act as a hole reservoir\cite{galanakis2002origin} compared to Ge. 

It can be noticed that there is also an optimum value of substitution {\em above} which the microstructure decomposes into multiple phases and the level of substitution required does depend on the substitution schemes. We find $\alpha_{Mn}$ $<$ 1.375 for (Co$_{2-\alpha}$Mn$_\alpha$)FeGe and $\alpha_{Fe}$ $<$ 1.50 for (Co$_{2-\alpha}$Fe$_\alpha$)MnGe, which corresponds to N$_\text{v}$ values of 27.25 and 27.50 respectively. Interestingly, these values are fairly close to one another (particularly given our relatively large steps in $\alpha$ and $\beta$) and thus suggest there is also a common {\em lower bound} on N$_\text{v}$ for obtaining single-phase compounds in the Co-Fe-Mn-Ge system. Also interesting is that both systems for low N$_\text{v}$ change to a hexagonal structure with endpoints Mn$_2$FeGe and Fe$_2$MnGe with 26 and 27 valence electrons, respectively. 

We speculate that the strict range of valence electron count for single phase behaviour is a result of the `4/2 rule'. First, consider the parent compound Co$_2$FeGe. As noted above, to follow the 4/2 rule Co would like to adopt an electron configuration of 5$\uparrow$ and 4$\downarrow$. While Co has only about 8.3 d electrons\cite{slater1937}, it can use lower-lying Ge $sp$ states rather than occupying its own higher-energy states to account for the remaining 0.7 electron per atom. The Fe atoms would need to adopt a 6$\uparrow$ and 2$\downarrow$ electron configuration. Not only does Fe only have about 7.3 d electrons, it also will have a difficult time supporting a moment of $4\,\mu_\text{B}$/atom when $\sim\!2-2.5\mu_{\text{B}}$/atom is more typical. Thus, both Co and Fe atoms need to occupy Ge-derived states\cite{galanakis2002origin} amounting to 2.5-3 electrons per unit cell and necessitating the use of higher-energy Ge states. Empirically, our substitution results show this is about 0.25 electrons too many. It seems some point the cost of accommodation is sufficiently high that it is more favourable as a whole for the system to simply alter its composition through phase segregation (or in related systems, perhaps adopt a different structure). 

For example, if the primary phase were to end up Co-poor, Fe atoms could move to the Co site and adopt the more favourable 4$\uparrow$ / 4$\downarrow$ configuration. A secondary Fe-poor phase would also serve to reduce the number of Fe atoms in an unfavourable configuration. Being Ge-poor is also advantageous: as we have shown earlier, substituting Fe for Ge in Co$_2$FeGe also stabilises a single phase.\cite{shambhu2022maxmoment} Taken together, this seems to provide multiple relatively easy avenues for the system to segregate into two (or more) more favourable phases. In fact, in looking carefully at the parent compound (see supplemental information here and in Ref.\ \cite{mahat2021possible}), one finds two phases: a main phase which is slightly Ge- and Co-poor, and a secondary phase which is very Fe-poor and Ge-rich. A small substitution in any scheme stabilises a single phase, and in the present case all three schemes serve to reduce the valence electron count. 

Looking at the substitutional series individually, we can see then that while each serves to reduce the electron count, the substitution of Co for Mn does it slightly more effectively. For (Co$_{2-\alpha}$Mn$_\alpha$)FeGe, dual sublattice substitution means that we have ($2-\alpha$) Co and $\alpha$ Fe on the A sublattice. Substituting $\alpha$ Mn for Co reduces the total electron count by 2$\alpha$ per unit cell while at the same time taking $\alpha$ Fe atoms out of the unfavourable 6$\uparrow$ / 2$\downarrow$ configuration, reducing the electron count on both sublattices. However, for Co$_2$(Fe$_{1-\beta}$Mn$_\beta$)Ge we have direct substitution, and are reducing the electron count by $\alpha$ and only on the B sublattice. Based on our electron counting argument, one would then expect the (Co$_{2-\alpha}$Mn$_\alpha$)FeGe scheme to require a slightly lower substitution level to achieve a single phase, as we observe. For (Co$_{2-\alpha}$Fe$_\alpha$)MnGe, the parent compound at $\alpha=0$ is already stable with a valence electron count of 29. In all cases the lower limit of 27.25-27.50 valence electrons is harder to speculate on, other than that it seems a lower electron count may allow the system to form a more densely packed hexagonal phase. Additional theoretical and experimental work on the hexagonal phases\cite{keshavarz2019fe2mnge,aryal2021} is probably necessary, particularly to assess if they have magnetocrystalline anisotropies relevant for applications. 

\textbf{Crystal structure and atomic order}: In all three substitution schemes, the samples which exhibited single-phase microstructure were found to crystallise in a cubic structure. For higher Mn and Fe concentration, that is $\alpha$ $\ge$ 1.50, phases crystallising in hexagonal structures are also realised, giving rise to phase separation at those concentration. The different substitution schemes are found to favour different structure types, as described in Table \ref{tab:atom_config2}: type II structures are favoured for (Co$_{2-\alpha}$Mn$_\alpha$)FeGe, whereas type I structures are favoured for Co$_2$(Fe$_{1-\beta}$Mn$_\beta$)Ge and (Co$_{2-\alpha}$Fe$_\alpha$)MnGe. The common feature is that Mn prefers to occupy the B$_T$ site in accordance with the 4/2 rule.

\textbf{Magnetic properties}: All the samples are observed to be soft ferromagnets, exhibiting very high saturation magnetisations and Curie temperatures. All schemes exhibit linear variation of M$_\text{s}$ and T$_\text{C}$ with N$_\text{v}$, and the moments also agree reasonably well with those expected from the Slater-Pauling rule, indicating possible half-metallic character. In table \ref{table:all_series}, one can observe alloys that have same N$_\text{v}$ exhibit nearly the same M$_\text{s}$ and T$_\text{C}$ independent of substitution schemes.

\textbf{Electronic properties}: The electronic properties, calculated for the most favourable structure, suggested upward shifts of bands after substitution. The details of the band shifting are found to depend on the type of substitution. In general, if the substitution is made on single sublattice as in Co$_2$(Fe$_{1-\beta}$Mn$_\beta$)Ge and (Co$_{2-\alpha}$Fe$_\alpha$)MnGe, only a few bands gain energy and are shifted. As a result a wider band gap, which continues to grow upon increasing the substituting element concentration, is achieved. On the other hand, in case of double sublattice substitution in (Co$_{2-\alpha}$Mn$_\alpha$)FeGe since there is a change in atomic orbitals in both sublattices, many bands gain or lose energy leading to narrower band gap. Interestingly, it is found that regardless of different sublattice substitution schemes, the onset of the half-metallic character is found at the same point N$_\text{v}=29.25$. It is, however, found that half-metallic character is lost a bit faster in case of double sublattice substitution. The range is wider and the half-metallic character more robust if the substitution is made on single sublattice. 

%%%%%%%%%%%%%%%
\begin{figure*}[h]
    \centering
    \includegraphics[width=1.5\columnwidth]{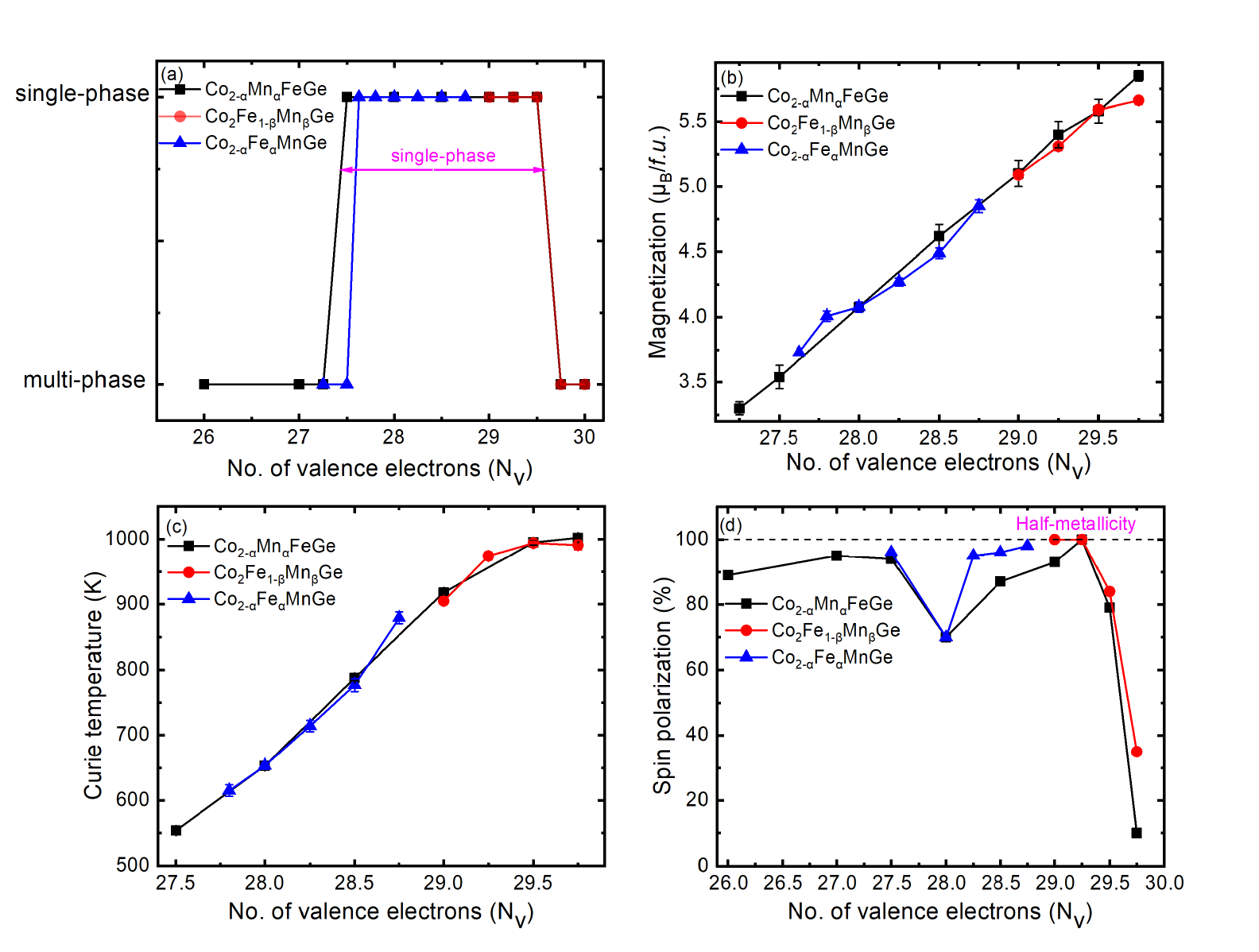}
    \caption{Variation of alloy properties; (a) experimental phase-stability, (b) experimental magnetisation, (c) experimental Curie temperature, and (d) calculated spin polarisation with number of valence electrons (N$_\text{v}$) for different substitution schemes. Interestingly the alloy properties are found to depend on N$_\text{v}$ rather than on substitution scheme.}
    \label{fig:three_together}
\end{figure*}
%%%%%%%%%%%%

The alloy properties discussed above and their dependence on the N$_\text{v}$ are apparent in figure \ref{fig:three_together}. As long as N$_\text{v}$ is same, alloys exhibit similar properties regardless of the scheme employed. This finding stresses the importance of N$_\text{v}$ in tailoring basic alloy properties, which likely stems from the 4-2 rule and helps simplify the rational design of functional materials. Several candidate alloys are discovered that have promise of exhibiting improved characteristics compared to existing compounds, namely, Co$_2$Fe$_{0.25}$Mn$_{0.75}$Ge for Co$_2$MnGe and Co$_{1.25}$Fe$_{0.75}$MnGe, Co$_{1.25}$Mn$_{0.75}$FeGe etc.\ for CoFeMnGe. Further exploration of these alloys could be an interesting subject of future work and may lead to the discovery of useful materials for spintronic applications.

The authors would like to warmly acknowledge helpful discussions and advice from Prof.\ I.\ Galanakis. This work utilised the facilities offered by the Alabama Analytical Research Center (AARC) of the University of Alabama. We are thus grateful to the members of AARC for helping us with measurements. The computational resources were provided by the MINT High Performance Computing Facility (MINT HPC). Many of the resources used in this work were originally provided through NSF DMREF Grant No.\ 1235396, and NSF DMR Grant No.\ 1508680. S.~KC thanks the Graduate Council at the University of Alabama for providing a fellowship while this work was completed.

\bibliographystyle{elsarticle-num}
\bibliography{myref}
\end{document}